\documentclass[a4paper]{report}
 
\usepackage{ifdraft}
\usepackage{amsmath}
\usepackage{amsfonts}
\usepackage{amssymb}
\usepackage{natbib}
\usepackage{rotating}
\usepackage{hyperref}  
\usepackage{appendix}
\usepackage{tipa}
\usepackage{setspace}
\usepackage[absolute]{textpos} 
\usepackage{booktabs}
\usepackage{algorithm}
\usepackage{algpseudocode}
\usepackage{enumitem}
\usepackage{graphicx}
\usepackage{xcolor}
\usepackage{subfigure}
\usepackage{longtable}
\usepackage{musicography}
\usepackage{multirow}
\usepackage{amssymb}
\usepackage{tablefootnote}

\newcommand{\highlight}[1]{%
  {#1}%
}

\begin{document}

\setlength{\TPHorizModule}{200mm} 
\setlength{\TPVertModule}{100mm} 
\textblockorigin{61mm}{19mm}


\onehalfspace{}


\providecommand{\OT}[1]{\operatorname{\Theta}\bigl(#1\bigr)}
\providecommand{\OOm}[1]{\operatorname{\Omega}\bigl(#1\bigr)}

\newcommand{\yixiao}[1]{\textcolor{blue}{\textit{#1}}}

\pagenumbering{arabic}

\title{Improving Controllability and Editability for Pretrained Text-to-Music Generation Models}
\author{Yixiao Zhang \\
\\
PhD thesis\\
\\
School of Electronic Engineering and Computer Science\\
Queen Mary University of London
}

\date{2024}

\maketitle

\chapter*{Statement of Originality}

I, Yixiao Zhang, confirm that the research included within this thesis is my own work or that where it has been carried out in collaboration with, or supported by others, that this is duly acknowledged below and my contribution indicated. Previously published material is also acknowledged below.

I attest that I have exercised reasonable care to ensure that the work is original, and does not to the best of my knowledge break any UK law, infringe any third party’s copyright or other Intellectual Property Right, or contain any confidential
material.

I accept that Queen Mary University of London has the right to use plagiarism detection software to check the electronic version of the thesis.

I confirm that this thesis has not been previously submitted for the award of a degree by this or any other university.

This work is copyright \copyright \ 2024 Yixiao Zhang, and is licensed under Creative Commons Attribution-Share Alike 4.0 Unported Licence. To view a copy of this licence, visit \\
\url{http://creativecommons.org/licenses/by-sa/4.0/} \\
or send a letter to Creative Commons, 171 Second Street, Suite 300, San Francisco, California, 94105, USA.

\vspace{2cm}

Signature: Yixiao Zhang

Date: 12 Oct 2024

\begin{abstract}

The field of AI-assisted music creation has made significant strides, yet existing systems often struggle to meet the demands of iterative and nuanced music production. These challenges include providing sufficient control over the generated content and allowing for flexible, precise edits. This thesis tackles these issues by introducing a series of advancements that progressively build upon each other, enhancing the controllability and editability of text-to-music generation models.

First, I introduce \textbf{Loop Copilot}, a system that tries to address the need for iterative refinement in music creation. Loop Copilot leverages a large language model (LLM) to coordinate multiple specialised AI models, enabling users to generate and refine music interactively through a conversational interface. Central to this system is the Global Attribute Table, which records and maintains key musical attributes throughout the iterative process, ensuring that modifications at any stage preserve the overall coherence of the music. While Loop Copilot excels in orchestrating the music creation process, it does not directly address the need for detailed edits to the generated content.

To overcome this limitation, \textbf{MusicMagus} is presented as a further solution for editing AI-generated music. MusicMagus introduces a zero-shot text-to-music editing approach that allows for the modification of specific musical attributes, such as genre, mood, and instrumentation, without the need for retraining. By manipulating the latent space within pre-trained diffusion models, MusicMagus ensures that these edits are stylistically coherent and that non-targeted attributes remain unchanged. This system is particularly effective in maintaining the structural integrity of the music during edits, but it encounters challenges with more complex and real-world audio scenarios.

Building on the advancements of the previous systems, \textbf{Instruct-MusicGen} tries to address the remaining limitations by incorporating instruction tuning into the MusicGen model. This approach enables precise and efficient editing of music through text-based instructions, such as adding, removing, or modifying specific musical stems. Instruct-MusicGen integrates a text fusion module and an audio fusion module, allowing the model to process instructions and audio inputs concurrently and produce high-quality edited music. This system not only achieves greater precision in edits but also broadens the applicability of music language models to more complex and dynamic production environments, offering a scalable and efficient solution.

Collectively, these contributions form a robust framework that significantly enhances the controllability and editability of AI systems in music production. By progressively addressing the limitations of each previous approach, this thesis advances the state of the art in AI-assisted music creation, enabling more flexible, precise, and dynamic music production processes.

\end{abstract}  

\setcounter{page}{3}

\chapter*{Acknowledgements}




This research would not have been possible without the help and support of many people~\footnote{This work was supported jointly by the China Scholarship Council, Queen Mary University of London, and Apple Inc.}. First, I express my deepest gratitude to my doctoral advisor, \textbf{Professor Simon Dixon}. Prof. Dixon is, in my view, one of the world's best supervisors. Although I often had a myriad of novel ideas, I lacked the experience to develop them into well-thought-out experiments and eventually into a qualified paper. He has always been supportive and understanding of my various ideas, consistently encouraging me to pursue them. He patiently guided me through my mistakes and ensured that I remained confident despite my inexperience during my PhD journey.

I am equally grateful to my PhD co-supervisor, \textbf{Dr. Mark Levy}. It has been a privilege to engage in insightful technical discussions with him on a regular basis. His intelligence and extensive experience enables him to quickly understand and effectively refine my ideas. He exemplifies the qualities I believe a top-tier industry researcher should possess and serves as an inspiring role model for my aspirations and future career in the field.

I am indebted to \textbf{Dr. Gus Xia}, who has been a constant collaborator and mentor throughout my PhD journey. A visionary scholar with an excellent taste for research, he introduced me to the field of Music AI, guided my first publication, and actively contributed to each subsequent paper. I will never forget the summer of 2019 spent in Shanghai, where we passionately discussed research ideas on whiteboards. That period marked the beginning of a completely different life path for me.

I would like to thank everyone at C4DM. The discussions on various MIR topics with professors and students often sparked new ideas. Special thanks go to \textbf{Lele Liu}, \textbf{Jiawen Huang}, \textbf{Yazhou Li}, \textbf{Yin-Jyun Luo}, \textbf{Ningzhi Wang}, \textbf{Chin-Yun Yu}, \textbf{Huan Zhang}, and \textbf{Yinghao Ma}, who were my warmest companions during this journey. The strong alumni network of C4DM also played a significant role in my internships and career development.

My appreciation extends to all my friends at Music X Lab. From Shanghai to Abu Dhabi, I can vividly recall every interaction. This young team is full of creativity and passion for research, and everyone has a genuine enthusiasm for their work, which has significantly shaped my perspective on academic research. I especially thank \textbf{Ziyu Wang}, \textbf{Junyan Jiang}, and \textbf{Liwei Lin}, who provided me with invaluable assistance and unwavering support when I felt lost, encouraging me to persevere through challenges from both work and life.

I am grateful to my mentor, \textbf{Dr. Akira Maezawa}, at Yamaha R\&D, along with \textbf{Dr. Kazuhiko Yamamoto} for his research advice, and to \textbf{Akane Noguchi}, \textbf{Aya Ogawa}, and other colleagues who offered their support. My three months in Hamamatsu were among the happiest times of my research, where I felt like part of a warm family, receiving ample care and assistance from highly skilled music experts who helped me realise my research concepts.

I am thankful to my mentor, \textbf{Yukara Ikemiya}, at Sony AI, and to \textbf{Naoki Murata}, \textbf{Dr. Marco Martínez}, \textbf{Dr. Wei-Hsiang Liao}, \textbf{Dr. Woosung Choi}, and \textbf{Dr. Yuki Mitsufuji}, who provided guidance during my internship. Sony AI is a world-class laboratory in Music AI, and my time in Tokyo was a period of rapid learning and growth, striving to become an outstanding researcher. I will never forget the enthusiastic support from everyone there.

I appreciate my mentor, \textbf{Dr. Jordi Pons}, and \textbf{Dr. Julian Parker} at Stability AI. They are exceptional scientists with extensive experience in audio processing. I thank them for their tolerance of my mistakes, their patient discussions about the experimental results, and their generous provision of research advice.

I thank \textbf{Prof. Robin Laney} and \textbf{Prof. Zhiyao Duan} for taking the time to examine my thesis and for their invaluable feedback and insightful suggestions, which have greatly enhanced the quality of the thesis.

My heartfelt thanks go to my father, \textbf{Mingshi Zhang}, my mother, \textbf{Zhaohui Xiao}, and my grandmother, \textbf{Jurong Zeng}, for their unwavering support and understanding of my decision to pursue a PhD. Their encouragement has been a cornerstone of my achievements, providing a strong sense of security and comfort throughout this journey.

I am grateful to friends around the world who have offered their assistance during my PhD life. Special thanks to \textbf{Dr. Xinhai Xie} for being a companion in my life, and to \textbf{Jiawei Wang} for our soulful mutual understanding. You both are incredibly important to me. I also thank \textbf{Yongyi Zang}, who has been the longest-standing colleague and friend in my research career, and \textbf{Sensen Gao}, \textbf{Kai Fu}, \textbf{Yuyang Li}, and \textbf{Yuhang Huang}, with whom I have shared countless joyful moments.

Finally, I want to thank my partner, \textbf{Liming Kuang}. Over the eight years we have known each other, he has been the most understanding and supportive person in my life. In my heart, he is always that cheerful and big-hearted boy. During these times, we have often been apart, and every meeting was accompanied by a farewell with tears.

\begin{center}
    \textit{``Un jour, je serai de retour près de toi."}
    \vspace{0.1cm}
    
    \includegraphics[width=0.4\linewidth]{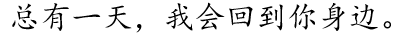}
\end{center}

\tableofcontents
\listoffigures
\listoftables

\chapter*{List of Abbreviations}

\begin{table}[htbp]
\begin{center}
\begin{tabular}{ll}

CLAP   & Contrastive Language-Audio Pretraining     \\
VQ-VAE & Vector Quantized Variational Autoencoder   \\
RVQ    & Residual Vector Quantization               \\
DDPM   & Denoising Diffusion Probabilistic Model    \\
DDIM   & Denoising Diffusion Implicit Models        \\
FAD    & Fréchet Audio Distance                     \\
PEFT   & Parameter-Efficient Fine-Tuning            \\
LoRA   & Low-Rank Adaptation                        \\
BERT   & Bidirectional Encoder Representations from Transformers \\
SFT    & Supervised Fine-Tuning                     \\
ODE    & Ordinary Differential Equation             \\
SDE    & Stochastic Differential Equation           \\
VAE    & Variational Autoencoder                    \\
DiT    & Diffusion Transformer                      \\
LLM    & Large Language Model                       \\
MLP  & Multi-Layer Perceptron\\
LoA & Language of Audio \\
CFG & Classifier-free Guidance\\
GAN & Generative Adversarial Network\\
TTA & Text-to-Audio \\
TTM & Text-to-Music \\

\end{tabular}
\end{center}
\end{table}%

\chapter{Introduction}
\label{ch:intro}

\section{Motivation}

The convergence of artificial intelligence (AI) and music creation is reshaping the landscape of artistic expression and production workflows. Over the past decade, advances in \textbf{AI-generated content (AIGC)}, driven by techniques like deep learning and large language models (LLMs), have enabled AI systems to compose, arrange, and even improvise music in ways that come close to human creativity. These systems can now generate full-length compositions~\citep{sd2}, assist in songwriting~\citep{songcomposer, singsong}, and offer personalised recommendations for sound design~\citep{iteratta}, marking a shift in how music is created and consumed.

Recent breakthroughs, such as \textbf{Stable Audio}~\citep{sd1, sd2}, \textbf{MusicLM}~\citep{musiclm}, and \textbf{MusicGen}~\citep{musicgen}, have demonstrated the power of large-scale music foundation models to generate high-quality music pieces. While these systems have made significant strides in enabling users with minimal formal musical training to generate complete music compositions, they are still in the early stages of being adopted by professionals. Several critical challenges remain, particularly when it comes to meeting the nuanced and iterative requirements of professional music production. While AI-generated music tools have made music creation more accessible and efficient, they fall short in key areas:

\begin{itemize}
    \item \textbf{Controllability}: Current AI music systems can generate high-quality music but often lack the fine-grained control required by musicians and producers. Professional music production demands precision, where users must control and modify specific musical elements such as rhythm, harmony, instrumentation, and emotional tone. However, AI systems frequently produce music that is difficult to adjust on a detailed level, making it hard to tailor the output to the artist's vision.
    \item \textbf{Editability}: Music production is an \textit{iterative process}. Artists seldom create a final product in a single step; they refine and rework their compositions over multiple iterations. However, most AI tools treat music generation as a one-off process, where users have little or no capacity to modify the generated content without starting from scratch. This lack of editability limits the practicality of these tools in real-world production environments, where iterative refinements are the norm.
    \item \textbf{Intuitive Interaction}: Although AI tools have advanced considerably, many still require users to engage with complex interfaces or possess technical knowledge, such as understanding model parameters or programming languages. For musicians focused on creativity, such technical barriers can be a major deterrent, preventing widespread adoption of AI technologies in music production. An intuitive, natural interface that allows musicians to interact with AI systems conversationally—rather than through technical inputs—could unlock the true potential of these tools.
\end{itemize}

These limitations underscore a gap between the current capabilities of AI music systems and the demands of musicians. As the boundaries of AI-generated content continue to expand, addressing these shortcomings is essential not only for increasing the accessibility of AI tools but also for ensuring that AI systems truly augment human creativity without compromising artistic control.

In light of these challenges, the motivation for this thesis is to explore how AI systems can be enhanced to offer more \textit{control}, \textit{editability}, and \textit{user-friendly interactions} in music creation. This research seeks to bridge the gap between current AI capabilities and the needs of professional and amateur musicians by proposing a series of solutions that push AI music systems closer to real-world applicability, ultimately empowering artists to harness AI as a collaborative creative tool.

\section{Aim}

The primary aim of this thesis is to develop AI-assisted music systems that overcome the limitations identified in \textit{controllability}, \textit{editability}, and \textit{user interaction}. These systems are intended to bridge the gap between the existing capabilities of AI-generated content tools and the practical needs of music production.

The overarching objective is to design systems that allow musicians and producers to interact with AI in a more \textit{intuitive} and \textit{seamless} manner, while retaining control over the \textit{fine-grained details} of their compositions. This research aims to extend the creative possibilities of AI, not just by generating new music, but by empowering users to \textit{refine}, \textit{edit}, and \textit{direct} the output in ways that align with their artistic intent. Specifically, this thesis investigates how AI systems can be structured to support \textit{iterative editing} and \textit{natural language control} within the music creation process, paving the way for more versatile and collaborative AI music tools.

\section{Research Questions}

To achieve this aim, the following key research questions will be addressed:

\subsection*{Research Question 1: How can we design \highlight{text-to-music models} that offer musicians fine-grained control over various musical elements during the creative process?}

\highlight{Current text-to-music (TTM) systems} often \highlight{lack} the ability to control specific musical attributes, such as \textit{rhythm}, \textit{harmony}, or \textit{instrumentation}, with the precision required by musicians. This research question explores how models can be designed to allow users to control and adjust these musical elements in a way that integrates naturally into their creative workflows. By embedding controllability into AI music models, the goal is to provide users with \textit{direct manipulation} of individual components, facilitating the creation of music that matches their vision.

\subsection*{Research Question 2: How can AI systems support iterative and flexible editing of AI-generated music, allowing musicians to refine their compositions seamlessly?}

Human music creation is inherently iterative, requiring continuous adjustments and refinements. This question examines how AI models can support \textit{iterative workflows}, where musicians can easily modify generated content without starting from scratch. By enabling flexible editing, the research seeks to develop systems that preserve the \textit{structural integrity} of the music while allowing specific attributes to be reworked in response to user feedback.

\subsection*{Research Question 3: How can AI music systems be made more intuitive, allowing users to interact with these technologies in a more natural manner?}

The technical complexity of current AI music tools often creates a barrier to their adoption. This question focuses on designing systems that allow users to interact with AI music tools in more \textit{natural and intuitive} ways. Possible directions include the use of \textit{conversational interfaces} or other simplified control mechanisms.  The goal is to lower the technical barrier for musicians, enabling more \textit{synergistic collaboration} between human creativity and AI-generated content. 

\section{Thesis Structure}

The thesis is organised as follows:

\begin{description}
\item[Chapter \ref{ch:intro}] Introduction

This chapter provides the motivation behind the research, outlines the aims and objectives of the thesis, and presents an overview of the structure and contributions. It sets the stage for the subsequent chapters by highlighting the challenges in current AI-assisted music creation systems and introducing the proposed solutions.

\item[Chapter \ref{ch:background}] Background

This chapter offers a comprehensive review of the existing literature and technologies in AI-assisted music creation. It covers fundamental concepts in music generation, text-to-music synthesis, and controllable music generation. The chapter also introduces key technologies underlying the work, such as large language models, diffusion models, and music language models. It concludes by identifying gaps in current research, providing context for the contributions.

\item[Chapter \ref{ch:loop}] Loop Copilot: Conducting AI Ensembles for Music Generation and Iterative Editing

This chapter introduces Loop Copilot, the first major contribution. It details the system's architecture, including the integration of multiple specialised AI models coordinated by a large language model. The chapter explains the concept of the Global Attribute Table and its role in maintaining musical coherence during iterative editing. It also presents a user study evaluating the system's effectiveness and limitations in real-world scenarios.

\item[Chapter \ref{ch:musicmagus}] MusicMagus: Zero-Shot Text-to-Music Editing via Diffusion Models

This chapter presents MusicMagus, the second major contribution. It explores the use of diffusion models for zero-shot text-guided music editing. The chapter details the methodology for manipulating the latent space to achieve specific musical attribute modifications while preserving overall structure. It discusses the system's strengths in music editing and its limitations with more complex operations.

\item[Chapter \ref{ch:instruct}] Instruct-MusicGen: Unlocking Text-to-Music Editing for Music Language Models via Instruction Tuning

This chapter introduces Instruct-MusicGen, the most advanced contribution. It describes the integration of instruction tuning with music language models to enable precise, text-guided editing of musical content. The chapter describes how the system combines both text and audio inputs to carry out complex edits. It also discusses the system's capabilities in handling a wide range of editing tasks and its potential impact on professional music production workflows.

\item[Chapter \ref{ch:conclusions}] Conclusions and Future Work

The final chapter summarises the key findings and contributions of the thesis. It reflects on how the presented work advances the field of AI-assisted music creation and discusses its potential impact on the music industry. The chapter also identifies remaining challenges and outlines promising directions for future research, including potential improvements to the proposed systems and new avenues for exploration in AI-music interaction.

\end{description}

\section{Contributions}

The contributions of this thesis represent significant advancements in the field of AI-assisted music creation, particularly in enhancing the controllability and editability of text-to-music generation systems. This thesis introduces novel approaches that progressively address the limitations of existing systems, culminating in a framework for flexible and precise music editing. The main contributions of this thesis are:

\begin{enumerate}
    \item \textbf{Loop Copilot: An LLM-Orchestrated Multi-Model System for Iterative Music Creation}
    \begin{itemize}
        \item Developed a system that leverages a large language model to coordinate multiple specialised AI music models.
        \item Introduced the Global Attribute Table (GAT) concept for maintaining musical coherence during iterative editing.
        \item Demonstrated the feasibility of using natural language interactions for complex music generation and editing tasks.
        \item Conducted a user study to evaluate the system's effectiveness in real-world music production scenarios.
    \end{itemize}
    
    \item \textbf{MusicMagus: Zero-Shot Text-to-Music Editing via Diffusion Models}
    \begin{itemize}
        \item Pioneered a novel approach for zero-shot text-guided music editing using pre-trained diffusion models.
        \item Developed techniques for manipulating the latent space of diffusion models to achieve specific musical attribute modifications.
        \item Demonstrated the ability to perform stylistically coherent edits without the need for additional training.
        \item Addressed limitations in stem level editing and preserving non-targeted musical elements during edits.
    \end{itemize}

    \item \textbf{Instruct-MusicGen: Text-to-Music Editing for Music Language Models via Instruction Tuning}
    \begin{itemize}
        \item Applied instruction tuning to the MusicGen model to enable precise and efficient editing through text-based instructions.
        \item Developed a dual-modality fusion process for concurrent processing of text instructions and audio inputs.
        \item Expanded the capabilities of music language models to handle complex editing tasks, including adding, removing, or modifying specific musical stems.
        \item Demonstrated the scalability and efficiency of the system in handling a wide range of music production scenarios.
    \end{itemize}
\end{enumerate}

\clearpage
\section{Associated publications}
\label{sec:publications}

The work presented in this thesis has been shared at international peer-reviewed conferences and workshops, published as preprints, or is currently under review. The following list categorizes all publications associated with this thesis by the author's contribution and the peer-review status.

\subsection{Peer-reviewed publications (First author)}

\begin{enumerate}

\item  \textbf{Yixiao Zhang}, Gus Xia, Mark Levy, and Simon Dixon, ``\textbf{COSMIC: A Conversational Interface for Human-AI Music Co-Creation}", in Proceedings of the 21st International Conference on New Interfaces for Musical Expression (NIME 2021), Shanghai, China, June 14--18, 2021 (\citep{cosmic}). 

In this work, I developed the system and led the design and evaluation of the experiments. Mark Levy and Simon Dixon, as my PhD supervisors, provided guidance on research design and manuscript preparation. Gus Xia contributed to the algorithm development and provided valuable feedback on the experimental setup.

\item  \textbf{Yixiao Zhang}, Yukara Ikemiya, Gus Xia, Naoki Murata, Marco A Martínez-Ramírez, Wei-Hsiang Liao, Yuki Mitsufuji, and Simon Dixon, ``\textbf{MusicMagus: Zero-Shot Text-to-Music Editing via Diffusion Models}", in Proceedings of the 33rd International Joint Conference on Artificial Intelligence (IJCAI 2024) - AI, Arts \& Creativity Track, Jeju, South Korea, 2024 (\citep{musicmagus}). 

I played a leading role in this project during my internship at Sony, focusing on system design and experimentation. Gus Xia and Simon Dixon provided substantial theoretical guidance and manuscript reviews. Yukara Ikemiya and Yuki Mitsufuji, from Sony, contributed to discussions and implementation details, while Naoki Murata, Marco A Martínez-Ramírez, and Wei-Hsiang Liao assisted with data processing and system evaluation.

\end{enumerate}

\subsection{Preprints and papers under review (First author)}

\begin{enumerate}[resume]

\item  \textbf{Yixiao Zhang}, Akira Maezawa, Gus Xia, Kazuhiko Yamamoto, and Simon Dixon, ``\textbf{Loop Copilot: Conducting AI Ensembles for Music Generation and Iterative Editing}", arXiv preprint arXiv: 2310.12404, 2023 (\citep{loopcopilot}). 

This work builds on my internship at Yamaha, where I was the lead contributor, developing the core system and conducting experiments. Akira Maezawa provided key insights during idea generation, experiment setup, and manuscript preparation. Gus Xia acted as an external collaborator and provided theoretical guidance. Kazuhiko Yamamoto assisted with experiments and collaborative efforts. Simon Dixon offered overall guidance and feedback on the project direction.

\item  \textbf{Yixiao Zhang}, Yukara Ikemiya, Woosung Choi, Naoki Murata, Marco A Martínez-Ramírez, Liwei Lin, Gus Xia, Wei-Hsiang Liao, Yuki Mitsufuji, and Simon Dixon, ``\textbf{Instruct-MusicGen: Unlocking Text-to-Music Editing for Music Language Models via Instruction Tuning}", arXiv preprint arXiv:2405.18386, 2024 (\citep{instructmusicgen}). 

This work represents the core of my research during my internship at Sony, where I took the lead in system design, experimentation, and writing. Simon Dixon and Gus Xia contributed to research direction and theoretical framing, while Liwei Lin joined in the manuscript preparation, leveraging their expertise from earlier works on content-based controls. Yukara Ikemiya, Yuki Mitsufuji, and other co-authors provided crucial support during the research meetings and discussions.

\end{enumerate}

\subsection{Peer-reviewed publications (Contributing author)}

\begin{enumerate}[resume]

\item  Liwei Lin, Gus Xia, Junyan Jiang, and \textbf{Yixiao Zhang}, ``\textbf{Content-based controls for music large language modeling}", in Proceedings of the 25th International Society for Music Information Retrieval Conference (ISMIR 2024), San Francisco, CA, USA, November 10--14, 2024 (\citep{cocomulla}). 

My contributions included participation in experimental design and discussions, data analysis, and assisting with the writing of the manuscript. I played a collaborative role in refining the concept of content-based controls for large music language models.

\item  Liwei Lin, Gus Xia, \textbf{Yixiao Zhang}, and Junyan Jiang, ``\textbf{Arrange, Inpaint, and Refine: Steerable Long-term Music Audio Generation and Editing via Content-based Controls}", in Proceedings of the 33rd International Joint Conference on Artificial Intelligence (IJCAI 2024) - AI, Arts \& Creativity Track, Jeju, South Korea, 2024 \citep{airgen}. 

This paper represents a follow-up to [5], where I contributed significantly to the conceptualization of experiments, offered critical revisions to the manuscript, and assisted in data analysis and discussion.

\end{enumerate}

Chapter 3 includes content from [1] and [3]. [1] introduced COSMIC, a conversational interface for human-AI music co-creation, where I was the main contributor, leading system development and experimentation, with Mark Levy and Simon Dixon offering research design and writing support, and Gus Xia assisting with the algorithm development. [3] presented Loop Copilot, a system for conducting AI ensembles for music generation and iterative editing, which was developed during my internship at Yamaha. I was the primary contributor, with Akira Maezawa providing support in idea generation and writing, Gus Xia offering theoretical guidance, and Simon Dixon giving feedback on project direction.

Chapter 4 is based on [2], which introduced MusicMagus, a system for zero-shot text-to-music editing via diffusion models. This work was conducted during my internship at Sony, where I was the lead researcher, developing the system and conducting experiments. Yukara Ikemiya, Yuki Mitsufuji, and other co-authors at Sony contributed to discussions and provided practical implementation support. Simon Dixon and Gus Xia played supervisory roles, giving guidance on theoretical aspects and the manuscript.

Chapter 5 includes contents from [4] and [5]. [5] and [6] are precursor works on content-based controls for music language modeling and long-term music audio generation and editing, respectively. While I was not the first author of these works, I contributed to idea discussions, data analysis, and writing support. [4] forms the core of this chapter, presenting Instruct-MusicGen for text-to-music editing, where I was the lead contributor, responsible for system design, experimentation, and manuscript preparation, with strong support from co-authors during my internship at Sony.

\chapter{Background}
\label{ch:background}

In this chapter, I explore the fundamental concepts and state-of-the-art methods central to this thesis, focusing on text-to-music backbone models and techniques for enhancing their controllability and editability. I begin by examining the basic knowledge underlying music and text representations, including symbolic, audio, and time-frequency representations, as well as modern music encoders and cross-modal representations.

I then delve into two main categories of text-to-music backbone models: diffusion models and audio language models. For diffusion models, I discuss Denoising Diffusion Probabilistic Models (DDPMs), Latent Diffusion Models (LDMs), and related innovations. In the realm of audio language modelling, I explore Vector Quantised Variational Autoencoders (VQ-VAEs), Residual Vector Quantisation (RVQ), and autoregressive models, with a particular focus on MusicGen and its architecture, training, and performance.

A comparison between diffusion models and audio language models is presented, highlighting the strengths and applications of each approach. I also introduce Parameter-Efficient Finetuning (PEFT) techniques, such as Low-Rank Adaptation (LoRA) and Llama-Adapter, which play a crucial role in adapting pretrained models for specific tasks.

Finally, I address the aspect of enriching controllability and editability for music generation models, which is critical to this thesis. This includes various approaches to enhancing control, specific techniques used in music generative models, and post-training generation models with control mechanisms. I also explore specialised editing models and agent-based methods for compositional music generation and editing.

This comprehensive background provides the foundation for understanding the advancements and innovations presented in subsequent chapters, setting the stage for our exploration of novel methodologies in text-to-music generation and control.

\section{Basic Knowledge}

 Imagine I have a music audio track and its corresponding text description. Given a text, $x$: ``A romantic pop song with piano and guitar accompaniment." The corresponding music is represented in waveform as $y$. Before diving into formal definitions, I will use this example to illustrate related terms and concepts.

\subsection{Music and Text Representations}

The processing of text and music audio presents unique challenges due to their inherently unstructured nature. To enable effective analysis and generation, it is usually necessary to transform both text and music into structured data representations, often in the form of high-dimensional vectors or matrices (some models are capable of operating directly on raw audio data). These representations serve as the foundation for various downstream tasks, including natural language processing, music information retrieval, and cross-modal learning.

\subsubsection{Text Representations}

At the most fundamental level, text data is stored as string objects within computational systems. However, the low-level nature of strings makes them unsuitable for the numerical operations required by machine learning algorithms, so they must first be transformed into more abstract, higher-level representations. Consequently, a series of transformations must be applied to convert raw text into a format amenable to computational analysis.

The initial step in this process is tokenisation, a linguistic technique that decomposes a continuous string of text into discrete units called tokens. These tokens may represent individual words, subwords, or characters, depending on the chosen tokenisation strategy. Subsequently, each token is mapped to a numerical index based on a predefined vocabulary, effectively translating the text into a sequence of integers.

Given the high-dimensional and sparse nature of natural language, it is advantageous to further transform these token sequences into dense vector representations. This process, known as encoding, maps the discrete tokens into a continuous high-dimensional space. Formally, this encoding can be expressed as a function $f: x \mapsto z_x$, where $x$ represents the original text and $z_x$ denotes its encoded representation in a high-dimensional vector space.

The encoding process is designed to capture the semantic relationships and contextual information present in natural language. By projecting text into a high-dimensional space, these encodings enable machine learning models to discern subtle linguistic patterns and relationships that may not be immediately apparent in the original text.

Several text encoders have been developed to perform this crucial transformation. Among the most prominent are T5~\citep{t5}, BERT~\citep{bert}, and RoBERTa~\citep{roberta}, each employing distinct architectural and training paradigms to generate rich, contextual representations of text:

\begin{enumerate}
    \item \textbf{T5 (Text-To-Text Transfer Transformer)~\citep{t5}:} T5 represents a paradigm shift in natural language processing by recasting all NLP tasks within a unified text-to-text framework. This approach enables T5 to address a diverse array of linguistic challenges, including but not limited to machine translation, text summarisation, and question answering. The architecture of T5 is predicated on a sequence-to-sequence model, which is trained to transmute input text into output text through a series of encoder and decoder transformations. This universal approach allows T5 to capture a wide spectrum of linguistic features and inter-textual relationships, making it exceptionally versatile for various NLP applications.

    \item \textbf{BERT (Bidirectional Encoder Representations from Transformers)~\citep{bert}:} BERT's innovation lies in its bidirectional training regime, which enables the model to comprehend the context of a word by considering both its preceding and subsequent context simultaneously. This bidirectional approach represents a significant advancement over unidirectional language models. BERT's architecture is based on the transformer model, which utilises self-attention mechanisms to weigh the importance of different words in a sequence when computing representations. The result is a set of deep, contextual embeddings for each word, which can be fine-tuned for a multitude of downstream NLP tasks, including sentiment analysis, named entity recognition, and text classification.
    
    \item \textbf{RoBERTa (A Robustly Optimised BERT Pretraining Approach) \citep{roberta}:} RoBERTa builds upon the foundational architecture of BERT while introducing several key optimisations to enhance its performance and robustness. These enhancements include training on significantly larger datasets, removing the next sentence prediction objective, and using larger batch sizes. Additionally, RoBERTa utilises longer sequences during training, allowing it to capture long-range dependencies more effectively. These modifications collectively result in a more powerful and versatile language model, capable of producing highly nuanced and context-aware representations.

\end{enumerate}

These encoders are part of a broader evolution in text embedding methods, which has progressed from static embeddings, such as GloVe~\citep{glove} and ELMo~~\citep{elmo}, to dynamic contextualised embeddings introduced by models like BERT. As the scale of these models has increased, so too has their ability to capture complex patterns in text, although this comes with higher computational demands. In recent years, large language models (LLMs) have shifted towards decoder-only architectures, such as GPT. Despite this trend, in domains like music text analysis, models from the second generation of encoders—T5, BERT, RoBERTa—continue to play a central role, since they keep a balance in both effectiveness and performance within the context of music-text multimodal learning.

These advanced text encoders transform raw text into dense vector representations that encapsulate semantic meanings, syntactic structures, and complex inter-word relationships. In the context of music generation, these sophisticated text representations can be leveraged to condition generative models, guiding the music creation process to align with specific textual descriptions or semantic concepts.

\subsubsection{Music Representations}

The representation of music for machine learning applications presents unique challenges due to its multi-faceted nature, encompassing temporal, spectral, and structural dimensions. Music representations can be broadly categorised into two main classes: symbolic representations and audio representations. Each of these categories offers distinct advantages and is suited to different types of musical analysis and generation tasks.

\paragraph{Symbolic Representations}

Symbolic music representations encode musical information in a discrete, event-based format, abstracting away from the continuous nature of audio signals. These representations typically capture high-level musical concepts such as notes, chords, rhythm, and dynamics. The most prevalent format for symbolic music representation is the Musical Instrument Digital Interface (MIDI) protocol, which encodes musical events as a series of messages specifying pitches, note onsets, durations (by calculating time deltas between note on and off messages), velocities, and other performance parameters.

Symbolic music representation has evolved significantly over time, starting with early formats like piano rolls, MusicXML, MIDI, and ABC notation. While these formats provided a foundation for digital music encoding, they also revealed limitations such as sparsity and difficulty in capturing nuanced musical details. Much like text tokenisation, where natural language is represented as sequences of discrete tokens, symbolic music representations face the challenge of converting non-sequential musical elements into serial forms that can be efficiently modelled. However, unlike language, where word order plays a dominant role, music inherently involves multiple, simultaneous streams of information (melody, harmony, rhythm), making its serialisation more complex and less straightforward.

\paragraph{Symbolic Domain Music Encoders}

Early symbolic representations like piano rolls and MIDI often suffer from sparsity and quantisation issues, particularly when it comes to representing expressive performances. Recent research has sought to address these limitations by developing more sophisticated encoding schemes that aim for tighter, more structured representations of music. For example, models such as PianoTree VAE \citep{pianotreevae} attempt to model music on a score level, focusing on hierarchical structures that reflect music's inherent nesting of notes within chords and phrases. On the other hand, performance-based models like REMI~\citep{remi} and CWT~\citep{cwt} focus on capturing the nuances of performance, introducing tokens that encode rhythmic, harmonic, and temporal information.

Despite these advancements, symbolic music tokenisation still faces inherent trade-offs. REMI, for example, offers a more compact and structured encoding compared to MIDI, but it relies on fixed temporal grids, which can result in a loss of expressive timing and rhythmic precision. This quantisation problem, as noted by PerTok~\citep{pertok} and other models, limits the ability to capture subtle performance details such as rubato and swing, which are crucial for realistic music generation. Tokenizers like CWT and PerTok have emerged in response to this challenge, aiming to balance compactness with performance fidelity by modeling both local and global musical structures.

While symbolic representations remain competitive in music processing tasks due to their compactness, editability, and interpretability, they have not dominated recent advances in music-text multimodal models. A key reason for this is the relative scarcity of symbolic music data compared to audio data. Audio-based approaches have benefited from the sheer volume of available data, which has driven recent breakthroughs in multimodal systems. This inherent limitation poses a challenge for the continued development of symbolic representations in the context of large-scale, data-driven machine learning models. Nevertheless, symbolic encoding remains a critical area of research for its advantages in specific tasks such as music generation, analysis, and manipulation.

\paragraph{Audio Representations}

Audio representations deal directly with the continuous, time-varying pressure waves that constitute musical sounds. While raw audio waveforms contain the full fidelity of a musical signal, their length and complex temporal structure make them challenging to process directly with many machine learning algorithms. Consequently, various transformations and feature extraction techniques are employed to convert raw audio into more tractable representations that capture salient musical characteristics.

Time-frequency representations aim to capture both the temporal evolution and spectral content of audio signals, providing a rich description of musical timbre, harmony, and rhythm.

\begin{enumerate}
    \item \textbf{Spectrogram:} A spectrogram is a visual representation of the spectrum of frequencies in a sound or other signal as they vary with time. It is typically computed using the Short-Time Fourier Transform (STFT), which applies the Fourier transform to windowed segments of the audio signal. The resulting spectrogram $S(t,f)$ represents the magnitude of the STFT, where $t$ denotes time and $f$ denotes frequency.
    
    Mathematically, the STFT can be expressed as:
    
    \begin{equation}
    STFT\{x(t)\}(\tau, \omega) = \int_{-\infty}^{\infty} x(t)w(t-\tau)e^{-j\omega t}dt,
    \end{equation}
    
    \noindent where $x(t)$ is the input signal, $w(t)$ is the window function, $\tau$ is the time shift, and $\omega$ is the angular frequency.
    
    Spectrograms offer a detailed view of the frequency content of a signal over time, making them useful for tasks such as instrument recognition, onset detection, and audio source separation.

    \item \textbf{Mel-spectrogram:} A mel-spectrogram is a variant of the spectrogram that applies a mel-scale filterbank to the power spectrum. The mel scale is a perceptual scale of pitches judged by listeners to be equal in distance from one another. This transformation aligns the frequency representation more closely with human auditory perception.
    
    The conversion from Hz to mel can be approximated by:
    
    \begin{equation}
    m = 2595 \log_{10}(1 + \frac{f}{700}),
    \end{equation}
    
    \noindent where $f$ is the frequency in Hz and $m$ is the corresponding mel value.
    
    Mel-spectrograms are widely used in music information retrieval tasks due to their ability to capture perceptually relevant features of music while reducing the dimensionality of the representation.

    \item \textbf{Mel-frequency Cepstral Coefficients (MFCCs):} MFCCs are a compact representation of the short-term power spectrum of a sound, based on a linear cosine transform of a log power spectrum on a nonlinear mel scale of frequency. The process of computing MFCCs involves:
    
    \begin{itemize}
        \item Computing the power spectrum of windowed frames of the audio signal.
        \item Applying a mel-scale filterbank to the power spectra.
        \item Taking the logarithm of the filterbank energies.
        \item Computing the discrete cosine transform (DCT) of the log filterbank energies.
    \end{itemize}
    
    Mathematically, the MFCCs can be expressed as:
    
    \begin{equation}
    MFCC[n] = \sum_{k=1}^{K} \log(S[k]) \cos[n(k-0.5)\frac{\pi}{K}],
    \end{equation}
    
    \noindent where $S[k]$ represents the log-energy output of the $k$-th filter in the mel-scale filterbank, and $K$ is the total number of filters.
    
    MFCCs are particularly effective at capturing the timbral characteristics of sound and have been widely used in speech recognition, music genre classification, and instrument identification tasks.
\end{enumerate}

\paragraph{Audio Domain Music Encoders}

From an informatics perspective, audio encoders efficiently compress audio data into rich, task-agnostic representations, enabling powerful downstream music analysis and generation tasks. Earlier models such as Variational Autoencoders (VAEs)~\citep{vae} and WaveNet~\citep{wavenet} laid the groundwork for compact representations, while large-scale pretrained models like Jukebox~\citep{jukebox} have demonstrated the ability to retain domain-specific knowledge. When probed using a simple two-layer MLP, these models reveal that their large-scale training allows them to capture a wealth of useful information for various tasks.

Audio encoders have developed along two main lines: one leverages prior knowledge of musical theory, while the other focuses on scaling up model parameters to gain advantages in performance. The latter approach, which relies on sheer model size and training data, has recently achieved greater success. The following models represent the latest advances in this field, each contributing to audio representation learning through different architectures and techniques.

\begin{enumerate}
    \item \textbf{HTS-AT (Hierarchical Token-Semantic Audio Transformer) \citep{htsat}:} HTS-AT is a transformer-based model designed specifically for sound classification and detection. Key features of HTS-AT include:
    \begin{itemize}
        \item A hierarchical structure to reduce model size and training time.
        \item A token-semantic module for mapping final outputs into class feature maps, enabling audio event detection and localisation.
    \end{itemize}
    HTS-AT achieves state-of-the-art results on AudioSet~\citep{audioset}, ESC50~\citep{esc50}, and Speech Command V2 datasets~\citep{speechcommandsv2}, demonstrating high performance and efficiency.

    \item \textbf{PaSST (Patchout Fast Spectrogram Transformer)~\citep{passt}:} PaSST applies a novel training method to transformers on audio spectrograms. The key innovations of PaSST include:
    \begin{itemize}
        \item Patch extraction and linear projection to form a sequence for the transformer.
        \item Patchout, a method to reduce computation and memory complexity during training, also functioning as a regulariser.
    \end{itemize}
    PaSST achieves state-of-the-art performance on Audioset with efficient training on a single consumer-grade GPU.

    \item \textbf{MERT (Music undERstanding model with large-scale self-supervised Training)~\citep{mert}:} MERT is a model that incorporates teacher models to provide pseudo labels in a masked language modelling style acoustic pre-training. Key aspects of MERT include:
    \begin{itemize}
        \item A combination of an acoustic teacher based on Residual Vector Quantisation Variational AutoEncoder (RVQ-VAE)~\citep{rvq} and a musical teacher based on the ConstantQ Transform (CQT).
        \item A multi-task paradigm balancing acoustic and musical representation learning.
    \end{itemize}
    MERT demonstrates strong performance on various music understanding tasks and attains state-of-the-art overall scores.
\end{enumerate}

While there are no clear superior or inferior models in this space, each of these modern audio encoders finds its own niche in different tasks, and they coexist in parallel, contributing to the broader landscape of audio representation learning.

\subsubsection{Cross-Modal Representation: CLAP and Its Variants}

The increasing interest in cross-modal learning has led to the development of models that can jointly process and align representations from different modalities. One of the most representative models in this domain is CLAP (Contrastive Language-Audio Pretraining), introduced by \citet{clap}, which extends the principles of CLIP (Contrastive Language-Image Pretraining)~\citep{clip} to the audio domain, enabling joint representation learning of music and text.

Multimodal alignment tasks inherently deal with the heterogeneity between different modalities, such as text and audio, which represent information in fundamentally different ways. Cross-modal machine learning aims to resolve this by aligning the representations of these distinct modalities into a shared space, where they can be processed together more effectively. In the context of this thesis, which focuses on text-to-music generation, cross-modal representations offer an improved framework for linking text-based inputs with music outputs, ultimately allowing for better multimodal integration.

Key features of CLAP include:

\begin{itemize}
    \item \textbf{Dual Encoder Architecture:} CLAP employs separate encoders for processing audio and text inputs. The audio encoder is typically based on a convolutional or transformer architecture, while the text encoder utilises pretrained language models such as BERT or RoBERTa.

    \item \textbf{Contrastive Learning Objective:} CLAP is trained using a contrastive learning approach, where the model learns to maximise the similarity between embeddings of matching audio-text pairs while minimising the similarity between non-matching pairs.

    \item \textbf{Large-Scale Pretraining:} CLAP is pretrained on large datasets (LAION-630K) \footnote{LAION-630K dataset: \url{https://github.com/LAION-AI/audio-dataset/tree/main/laion-audio-630k}.} of audio-text pairs, allowing it to learn general-purpose audio and text representations that can be fine-tuned for various downstream tasks.
\end{itemize}

The contrastive learning objective of CLAP can be formalised as:

\begin{equation}
\mathcal{L} = -\log \frac{\exp(sim(f_a(x_i), f_t(y_i)) / \tau)}{\sum_{j} \exp(sim(f_a(x_i), f_t(y_j)) / \tau)},
\end{equation}

\noindent where $f_a$ and $f_t$ are the audio and text encoders respectively, $x_i$ and $y_i$ are corresponding audio and text inputs, $sim$ is a similarity function (typically cosine similarity), and $\tau$ is a temperature parameter.

Although CLAP is one of the most representative multimodal models, it is not the only one in the field. Several variants and other approaches~\citep{compa, clapnew} have recently emerged, each pushing the performance of multimodal alignment models further. The development of CLAP and its successors represents a significant step towards achieving unified multimodal understanding of music and language, with applications ranging from music cognition to creative AI applications. In the context of text-to-music generation, these models offer a more coherent way to align text and music modalities, addressing the inherent heterogeneity in multimodal tasks and driving forward the potential for cross-modal translation and generation.

\subsection{Pretraining, Continued Pretraining, Post-training, and Finetuning}

The processes of pretraining, continued pretraining, post-training, and finetuning form a cohesive workflow commonly used in large-scale model training. 

Each phase builds on the previous one to incrementally refine the model's versatility and task-specific performance. While continued pretraining is less frequently applied, it can significantly enhance a model's ability to handle complex or specialised data. Post-training and finetuning further optimise the model for adaptability and precision, ensuring its effectiveness in specific applications.

\begin{enumerate}
    \item \textbf{Pretraining}:
    This is the initial phase where a model is trained on a vast and varied dataset to discern general patterns and extract fundamental features. In our research, I concentrate on pretraining text-to-music models to grasp the intrinsic connections between textual inputs and their musical counterparts. This foundational step utilises an extensive corpus of paired text and music to construct a versatile base model that can interpret and produce music from textual descriptions.
    
    \item \textbf{Continued Pretraining}:
    Following the initial pretraining, continued pretraining involves further training on additional data to enhance the model's generalisation capabilities. Although not that common, this step is helpful to strengthen the model's understanding with more data.

    \item \textbf{Post-training}:
    Post-training is a transitional phase where the model undergoes adjustments and optimisations based on the insights gained from pretraining. This step might involve techniques such as knowledge distillation or model compression, aiming to make the model more efficient or suitable for deployment in specific environments.

    \item \textbf{Finetuning}:
    The final stage involves finetuning the model on a smaller, task-specific dataset. This step tailors the model to meet specific requirements, such as generating music that adheres to particular styles or incorporating user-defined constraints. Finetuning often introduces new control mechanisms, enabling the model to generate music that is more aligned with detailed textual descriptions and additional user inputs.

    \item \textbf{Supervised Fine-tuning (SFT)}:
    As a subset of finetuning, SFT focuses on further honing the model's performance under the guidance of labeled data. This supervised approach ensures that the model not only adapts to the task but also maintains a high level of accuracy and consistency in its outputs.
\end{enumerate} 

\section{Text-to-Music Backbone Models}

In this section, I will explore the backbone models that are fundamental to this thesis in text-to-music generation. These models provide the underlying architecture and mechanisms through which text inputs are transformed into music outputs. 

\subsection{Diffusion Models}

\subsubsection{Denoising Diffusion Probabilistic Models (DDPMs)}

DDPMs~\citep{ddpm} operate by gradually adding noise to the data and then learning to reverse this process, effectively denoising the data to generate new samples. The model is trained on a forward diffusion process that progressively adds Gaussian noise to the data, and a reverse process that learns to denoise the data. 

\begin{figure}[tb]
    \centering
    \includegraphics[width=\linewidth]{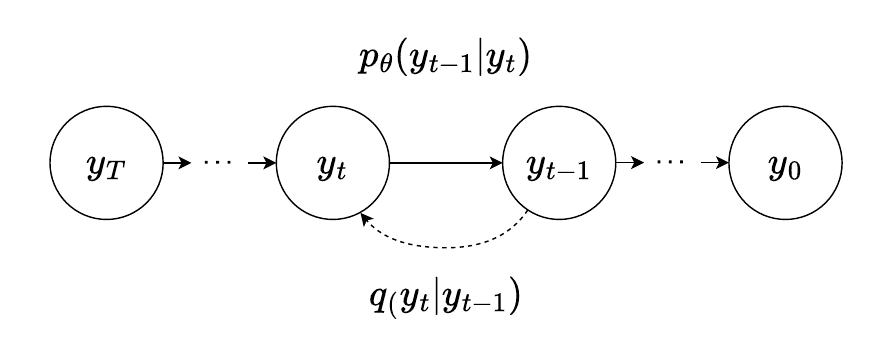}
    \caption{The forward and backward process of the diffusion model.}
    \label{fig:ddpm}
\end{figure}

\paragraph{Forward Process.} The forward process \( q \) adds noise to the data \( y \) over \( T \) time steps:

\begin{equation}
  q(y_t | y_{t-1}) = \mathcal{N}(y_t; \sqrt{\alpha_t} y_{t-1}, (1 - \alpha_t). \mathbf{I}),
\end{equation}

\noindent where \( \alpha_t \) is a schedule parameter that controls the variance of the noise added at each step. Here, \( \mathcal{N} \) denotes a normal distribution. The term \( \sqrt{\alpha_t} y_{t-1} \) represents the mean of the noisy data at step \( t \), and \( (1 - \alpha_t) \mathbf{I} \) represents the variance, which is controlled by \( \alpha_t \). This process adds a small amount of noise at each step.

This process can be iteratively applied starting from the original data \( y_0 \):

\begin{equation} 
q(y_{1:T} | y_0) = \prod_{t=1}^T q(y_t | y_{t-1}).
\end{equation}

After enough steps, \( y_T \) becomes almost pure noise. The idea is that by the end of the forward process, the data should be sufficiently noisy, resembling a Gaussian distribution.

\paragraph{Reverse Process.} The reverse process \( p \) aims to denoise \( y \) by estimating the original data distribution. It is defined as:

\begin{equation} 
p_\theta(y_{t-1} | y_t) = \mathcal{N}(y_{t-1}; \mu_\theta(y_t, t), \Sigma_\theta(y_t, t)),
\end{equation}

\noindent where \( \mu_\theta(y_t, t) \) and \( \Sigma_\theta(y_t, t) \) are parameterised by neural networks and are learned during training. In the reverse process, the model predicts the previous step \( y_{t-1} \) from the current step \( y_t \). The mean \( \mu_\theta \) and the variance \( \Sigma_\theta \) are learned functions that depend on \( y_t \) and the time step \( t \).

The reverse process can be written as:

\begin{equation} 
p_\theta(y_{0:T}) = p(y_T) \prod_{t=1}^T p_\theta(y_{t-1} | y_t). 
\end{equation}

This represents the application of the reverse process over \( T \) steps, starting from almost pure noise \( y_T \). The reverse process is essentially a learned denoising pathway that recovers the original data distribution from the noisy version.

\paragraph{Training Objective.} The training objective is to minimise the difference between the forward and reverse processes, which can be expressed as a variational lower bound:

\begin{equation}
L = \mathbb{E}_{q} \left[ \sum_{t=1}^T D_\text{KL}(q(y_{t-1} | y_t, y_0) | p_\theta(y_{t-1} | y_t)) \right],
\end{equation}

\noindent where \( D_\text{KL} \) denotes the Kullback-Leibler divergence between the true forward process and the learned reverse process. The objective function \( L \) is the expected sum of the Kullback-Leibler divergences between the true distribution and the predicted distribution at each step. Minimising this loss function trains the model to accurately reverse the diffusion process.

\subsubsection{Latent Diffusion Models (LDMs)}

Latent diffusion models~\citep{ldm} build on the principles of DDPM but operate in a latent space rather than directly on the data space. This approach can reduce computational complexity and improve efficiency, making it particularly suitable for high-dimensional data like music. The key advantage of LDMs is that they handle the diffusion process in a compressed representation of the data, which significantly reduces the computational burden while preserving the essential characteristics of the data.

\paragraph{Encoding and Decoding.} In latent diffusion models, the data \( y \) is first mapped to a lower-dimensional latent space \( z \) using an encoder:

\begin{equation} 
z = E(y).
\end{equation}

The encoder \( E \) compresses the high-dimensional data \( y \) into a lower-dimensional latent representation \( z \), which retains the essential features of the original data.

After the diffusion process is applied in the latent space, the denoised latent variables are mapped back to the data space using a decoder:

\begin{equation} 
\hat{y} = D(z). 
\end{equation}

The decoder \( D \) reconstructs the data \( \hat{y} \) from the denoised latent variables \( z \), producing an output that is similar to the original data.

\paragraph{Forward and Reverse Processes in Latent Space.} The forward process \( q \) and reverse process \( p \) in the latent space are defined similarly to DDPM:

\begin{equation} 
q(z_t | z_{t-1}) = \mathcal{N}(z_t; \sqrt{\alpha_t} z_{t-1}, (1 - \alpha_t) \mathbf{I}) 
\end{equation}

The forward process in the latent space adds Gaussian noise to the latent variables \( z \) at each step, similar to the forward process in DDPM but operating on the compressed representation.

\begin{equation} 
p_\theta(z_{t-1} | z_t) = \mathcal{N}(z_{t-1}; \mu_\theta(z_t, t), \Sigma_\theta(z_t, t)).
\end{equation}

The reverse process in the latent space predicts the previous latent variable \( z_{t-1} \) from the current latent variable \( z_t \), using learned functions \( \mu_\theta \) and \( \Sigma_\theta \).

The objective remains to minimise the difference between the forward and reverse processes, but this time in the latent space:

\begin{equation} 
L = \mathbb{E}_{q} \left[ \sum_{t=1}^T D_{KL}(q(z_{t-1} | z_t, z_0) | p_\theta(z_{t-1} | z_t)) \right]. 
\end{equation}

The training objective is to minimise the Kullback-Leibler divergence between the true forward process and the learned reverse process in the latent space, ensuring accurate reconstruction of the original data from the latent variables.

\subsubsection{Denoising Diffusion Implicit Models (DDIM) and Fast Sampling}

Denoising Diffusion Implicit Models (DDIMs)~\citep{ddim} enhance the efficiency of sampling in diffusion models by introducing a deterministic alternative to the stochastic reverse process found in Denoising Diffusion Probabilistic Models (DDPMs). This deterministic approach allows for faster sampling while maintaining the quality of the generated outputs.

\begin{figure}
    \centering
    \includegraphics[width=\linewidth]{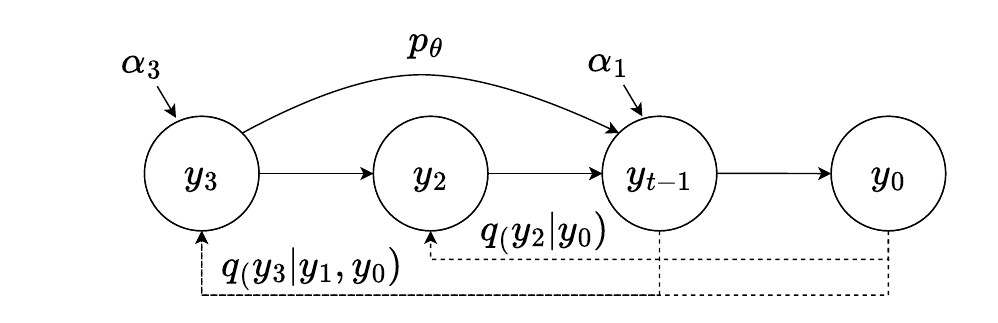}
    \caption{The diagram of DDIM models.}
    \label{fig:ddim}
\end{figure}

DDPMs require many steps to gradually denoise data, which can be computationally expensive and slow. DDIMs address this limitation by modifying the reverse process to follow a deterministic path, eliminating the need for sampling from a Gaussian distribution at each step. This deterministic reverse process can be described as:

\begin{equation}
y_{t-1} = \sqrt{\alpha_{t-1}} \left( \frac{y_t - \sqrt{1 - \alpha_t} \epsilon_\theta(y_t, t)}{\sqrt{\alpha_t}} \right) + \sqrt{1 - \alpha_{t-1}} \epsilon_\theta(y_t, t).
\end{equation}

In this equation, \( \epsilon_\theta(y_t, t) \) represents the predicted noise component at step \( t \), obtained from a neural network, and \( \alpha_t \) is the noise schedule parameter. By following this deterministic path, DDIMs can generate samples with fewer steps compared to the stochastic sampling of DDPMs.

To understand DDIMs in detail, it is crucial to grasp several key concepts:

\begin{enumerate}
    \item \textbf{Noise Schedule Parameter (\( \alpha_t \))}: This parameter controls the amount of noise added at each step of the forward diffusion process. It typically follows a schedule that starts with large values and gradually decreases, allowing the model to add less noise as it progresses. This schedule is critical in ensuring that the model can effectively learn the denoising process.
    \item \textbf{Predictive Noise Model (\( \epsilon_\theta \))}: The neural network \( \epsilon_\theta \) is trained to predict the noise that was added to the data at each step. 
    \item \textbf{Deterministic Reverse Process}: Unlike the stochastic nature of DDPMs, where each step involves sampling from a Gaussian distribution, DDIMs use a deterministic formula to revert the noisy data back to its original state. This deterministic nature reduces the variability and randomness in the generated samples, leading to more consistent outputs.
\end{enumerate}

To implement DDIMs, the training phase involves a forward process that progressively adds Gaussian noise to the data, defined by:

\begin{equation}
q(y_t | y_{t-1}) = \mathcal{N}(y_t; \sqrt{\alpha_t} y_{t-1}, (1 - \alpha_t) \mathbf{I}).
\end{equation}

A neural network \( \epsilon_\theta \) is trained to predict the noise added at each step by minimising the difference between the predicted noise and the actual noise:

\begin{equation}
L_{train} = \mathbb{E}_{q} \left[ | \epsilon_\theta(y_t, t) - \epsilon |^2 \right].
\end{equation}

During sampling, the trained model applies the deterministic reverse process, starting from pure noise \( y_T \) and iteratively generating the final sample \( y_0 \). For text-to-music generation, text embeddings are incorporated into the neural network \( \epsilon_\theta \), conditioning the generation process on the text input. The modified reverse process becomes:

\begin{equation}
y_{t-1} = \sqrt{\alpha_{t-1}} \left( \frac{y_t - \sqrt{1 - \alpha_t} \epsilon_\theta(y_t, t, x)}{\sqrt{\alpha_t}} \right) + \sqrt{1 - \alpha_{t-1}} \epsilon_\theta(y_t, t, x).
\end{equation}

\subsubsection{Stochastic Differential Equation (SDE) Interpretation of DDPMs}

The forward diffusion process in DDPMs can be viewed through the lens of a Stochastic Differential Equation (SDE). An SDE provides a continuous-time perspective on the discrete-time forward diffusion process. In this framework, the forward process is seen as a continuous-time stochastic process governed by the SDE:

\begin{equation}
dX_t = -\frac{1}{2}\beta(t) X_t dt + \sqrt{\beta(t)} dW_t,
\end{equation}

\noindent where \(X_t\) is the state of the system at time \(t\), \(\beta(t)\) is a time-dependent diffusion coefficient, and \(W_t\) is a Wiener process representing the source of random noise. This equation describes how noise is continuously added to the data over time, leading to a gradual loss of information.

\subsubsection{Ordinary Differential Equation (ODE) Interpretation of DDIMs}

DDIMs, on the other hand, offer a deterministic counterpart to the stochastic diffusion process. Instead of an SDE, the reverse process in DDIMs can be interpreted as an Ordinary Differential Equation (ODE), which governs a deterministic trajectory from noise back to the original data:

\begin{equation}
\frac{dX_t}{dt} = -\frac{1}{2}\beta(t) X_t + \sqrt{\beta(t)} \epsilon_\theta(X_t, t),
\end{equation}

\noindent where \(\epsilon_\theta(X_t, t)\) is the predicted noise component at time \(t\) obtained from the trained neural network. This ODE formulation allows for fast and deterministic sampling, as it removes the stochasticity present in the reverse process of DDPMs.

\subsubsection{Connecting DDPMs and DDIMs Through SDE and ODE}

The SDE interpretation of DDPMs and the ODE interpretation of DDIMs highlight the transition from a stochastic to a deterministic framework. Both models start with the same goal: to denoise data that has been corrupted by additive Gaussian noise. However, they achieve this goal through different methods.

In DDPMs, the SDE provides a probabilistic framework that captures the inherent uncertainty in the diffusion process. The reverse process in DDPMs is inherently stochastic, aiming to estimate the posterior distribution of the data at each step. This stochastic nature leads to slower sampling as each step requires a draw from a probability distribution.

In contrast, DDIMs employ an ODE to guide the reverse process deterministically. By removing the stochastic component, DDIMs can significantly speed up the sampling process without compromising the quality of the generated outputs. This deterministic approach makes DDIMs more efficient for applications requiring real-time or rapid generation.

\subsubsection{Fast Sampling in DDIMs}

The fast sampling in DDIMs is achieved by utilising the deterministic reverse process derived from the ODE. This process can be summarised as:

\begin{equation}
X_{t-\Delta t} = X_t - \frac{1}{2}\beta(t) X_t \Delta t + \sqrt{\beta(t)} \epsilon_\theta(X_t, t) \Delta t,
\end{equation}

\noindent where \(\Delta t\) is the time step size. By choosing appropriate values for \(\Delta t\), the number of steps required for sampling can be reduced, leading to faster generation times.

\subsubsection{DPM-Solver: A Fast ODE Solver for Diffusion Probabilistic Model Sampling}

Diffusion Probabilistic Models (DPMs) typically require a large number of steps for accurate sampling, which can be slow and computationally expensive. DPM-Solver~\citep{dpmsolver} is an efficient method that addresses this issue by using a fast Ordinary Differential Equation (ODE) solver, significantly reducing the number of steps required for sampling.

DPM-Solver leverages the fact that the reverse diffusion process can be viewed as solving an ODE. By applying advanced ODE solvers, DPM-Solver can generate high-quality samples in as few as ten steps. The core idea is to approximate the reverse diffusion process using a discretised ODE, which can be solved efficiently using numerical methods.

The forward diffusion process can be described by the following Stochastic Differential Equation (SDE):

\begin{equation}
d\mathbf{y} = f(\mathbf{y}, t) dt + g(t) d\mathbf{w},
\end{equation}

\noindent where \( \mathbf{y} \) is the data, \( f \) and \( g \) are functions defining the drift and diffusion coefficients, and \( d\mathbf{w} \) represents the Wiener process. The reverse process, which aims to denoise the data, can be reformulated as an ODE:

\begin{equation}
d\mathbf{y} = \left( f(\mathbf{y}, t) - g(t)^2 \nabla_{\mathbf{y}} \log p_t(\mathbf{y}) \right) dt.
\end{equation}

In this formulation, \( f(\mathbf{y}, t) \) captures the deterministic part of the process, while \( g(t)^2 \nabla_{\mathbf{y}} \log p_t(\mathbf{y}) \) represents the gradient of the log probability, which acts as a correction term to guide the data back to its original distribution.

DPM-Solver approximates this ODE using advanced numerical solvers, such as Runge-Kutta methods, which can efficiently handle the discretisation of the reverse process. These solvers can adapt the step size dynamically, ensuring that the approximation remains accurate while minimising the computational burden.

In practice, DPM-Solver can be integrated into existing diffusion models by replacing the standard reverse diffusion process with the ODE-based solver. This integration results in faster sampling times and reduced computational costs, making it suitable for real-time applications like text-to-music generation.

\subsubsection{DDIM Inversion and Text Inversion}

Denoising Diffusion Implicit Models (DDIMs) introduce an efficient deterministic sampling method for diffusion models. A particularly useful feature of DDIMs is their ability to invert the generative process, which means one can map generated samples back to latent variables. This inversion is crucial for applications like text-to-music generation, where conditioning on text input must be precise and controllable.

\paragraph{DDIM Inversion}

DDIM inversion is the process of mapping generated samples \( \hat{y}_0 \) back to their latent representations \( z \). This inversion allows for editing or modifying the generated samples by manipulating their latent representations and then re-generating the samples using the reverse DDIM process.

Given the deterministic nature of DDIMs, the inversion process can be described as follows. The forward process starts from the initial data \( y_0 \) and adds noise according to the schedule parameter \( \alpha_t \):

\begin{equation}
q(y_t | y_{t-1}) = \mathcal{N}(y_t; \sqrt{\alpha_t} y_{t-1}, (1 - \alpha_t) \mathbf{I}).
\end{equation}

In the reverse process, using the learned neural network \( \epsilon_\theta(y_t, t) \), the deterministic reverse process predicts \( y_{t-1} \) from \( y_t \):

\begin{equation}
y_{t-1} = \sqrt{\alpha_{t-1}} \left( \frac{y_t - \sqrt{1 - \alpha_t} \epsilon_\theta(y_t, t)}{\sqrt{\alpha_t}} \right) + \sqrt{1 - \alpha_{t-1}} \epsilon_\theta(y_t, t).
\end{equation}

To invert the generative process, it is needed to estimate the latent variable \( z \) that corresponds to the generated sample \( \hat{y}_0 \). This involves applying the forward diffusion process in reverse:

\begin{equation}
z_t = \frac{\hat{y}_0 - \sqrt{1 - \alpha_t} \epsilon_\theta(\hat{y}_0, t)}{\sqrt{\alpha_t}}
\end{equation}

This equation estimates the latent variable \( z_t \) from the generated sample \( \hat{y}_0 \) by removing the noise component predicted by the neural network.

\paragraph{Text Inversion}

Text inversion in the context of DDIMs involves incorporating text embeddings into the diffusion model to guide the generation process. This conditioning allows for more accurate and contextually relevant generation, particularly in tasks like text-to-music generation.

To incorporate text \( x \) into the DDIM framework, I use text embeddings \( \phi(x) \), which are integrated into the neural network \( \epsilon_\theta(y_t, t, \phi(x)) \). This conditioning ensures that the generated samples align with the textual input.

First, represent the text \( x \) using a pretrained text encoder, producing embeddings \( \phi(x) \):

\begin{equation}
\phi(x) = \text{TextEncoder}(x).
\end{equation}

Next, modify the noise prediction network to incorporate text embeddings, producing a conditioned noise estimate:

\begin{equation}
\epsilon_\theta(y_t, t, \phi(x)) = \text{NeuralNetwork}(y_t, t, \phi(x)).
\end{equation}

Finally, integrate the text-conditioned noise prediction into the reverse process:

\begin{equation}
y_{t-1} = \sqrt{\alpha_{t-1}} \left( \frac{y_t - \sqrt{1 - \alpha_t} \epsilon_\theta(y_t, t, \phi(x))}{\sqrt{\alpha_t}} \right) + \sqrt{1 - \alpha_{t-1}} \epsilon_\theta(y_t, t, \phi(x)).
\end{equation}

This modified reverse process uses text embeddings to influence the denoising steps, producing samples that reflect the input text \( x \).

The inversion process with text conditioning involves estimating the latent variables that correspond to a generated sample conditioned on text input. This can be achieved by incorporating the text embeddings into the inversion equation:

\begin{equation}
z_t = \frac{\hat{y}_0 - \sqrt{1 - \alpha_t} \epsilon_\theta(\hat{y}_0, t, \phi(x))}{\sqrt{\alpha_t}}.
\end{equation}

This formula estimates the latent variable \( z_t \) from the generated sample \( \hat{y}_0 \) and the text-conditioned noise prediction \( \epsilon_\theta(\hat{y}_0, t, \phi(x)) \).

\subsubsection{Related Work on Diffusion-Based Text-to-Music Backbone Models} 

The evolution of diffusion-based text-to-music models reflects a series of incremental innovations and refinements, each building on the strengths of its predecessors while addressing specific limitations. These models can be grouped into several categories based on their shared architectures, conditioners, and improvements in output quality. Table~\ref{tab:diffusion_list} provides an overview of these models.

The journey began with \textbf{Riffusion} (2022), which applied diffusion processes to music using methods initially designed for image generation, without making specific optimisations for musical outputs. This set the stage for subsequent models like \textbf{Mousai} (2023) and \textbf{Noise2Music} (2023), which moved beyond Riffusion by employing cascaded 1D UNet architectures, specifically conditioned on T5, to generate longer, higher-quality waveforms. These models demonstrated a significant step forward by directly targeting music generation with dedicated architectures.

In the same period, \textbf{TANGO} (2023) and its refined version, \textbf{Mustango} (2023), emerged, sharing the same VAE and 2D UNet architecture while utilising the FLAN-T5 conditioner. Mustango introduced subtle improvements over TANGO, maintaining the same core framework but optimising performance and generation efficiency. Similarly, the AudioLDM family, consisting of \textbf{Audio\-LDM}, \textbf{MusicLDM}, and \textbf{AudioLDM 2}, followed a trajectory of continuous refinement. MusicLDM is essentially a fine-tuned version of AudioLDM, while AudioLDM 2 brought separate improvements specifically to the encoder, enhancing flexibility and adaptability across various tasks.

Meanwhile, \textbf{Stable Audio} (2024) and \textbf{Stable Audio 2} further advanced the field. Stable Audio employed a VAE and 1D UNet architecture conditioned on CLAP, generating up to 95-second audio outputs. Stable Audio 2 extended this capability by introducing DiT (Diffusion Transformers), significantly increasing the length of generated outputs to up to 4 minutes and 45 seconds, marking a notable leap in generation length and quality.

Lastly, \textbf{Jen-1} and \textbf{Jen-1 Composer} (2023) introduced more sophisticated features, including support for multitrack generation in Jen-1 Composer, built on the foundation of MAE and 1D UNet architectures. These models leveraged FLAN-T5 for conditioning, focusing on producing high-quality stereo waveforms, with Jen-1 Composer expanding capabilities to multitrack outputs for more complex compositions.

\begin{sidewaystable}[htbp]
\caption{An overview of models for diffusion-based text-to-music generation.}
\begin{tabular}{lllllll}
\toprule
               & Date    & Architecture     & Conditioner   & Output              & Length      & Sample Rate and Channels      \\
               \midrule
Riffusion      & 2022.11 & VAE + 2D UNet    & CLIP          & Audio (Spectrogram) & 5.12s       & 16kHz mono   \\
Mousai         & 2023.01 & Cascaded 1D UNet & T5            & Audio (Waveform)    & 90s         & 48kHz stereo \\
Noise2Music    & 2023.02 & Cascaded 1D UNet & T5            & Audio (Waveform)    & 30s         & 24kHz mono   \\
TANGO          & 2023.04 & VAE + 2D UNet    & FLAN-T5       & Audio (Spectrogram) & 10s         & 16kHz mono   \\
Mustango       & 2023.11 & VAE + 2D UNet    & FLAN-T5       & Audio (Spectrogram) & 10s         & 16kHz mono   \\
AudioLDM       & 2023.01 & VAE + 2D UNet    & CLAP          & Audio (Spectrogram) & 10s         & 16kHz mono   \\
MusicLDM       & 2023.08 & VAE + 2D UNet    & CLAP          & Audio (Spectrogram) & 10s         & 16kHz mono   \\
AudioLDM 2     & 2023.08 & VAE + 2D UNet    & FLAN-T5 + GPT & Audio (Spectrogram) & variable    & 16kHz mono   \\
Stable Audio   & 2024.02 & VAE + 1D UNet    & CLAP          & Audio (Waveform)    & up to 95s   & 16KHz mono   \\
Stable Audio 2 & 2024.07 & AE + DiT         & CLAP          & Audio (Waveform)    & up to 4m45s & 48kHz stereo \\
Jen-1          & 2023.08 & MAE + 1D UNet    & FLAN-T5       & Audio (Waveform)    & 10s         & 48kHz stereo \\
Jen-1 Composer & 2023.10 & AE + 1D UNet     & FLAN-T5       & Audio (Waveform)    & 10s         & 48kHz stereo \\
MusicFlow      & 2024.05 & EnCodec + Flow   & HuBERT        & Audio (Spectrogram) & 10s         & 32kHz mono  \\
\bottomrule
\end{tabular}
\label{tab:diffusion_list}
\end{sidewaystable}

\paragraph{Riffusion.} Riffusion~\citep{riffusion} is one of the pioneering projects in the field of text-to-music generation using diffusion models. The core component of Riffusion is its use of a latent text-to-image diffusion model, specifically a fine-tuned Stable-Diffusion-v1-5 checkpoint, to create spectrograms from text. These spectrograms visually represent the audio and are subsequently converted into sound using the Griffin-Lim algorithm. This method allows for real-time music generation, showcasing the potential of diffusion models in audio synthesis. However, the reliance on Griffin-Lim limits the audio quality waveform reconstruction; the length of audio is also limited to only 5 seconds.

Riffusion was trained on a dataset composed of paired text and spectrogram images, leveraging the LAION-5B dataset for text-image pairs and additional audio data. Despite its foundational nature and certain limitations in audio fidelity, Riffusion has significantly influenced subsequent research by demonstrating the feasibility of using diffusion models for generating music from textual descriptions. The project is available for public use and experimentation through its web app and model repository.

\paragraph{Mo\^usai.} Moûsai~\citep{mousai} is another pioneering research model in the field of text-to-music generation, marking the first significant exploration into diffusion-based methods for this purpose. It bridges the gap between textual descriptions and music by employing a highly efficient, expressive, and long-context handling approach. Moûsai leverages a cascading two-stage latent diffusion model that can generate high-quality stereo music at 48kHz, spanning multiple minutes from textual inputs. The core component of Moûsai is its diffusion autoencoder, which first compresses the audio into a magnitude-only spectrogram and then decodes it back to waveform using a custom 1D U-Net. This structure enables the model to efficiently handle long-term musical structure and produce real-time music generation on consumer GPUs. The dataset used for training Moûsai includes a large collection of high-quality music samples paired with textual descriptions, facilitating the model's capability to generate music that reflects the provided text.

\paragraph{Noise2Music.} Noise2Music~\citep{noise2music} is one of the pioneering models in the field of text-to-music generation, particularly notable for its use of diffusion models to directly generate audio waveforms from textual descriptions. This model addresses the challenge of generating high-fidelity music by employing a two-stage diffusion process. The first stage involves a generator model that produces an intermediate representation conditioned on the text input, while the second stage involves a cascader model that enhances the fidelity of this intermediate representation to produce high-quality audio.

The core component of Noise2Music is its diffusion-based architecture, which leverages a 1D Efficient U-Net for the generator and cascader models. The generator model first generates a low-fidelity audio waveform at 3.2kHz conditioned on the text. This low-fidelity audio is then upsampled and refined by the cascader model to produce a high-fidelity audio waveform at 16kHz. This two-stage approach allows the model to effectively compress and refine audio information, leading to improved audio quality.

For its training, Noise2Music utilises a large-scale dataset that includes MusicCaps~\citep{musiclm}, AudioSet~\citep{audiocaps}, and MagnaTagATune~\citep{mtat}. These datasets provide a diverse range of music samples paired with detailed textual descriptions, enabling the model to learn complex mappings between text and music. The model also employs pretrained large language models to generate paired text for the audio in the training set and to extract text embeddings for conditioning the diffusion models.

\paragraph{TANGO.} TANGO~\citep{tango} is a versatile text-to-audio generation model that leverages a latent diffusion model (LDM) to generate high-quality audio, including music, from textual descriptions. Initially designed for general text-to-audio tasks, TANGO was later adapted to include music generation capabilities.

The core component of TANGO is its use of the instruction-tuned large language model FLAN-T5~\citep{flant5} as the text encoder, which generates detailed text embeddings from input descriptions. These embeddings are then used to guide the LDM in constructing a latent audio representation. The model utilises a 1D U-Net architecture within the LDM for both the forward and reverse diffusion processes, enabling it to handle the complexities of audio generation effectively.

TANGO employs a cascading approach for diffusion, with a generator model producing an intermediate audio representation at a lower fidelity (e.g., 3.2kHz), which is then refined by a cascader model to a higher fidelity (e.g., 16kHz). This two-stage process helps in compressing and then enhancing the audio quality, ensuring that the final output is both high-fidelity and true to the textual prompt.

\paragraph{Mustango.} Mustango~\citep{mustango} extends the capabilities of the TANGO model by incorporating music-specific metadata to enhance controllability over the generated music. This model allows users to provide detailed musical instructions in addition to general text descriptions, enabling control over aspects such as chords, beats, tempo, and key.

The core component of Mustango is its use of MuNet, a Music-Domain-Knowledge-Informed UNet module. MuNet integrates these music-specific features, predicted from the text prompt, into the diffusion denoising process. This approach ensures that the generated music adheres to the specified musical properties, offering a high degree of control over the output.

For training, TANGO utilises datasets such as AudioCaps, which contain extensive audio clips paired with human-written captions. Despite training on a relatively smaller dataset compared to other models, TANGO achieves impressive performance, owing to its robust architecture and the efficiency of its instruction-tuned text encoder.

\paragraph{AudioLDM and MusicLDM.} AudioLDM \citep{audioldm} and MusicLDM \citep{musicldm} are innovative models in the field of text-to-audio and text-to-music generation. Both models leverage Latent Diffusion Models (LDMs) and the Contrastive Language-Audio Pretraining (CLAP) framework to bridge the gap between text and audio representations. Here I discuss each model and compare their approaches. 

AudioLDM generates audio from textual descriptions using a diffusion model operating in a latent space. It aims to create high-fidelity audio samples by leveraging the pretrained CLAP model to align text and audio embeddings.

The core of AudioLDM involves two main components: the CLAP model and the Latent Diffusion Model (LDM). The CLAP model provides a shared embedding space for both text and audio. During training, audio embeddings from the CLAP model are used to condition the LDM. This conditioning is achieved through a Feature-wise Linear Modulation (FiLM) mechanism, which modulates the latent diffusion process using the text embeddings during inference. This approach allows the model to learn from audio-only data during training and use text embeddings for generation.

AudioLDM is trained on a variety of datasets, including AudioSet and FSD50K, which provide diverse audio samples paired with textual descriptions. The model benefits from a large amount of audio data to learn robust audio representations.

MusicLDM builds on the architecture of AudioLDM but introduces two significant modifications to enhance its performance for music generation. First, it retrains the CLAP model on text-music pair datasets to improve its understanding of musical contexts. Second, MusicLDM employs beat-synchronous mixup strategies for data augmentation. These strategies involve mixing audio samples at synchronised beats, ensuring the resultant training data is musically coherent. This beat-synchronous augmentation helps the model generate more diverse and rhythmically accurate music.

MusicLDM is trained on specialised music datasets such as MusicCaps and MusicBench. These datasets contain rich textual descriptions of music, which help the model learn to generate music that aligns closely with detailed text prompts.

While both AudioLDM and MusicLDM use LDMs and the CLAP framework, MusicLDM's enhancements make it particularly suited for generating music. The beat-synchronous data augmentation and retraining of the CLAP model on music-specific datasets allow MusicLDM to generate more musically coherent and varied outputs compared to the more general audio generation capabilities of AudioLDM. However, both models face limitations in precise control over the generated audio due to the global nature of CLAP embeddings and the gap between training (audio embeddings) and inference (text embeddings).

\paragraph{AudioLDM 2.} AudioLDM 2~\citep{audioldm2} is an advanced text-to-music generation model that improves upon its predecessor, AudioLDM, by integrating more detailed text representations and refining the conditioning process to enhance the quality and controllability of the generated audio. 

AudioLDM 2 generates high-fidelity audio, including music, from textual descriptions by fusing comprehensive text representations with advanced diffusion models. This model is designed to address the limitations in earlier models regarding the precision and detail of the generated audio.

The core components of AudioLDM 2 include the use of Contrastive Language-Audio Pretraining (CLAP) embeddings and a Latent Diffusion Model (LDM). The model operates through a two-step process: 1, Conditioning Information to LOA Translation: AudioLDM 2 introduces the concept of the ``Language of Audio" (LOA), an intermediate feature that bridges the gap between text and audio representations. The LOA is derived using a self-supervised pretrained AudioMAE~\citep{audiomae} model, which captures both semantic and acoustic details of audio signals; 2, LOA to Audio Generation: The model uses the LOA to condition the latent diffusion model, which generates the final audio output. By employing a cross-attention mechanism, AudioLDM 2 can incorporate detailed text representations directly into the diffusion process. This allows the model to utilise CFG, enhancing the precision and fidelity of the generated audio.

AudioLDM 2 is trained on comprehensive datasets such as AudioSet and MusicCaps, which provide a diverse range of audio samples paired with textual descriptions. This extensive training data enables the model to learn robust mappings between text and audio, improving its generalisation and performance across various audio generation tasks.

\paragraph{Stable Audio 1 and 2.} Stable Audio 1~\citep{sd1} is a diffusion model designed for generating long-form, variable-length stereo audio, including music, from text prompts. Unlike earlier models that produce fixed-length outputs, Stable Audio 1 introduces timing conditioning to generate audio of varying lengths, making it suitable for creating long and complex audio sequences. 

Stable Audio 1 employs a 907M parameter U-Net, incorporating residual, self-attention, and cross-attention layers to denoise the latent representations conditioned on text and timing embeddings. Memory-efficient attention mechanisms enable the handling of longer audio sequences.

Stable Audio 1 is trained on a dataset comprising over 800,000 audio files, totaling approximately 19,500 hours of music, sound effects, and single-instrument stems. The dataset, sourced from AudioSparx, includes detailed text metadata for effective text-to-audio generation.

Stable Audio 2~\citep{sd2} is an advanced model designed for long-form music generation using latent diffusion techniques. It builds upon the foundations of Stable Audio 1, enhancing its capabilities to generate longer, more complex music tracks with coherent structure. Stable Audio 2 can generate music up to 4 minutes and 45 seconds in length, a significant improvement over its predecessor.

The core components of Stable Audio 2 include: 1. Diffusion-Transformer (DiT)~\citep{dit}: Unlike the convolutional U-Net structure used in Stable Audio 1, Stable Audio 2 employs a Diffusion-Transformer (DiT). The DiT integrates attention mechanisms and gated multi-layer perceptrons (MLPs) with cross-attention layers to incorporate conditioning signals. This setup allows for efficient handling of long sequences and the integration of detailed text and timing information. The model uses rotary positional embeddings and efficient block-wise attention, combined with gradient checkpointing, to manage the computational and memory demands of processing long temporal contexts.

The training process involves a multi-stage approach. Initially, the model is pretrained to generate music up to 3 minutes and 10 seconds long, followed by fine-tuning to extend the maximum length to 4 minutes and 45 seconds. This phased training ensures the model can handle extended sequences effectively.

\paragraph{Jen-1.} JEN-1~\citep{jen1} is a high-fidelity model designed for text-to-music generation, leveraging an omnidirectional diffusion model that integrates both autoregressive (AR) and non-autoregressive (NAR) training paradigms. This dual approach allows JEN-1 to capture sequential dependencies in music (via AR training) while also benefiting from the efficiency of parallel generation (via NAR training).

The core component of JEN-1 is its hybrid diffusion model. This model operates within a noise-robust latent embedding space obtained from a masked audio autoencoder. The diffusion model utilises bidirectional modes to gather comprehensive context and unidirectional modes to capture sequential dependencies, providing a balanced approach to generating coherent and high-fidelity music.

JEN-1 is trained on datasets such as MusicCaps, which contain extensive audio clips paired with textual descriptions. These datasets enable the model to learn the intricate mappings between text and music, enhancing its ability to generate music that aligns well with textual prompts.

\paragraph{Jen-1 Composer.} JEN-1 Composer~\citep{jen1composer} extends the capabilities of the original JEN-1 model to facilitate high-fidelity multi-track music generation. Building on JEN-1’s foundation, which integrates both autoregressive and non-autoregressive training for text-guided music generation, JEN-1 Composer introduces a unified framework that efficiently models the marginal, conditional, and joint distributions over multi-track music.

JEN-1 Composer is designed for versatile and controllable multi-track music generation. This model extends the single-track architecture of JEN-1 by concatenating multi-track latent representations, thus enabling the explicit modelling of interdependencies between different tracks. The model employs a curriculum training strategy that incrementally transitions from single-track generation to multi-track generation, improving its capacity to generate coherent multi-track music.

JEN-1 Composer is also trained on the same dataset. Additionally, the model leverages source separation tools like Spleeter and Demucs to augment its training data, allowing it to learn from a diverse range of musical compositions and enhancing its ability to generate high-fidelity music across multiple tracks.

\paragraph{MusicFlow. } MusicFlow~\citep{musicflow} explores a variant of diffusion models called flow matching for text-to-music generation. This method aims to improve training stability and efficiency while generating high-quality music. Unlike diffusion models, flow matching focuses on modelling the conditional distribution of semantic and acoustic features using two flow matching networks. This approach achieves faster and more stable training with minor revisions to existing diffusion techniques. MusicFlow does not represent a significant breakthrough in text-to-music quality but opens a pathway for efficient model training.

MusicFlow utilises a cascaded flow matching network for its generation process. This involves two networks: one for modelling the semantic features extracted from text descriptions and another for acoustic features. The training objective incorporates masked prediction, which enhances the model’s ability to generalise to tasks like music infilling and continuation. By employing a flow matching strategy, the model operates with fewer iterative steps and a smaller parameter size compared to conventional diffusion models.

MusicFlow is trained on an internal dataset comprising a large collection of music audio paired with textual descriptions. While specific datasets used for training are not publicly detailed, it is noted that MusicCaps was used for evaluation purposes, not training. This internal dataset enables the model to learn complex associations between text and music, allowing for efficient and effective music generation.

\subsection{Music Language Modelling}

Music language modelling plays a crucial role in text-to-music generation by enabling the system to understand and generate audio sequences. Several advanced models and techniques are employed to achieve this, including Vector Quantised Variational Autoencoders (VQ-VAEs), Residual Vector Quantisation (RVQ), and autoregressive models. Here, I will discuss these methods and their contributions to text-to-music generation.

\subsubsection{Vector Quantised Variational Autoencoders (VQ-VAEs)}

VQ-VAEs~\citep{vqvae} are a type of autoencoder that incorporates discrete latent variables, making them particularly effective for modelling high-dimensional data such as audio. In the context of music generation, VQ-VAEs encode the audio into a discrete latent space, which is then used to generate new musical sequences by generating new latent codes.

\paragraph{Encoding and Decoding.} In a VQ-VAE, the encoder \( E \) maps the input audio \( y \) to a discrete latent space \( z \):

\begin{equation} \label{eq:encoder}
z = E(y). 
\end{equation}

The encoder transforms the high-dimensional audio input into a lower dimensional latent representation. This step is crucial for reducing the complexity of the data while preserving its essential characteristics.

The decoder \( D \) then reconstructs the audio from the discrete latent variables:

\begin{equation} \label{eq:decoder}
\hat{y} = D(z). 
\end{equation}

The decoder takes the compressed latent representation and reconstructs it back into the audio signal, attempting to reproduce the original input as closely as possible.

\paragraph{Quantisation.} The latent space is quantised to a finite set of possible values, which allows the model to capture the essential features of the audio:

\begin{equation} 
z_q = \text{Quantize}(z). 
\end{equation}

Quantisation involves mapping the continuous latent variables to a discrete set of values, making it easier for the model to learn and generate accurate representations of the audio data.

\subsubsection{Residual Vector Quantisation (RVQ) and EnCodec}

RVQ~\citep{rvq} is an extension of VQ-VAEs that improves the representation capacity by using multiple quantisation stages. EnCodec is an advanced implementation of RVQ, developed by Meta AI, which leverages neural networks for high-fidelity audio compression and generation.

\paragraph{Multi-stage Quantisation.} In RVQ, the latent space is quantised in multiple stages, each refining the representation:

\begin{equation} 
z_{q_1} = \text{Quantize}(z_1), \quad z_{q_2} = \text{Quantize}(z_2 | z_{q_1}), \quad \ldots .
\end{equation}

In the equation, \( z_{q_1} \) represents the quantised version of the latent variable \( z_1 \), obtained after the first quantisation stage. \( z_{q_2} \) is the quantised latent variable after the second stage, where the quantisation is conditioned on \( z_{q_1} \). This pattern continues for subsequent stages, where each quantised latent variable \( z_{q_i} \) is refined by the previous stage’s quantised output, progressively improving the fidelity of the representation. 

\paragraph{EnCodec Architecture.} EnCodec~\citep{audiogen} utilises a convolutional encoder-decoder architecture with residual vector quantisation. It compresses the audio into a lower-dimensional representation and then reconstructs it, maintaining high fidelity.

Encoder:
  \begin{equation} \tag{\ref{eq:encoder}}
  z = E(y). 
  \end{equation}
  The encoder consists of multiple convolutional layers followed by a residual vector quantisation bottleneck.
  
Decoder:
  \begin{equation} \tag{\ref{eq:decoder}}
  \hat{y} = D(z). 
  \end{equation}
  The decoder mirrors the encoder, reconstructing the audio from the quantised latent representation.

Performance: EnCodec is capable of compressing audio to various bitrates while preserving high audio quality. It supports both monophonic and stereophonic audio at different sample rates.

\subsubsection{Autoregressive Models}

\begin{figure}[tb]
    \centering
    \includegraphics[width=\linewidth]{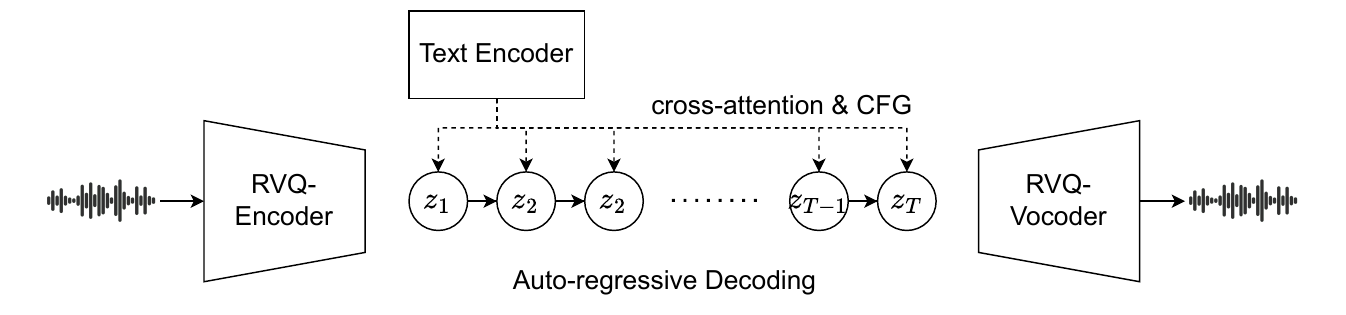}
    \caption{A simplified diagram of an audio language model. The encoder compresses audio waveforms into discrete tokens, and a auto-regressive model is responsible for generating new tokens. Text conditions can be fused through the cross-attention module.}
    \label{fig:audiolm}
\end{figure}

Autoregressive models are fundamental in generating sequences where each step depends on the previous ones, making them particularly well-suited for music generation tasks. These models are designed to predict the next element in a sequence based on the previous elements, ensuring temporal coherence and consistency in the generated music.

In the context of text-to-music generation, autoregressive models are often built on RVQ tokens. RVQ tokens are generated by compressing audio data into a lower-dimensional, quantised form using Residual Vector Quantisation. This hierarchical quantisation process allows for efficient and detailed audio representation, which is crucial for high-fidelity music generation.

Residual Vector Quantisation involves multiple stages of quantisation. Each stage refines the previous stage's output, allowing for a detailed and accurate representation of the audio data. The quantised tokens generated through RVQ serve as the building blocks for the autoregressive models.

\begin{equation} 
z_{q_1} = \text{Quantize}(z_1), \quad z_{q_2} = \text{Quantize}(z_2 | z_{q_1}), \quad \ldots. 
\end{equation}

Here, \( z_{q_i} \) represents the quantised latent variables at each stage \( i \).

\paragraph{Autoregressive Model Architecture}

Autoregressive models in text-to-music generation typically employ transformer architectures due to their ability to handle long-range dependencies and sequential data efficiently. These models are conditioned on the quantised tokens generated by RVQ.

\begin{itemize}
  \item \textbf{Token Embedding:}
    \begin{equation}
    \text{Token Embedding} = \text{Embed}(z_q)
    \end{equation}
    The quantised tokens \( z_q \) are embedded into a higher-dimensional space suitable for the transformer model.

  \item \textbf{Sequence Generation:}
    \begin{equation}
    p(y) = \prod_{t=1}^T p(y_t | y_{1:t-1}).
    \end{equation}
    The autoregressive model generates the sequence one step at a time, with each step \( y_t \) conditioned on all previous steps \( y_{1:t-1} \).

  \item \textbf{Transformer Decoder:}
    \begin{equation}
    \text{Decoder Output} = \text{TransformerDecoder}(\text{Token Embedding})
    \end{equation}
    The embedded tokens are passed through the transformer decoder, which predicts the next token in the sequence.
\end{itemize}

\paragraph{Conditioning and Training}

Autoregressive models for text-to-music generation are often conditioned on text embeddings to align the generated music with the given textual descriptions. This conditioning is achieved by incorporating text embeddings into the model's architecture.

\begin{equation}
\text{Text Embedding} = \text{T5}(x).
\end{equation}

Here, \( x \) represents the textual description, and T5 is a pretrained text encoder that generates text embeddings.

The model is trained end-to-end, optimising the likelihood of the observed sequences. During training, the model learns to predict the next token in the sequence based on the previous tokens and the text conditioning.

\begin{equation}
L = -\sum_{t=1}^T \log p(y_t | y_{1:t-1}, x).
\end{equation}

The training objective is to maximise the log-likelihood of the observed sequences, ensuring that the model generates sequences that are consistent with both the previous tokens and the text descriptions.

\paragraph{Model Developments}

Recent breakthroughs like \textbf{MusicLM} (2023) established the potential of language models for generating high-quality music, leveraging a hierarchical combination of SoundStream for audio synthesis and w2v-BERT for maintaining long-term coherence. The use of MuLan embeddings enabled the model to align music and text representations effectively, laying the groundwork for future advancements in generating coherent, stylistically rich music directly from textual descriptions. \textbf{MusicGen} (2023) continued this trend by integrating EnCodec and T5 into a single-stage transformer framework, focusing on efficient token processing to generate longer, high-fidelity music sequences.

More recent models, such as \textbf{MAGNeT} (2024), introduced non-AR transformers to enhance generation speed without sacrificing quality. MAGNeT uses a masked generative approach, significantly reducing the time required to generate audio while still maintaining reasonable fidelity, thanks to its reliance on the EnCodec model for high-quality tokenised representations. \textbf{StemGen} (2024) took a different path by focusing on stem generation, offering greater flexibility in music production by generating individual instrumental tracks conditioned on existing stems and specific instrument categories.

In parallel, symbolic-domain language models have made strides in generating music using discrete musical representations. \textbf{MuseCoCo} (2023) introduced a two-stage framework to separate text-to-attribute understanding and attribute-to-music generation, offering precise control over the musical attributes described in text. Models like \textbf{ChatMusician} (2024) extended these ideas by leveraging large language models (LLMs) and training them on text-based music representations such as ABC notation, further demonstrating the versatility of language models in symbolic music generation. Despite limitations in representing complex musical structures with character-level notation, these models have shown significant potential in treating music generation as a natural language task.

\begin{sidewaystable}[htbp]
\caption{An overview of models for language modelling based text-to-music generation.}
\begin{tabular}{llllll}
\toprule
 & Dates   & Architecture                       & Conditioner & Output           & Sample Rate and Channels \\
\midrule
MusicLM                              & 2023.01 & SoundStream codec + Transformer    & MuLan       & Audio (Waveform) & 24kHz mono    \\
MusicGen                             & 2023.06 & EnCodec codec + Transformer        & T5          & Audio (Waveform) & 32kHz mono    \\
Magnet                               & 2024.02 & EnCodec codec + non-AR Transformer & T5          & Audio (Waveform) & 32kHz mono    \\
StemGen                              & 2024.01 & EnCodec codec + Transformer        & CLAP        & Audio (Waveform) & 32kHz mono    \\
\midrule
MuseCoCo                             & 2023.05 & Seq2seq Transformer                & Prefix      & Symbolic         &               \\
\citet{shangda2023} & 2023.06 & Seq2seq Transformer                & Prefix      & Symbolic         &               \\
ChatMusician                         & 2024.02 & Seq2seq Transformer                & Prefix      & Symbolic         &  \\
\bottomrule
\end{tabular}
\end{sidewaystable}

\subsubsection{Audio-domain Language Models}

\paragraph{MusicLM. } 

MusicLM~\citep{musiclm} is capable of generating music at 24 kHz that remains coherent over extended durations. MusicLM generates raw audio, ensuring a natural and fluid progression in the music it creates. The core architecture of MusicLM utilises a combination of SoundStream~\citep{rvq} for high-quality audio synthesis, w2v-BERT~\citep{w2vbert} for maintaining long-term semantic coherence, and MuLan~\citep{mulan} for joint music-text embeddings. The SoundStream model provides acoustic tokens, while w2v-BERT offers semantic tokens, ensuring that both fine acoustic details and overall musical structure are captured. MuLan embeddings are used to condition the model, with music embeddings employed during training and text embeddings used during inference. This hierarchical approach allows MusicLM to balance the generation of detailed and coherent music with adherence to textual descriptions.

MusicLM is trained on a large and diverse dataset consisting of 280,000 hours of music, enabling it to capture a wide range of musical styles and contexts. This extensive dataset allows the model to learn complex mappings between textual descriptions and musical outputs, ensuring high fidelity and stylistic accuracy in the generated music.

\paragraph{MusicGen}

MusicGen~\citep{musicgen} integrates a single-stage transformer LM with efficient token interleaving patterns. The transformer model processes sequences of tokens (discrete audio representations) efficiently, capturing long-range dependencies and enabling high-quality music generation from text inputs.

The model is trained end-to-end with text embeddings incorporated into the neural networks that parameterise the diffusion and autoregressive processes. This allows the model to learn the complex relationships between text and music. MusicGen models are trained on 30-second chunks of audio but can generate longer sequences with a windowing approach.

\paragraph{MAGNeT. }

MAGNeT~\citep{magnet} is designed to generate high-quality audio, including music, from textual descriptions using a single non-autoregressive transformer model. Unlike autoregressive models that generate audio sequentially, MAGNeT predicts spans of masked tokens in parallel, allowing for faster audio generation. This parallel approach makes MAGNeT significantly faster than models like MusicGen, though it does not achieve the same level of audio quality. MaGNeT's approach is similar to VampNet~\citep{vampnet}, while VampNet focuses more on unconditional music inpainting, and MaGNeT supports text conditioning.

The core component of MAGNeT is its non-autoregressive transformer architecture, which operates over multiple streams of audio tokens generated by EnCodec. This approach leverages the EnCodec model, which provides a high-quality latent representation of audio by using vector quantisation to encode the audio into discrete tokens. MAGNeT employs a unique masking strategy that focuses on spans of tokens rather than individual tokens, improving the efficiency and coherence of the generated audio. This approach contrasts with MusicGen, which uses an autoregressive transformer model, leading to differences in generation speed and quality.

MAGNeT is trained on a substantial dataset of 16,000 hours of audio, sourced from licenced datasets such as the Meta Music Initiative Sound Collection, Shutterstock music collection, and Pond5 music collection. However, the exact datasets used for training are not publicly available due to legal agreements with the rights holders, and only a dummy dataset is provided for illustrative purposes. 

\paragraph{StemGen. } StemGen~\citep{stemgen} is a music generation model that focuses on generating individual stems based on other stems, differing from other foundation models that typically aim to generate complete music tracks. This approach allows for more granular control and flexibility in music production, making it suitable for applications where individual instrumental tracks need to be manipulated independently.

StemGen employs a LLaMa-type transformer model, leveraging a context-mix and an instrument category as conditioning sources. This dual conditioning approach helps ensure that the generated stems align well with the existing context and specified instrumentation. The model operates on 32kHz audio, using EnCodec to encode the audio into discrete tokens, which are then processed by the transformer model. A trainable codebook with 18 entries, corresponding to the General MIDI instrument categories, is used to translate conditioning information into embeddings.

StemGen is trained on a combination of the publicly available Slakh dataset, which consists of 145 hours of synthetic musical audio separated into stems, and an internal dataset of 500 hours of licenced human-played music, also separated into stems. The datasets include metadata specifying the instrument type of each stem, allowing the model to learn detailed mappings between the context-mix, instrument category, and the generated stems. This extensive training setup enables StemGen to handle a wide range of musical contexts and instrumentation accurately.

\subsubsection{Symbolic-domain Language Models}

As symbolic music generation models, the large-scale models below represent the latest methods for text-to-music generation. I list and discuss them as a comparison with the audio language modelling methods.

\paragraph{MuseCoCo. } MuseCoCo~\citep{musecoco} is a model designed to generate symbolic music from text descriptions by employing a two-stage framework. This approach separates the task into text-to-attribute understanding and attribute-to-music generation stages, each trained independently. This separation allows MuseCoCo to effectively interpret and control various musical attributes specified in the text, such as genre, mood, tempo, and key, enhancing the model's ability to generate music that aligns well with user specifications.

The core components of MuseCoCo include a robust attribute extraction mechanism and a music generation model. The first stage, text-to-attribute understanding, uses a pretrained language model to extract diverse musical attributes from textual descriptions. These attributes are then transformed into prefix tokens, which guide the subsequent music generation process. In the second stage, attribute-to-music generation, the model leverages these tokens to produce symbolic music. This two-stage approach allows the model to handle large amounts of unlabeled data efficiently, using self-supervised methods to extract musical attributes and combining them to achieve complex musical expressions.

MuseCoCo is trained on extensive symbolic music datasets, allowing it to learn a wide variety of musical styles and structures. The training data consists of a large collection of symbolic music notations paired with textual descriptions, which include both objective attributes like tempo and key, and subjective attributes like mood and genre. This dataset is primarily constructed using publicly available symbolic music data, combined with synthesised text-attribute pairs to enrich the training set. 

\paragraph{\citet{shangda2023}} This study investigates the use of pretrained language models (BERT, GPT-2, and BART) for the task of text-to-music generation. It aims to generate complete and semantically consistent symbolic music scores directly from natural language descriptions. By leveraging pretrained checkpoints, the study explores how these models can be adapted for generating music notation from text, assessing their performance and the benefits they bring to the text-to-music generation process.

The core component of the study is the use of pretrained language models to generate music in ABC notation. This choice of notation, however, presents certain limitations due to its character-level tokenisation, which can struggle with capturing more complex musical nuances and dependencies. The models are fine-tuned on a dataset of 282,870 English text-music pairs, using publicly available checkpoints for BERT, GPT-2, and BART. The training involves initialising the model parameters from these checkpoints and adapting them for music generation tasks. Despite these efforts, the study acknowledges the inherent weaknesses of ABC notation in accurately representing the intricate details of music, which can limit the expressiveness and quality of the generated music.

The training data for this study is derived from a combination of public datasets, but the specific dataset used for training the models has not been publicly released due to copyright restrictions. Instead, the authors have made the WikiMusicText (WikiMT) dataset available, which includes 1,010 pairs of text-music data for evaluation purposes. This dataset helps in evaluating the performance of the models and provides a benchmark for future research. The study highlights the potential of pretrained language models in the music generation domain while also pointing out the challenges and limitations posed by the choice of music notation and the availability of high-quality training data.

\paragraph{ChatMusician.} ChatMusician~\citep{chatmusician} is a novel large language model (LLM) developed to understand and generate music using a text-compatible music representation known as ABC notation. This model integrates musical abilities intrinsically within the LLM framework, allowing it to process and generate music through pure text tokenisation without relying on external multi-modal structures. The primary aim of ChatMusician is to demonstrate that LLMs can be adapted to handle the language of music, treating it as a second language and enabling the generation of structured, full-length musical compositions conditioned on various musical attributes like chords, melodies, motifs, and forms.

The core components of ChatMusician include continual pre-training and fine-tuning of the LLaMA2 model on a curated dataset called MusicPile. MusicPile consists of 4.16 billion tokens, encompassing a variety of music scores, music-related knowledge, and general language data. The dataset includes public datasets such as Pile, Falcon-RefinedWeb, and Wikipedia, alongside specialised music data like KernScores and Bach music scores. Additionally, synthetic data generated with the help of GPT-4 supplements the training corpus, enhancing the model's musical understanding and generation capabilities.

Despite its innovative approach, ChatMusician's reliance on ABC notation presents limitations. ABC notation, a character-level music representation, can struggle to capture complex musical nuances and dependencies, potentially limiting the expressiveness and detail of the generated music. However, the model demonstrates significant potential by integrating musical capabilities into LLMs, showing that these models can serve as effective tools for music composition and understanding, even if they do not achieve state-of-the-art performance in symbolic and audio music generation tasks. ChatMusician offers a novel approach in the landscape of text-to-music generation by leveraging the capabilities of LLMs to process and generate music through text alone.


\subsection{Comparison Between Two Types of Backbone Models}

In the field of text-to-music generation, two primary model types have emerged: diffusion models and audio language models, each with unique advantages and challenges.

Diffusion models, such as those used in Moûsai, Noise2Music, TANGO, Mustango, AudioLDM, and MusicFlow, focus on iteratively refining an initial noisy signal to generate high-fidelity audio outputs. They excel in capturing fine-grained audio details and producing high-quality, coherent music. Advanced sampling methods like DDIM and DPM-Solver have significantly reduced the time required for generating samples, making diffusion models more efficient. Furthermore, techniques like progressive distillation can convert diffusion models to more efficient generative adversarial networks (GANs), balancing quality and speed. These models are well-suited for applications where audio fidelity and nuanced musical detail are paramount, despite requiring significant computational resources.

Audio language models like MusicGen and MusicLM leverage pretrained language models to understand and generate music. These models often use hierarchical structures to balance long-term coherence with detailed audio generation.  Techniques such as non-autoregressive sampling, as seen in MAGNeT, have further enhanced the efficiency of these models, making them faster and more suitable for real-time applications. These models are generally more efficient than diffusion models, providing reasonable audio quality with faster generation times. They are ideal for interactive music creation and scenarios requiring quick iterations.

The choice between these two types depends on the specific application requirements. Diffusion models are preferable for tasks that demand high-fidelity audio and detailed musical nuances, especially in offline settings where generation quality is more important than speed. On the other hand, audio language models are better suited for real-time applications and interactive scenarios due to their efficiency and quicker generation capabilities. The integration of fast sampling methods and efficient tokenisation techniques in both types of models continues to push the boundaries, making it crucial to consider the specific needs and constraints of the application when choosing the appropriate model type.

\section{Parameter-Efficient Finetuning (PEFT)}

Fine-tuning pretrained models is a critical process in adapting powerful machine learning systems to specific tasks, such as music generation. This process allows the models to leverage their general understanding while acquiring specialised knowledge tailored to the task at hand. Parameter-Efficient Fine-tuning (PEFT) is a subset of this process that seeks to achieve this adaptation with minimal increase in the number of parameters, thus maintaining computational efficiency and reducing the risk of overfitting.

One of the key challenges in music generation is capturing the complex and nuanced structures inherent in musical compositions. Traditional fine-tuning methods can be computationally expensive and may require a large number of additional parameters. PEFT techniques, such as LoRA and Llama-adapter, offer innovative solutions to this problem by introducing small, yet effective, modifications to the model's existing architecture.

LoRA, for instance, employs low-rank matrices to adapt the attention mechanisms of transformers. By fine-tuning the cross-attention and self-attention layers, LoRA allows the model to better understand and generate music with a nuanced grasp of its structure.

The Llama-adapter, although a conceptual technique, presents an intriguing approach to adapting the latent space of models like LLaMA, which are known for their efficiency in language tasks. By transforming the latent vectors with a matrix that represents musical characteristics, the Llama-adapter could potentially enhance the model's ability to generate music that aligns with specific stylistic or structural requirements.

\subsection{LoRA: Low-Rank Adaptation for Transformers}
LoRA~\citep{lora} is a technique designed to fine-tune pretrained models with minimal additional parameters. In the context of music generation, LoRA can be particularly effective for adapting the model's cross-attention and self-attention mechanisms to better capture the musical structure.

\begin{figure}
    \centering
    \includegraphics[width=0.5\linewidth]{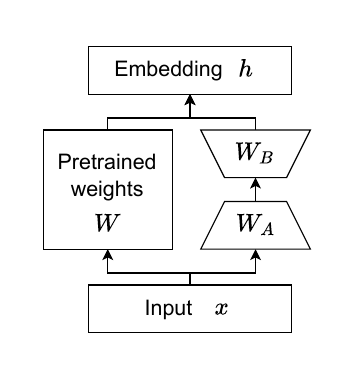}
    \caption{Diagram of the Low-Rank Adaptation (LoRA) algorithm. LoRA is a PEFT method that decomposes a large matrix into two smaller low-rank matrices in the attention layers, which facilitates model finetuning.}
    \label{fig:lora}
\end{figure}

\begin{itemize}
    \item \textbf{Cross-Attention in Transformers:} The cross-attention mechanism in transformers allows the model to focus on different parts of the input sequence when generating each token. By applying LoRA to the cross-attention layers, we can introduce low-rank matrices \( A \) that adapt the attention weights \( w \) as follows:
    \begin{equation}
    w' = A \cdot w,
    \end{equation}
    \noindent where \( A \) is a low-rank matrix that captures the fine-grained control needed for music generation.

    \item \textbf{Self-Attention in Transformers:} Self-attention allows the model to focus on different parts of the generated sequence itself. Fine-tuning this with LoRA involves adapting the self-attention weights \( v \) and \( k \) using a low-rank matrix \( B \) as follows:
    \begin{equation}
    v' = B \cdot v, \quad k' = B \cdot k
    \end{equation}
    This adaptation enables the model to better represent the internal structure of the music.

    \item \textbf{Diffusion Cross-Attention:} In models that combine diffusion processes with attention mechanisms, LoRA can be applied to the cross-attention layers to control the diffusion process. The diffusion cross-attention can be formulated as:
    \begin{equation}
    w' = D \cdot A \cdot w,
    \end{equation}
    \noindent where \( D \) represents the diffusion process matrix and \( A \) is the low-rank adaptation matrix.
\end{itemize}
\subsection{Llama-Adapter: Adaptation for Latent Space Control}

\highlight{The Llama-adapter is a fine-tuning module that could be used to adapt the latent space of a pretrained model like LLaMA~\citep{llama}}, which is known for its efficiency and effectiveness in language tasks. 

\begin{figure}[tb]
    \centering
    \includegraphics[width=0.7\linewidth]{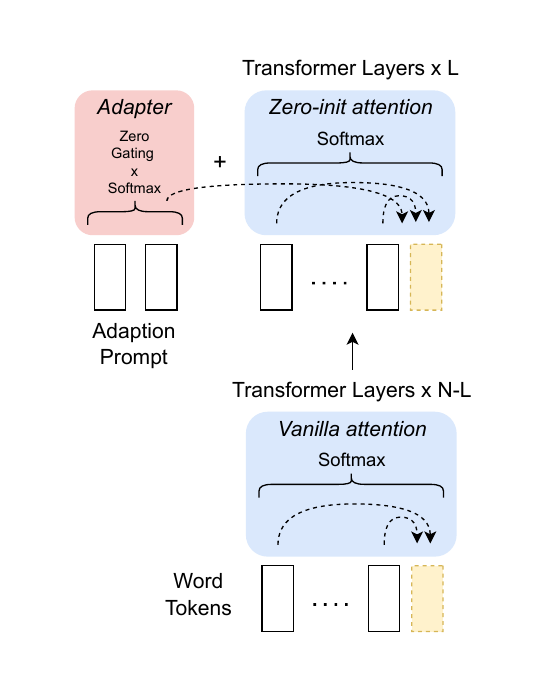}
    \caption{The diagram of Llama-adapter architecture.}
    \label{fig:llama-adapter}
\end{figure}

\begin{itemize}
    \item \textbf{Adapting Latent Space:} The Llama-adapter could be used to adapt the latent space of the LLaMA model to better represent musical features. This could involve modifying the latent vectors \( z \) with a transformation matrix \( T \) as follows:
    \begin{equation}
    z' = T \cdot z,
    \end{equation}
    \noindent where \( T \) is a matrix that captures the musical characteristics to be emphasised.

    \item \textbf{Application in Music Generation:} The adapted latent space could then be used to influence the generation process, ensuring that the generated music adheres to specific stylistic or structural requirements. This could be particularly useful for generating music in a specific genre or style.

    \item \textbf{Fine-Tuning with Music Data:} After adapting the latent space, the model could be further fine-tuned on a dataset of music to refine its understanding of musical patterns and structures. This would involve optimising the parameters of the Llama-adapter to minimise the difference between the generated music and the target music.
\end{itemize}

\section{Controllability and Editability of Pretrained Models}

The focus of this section is on how to enable and extend the controllability and editability of pretrained music generation models, making them more aligned with real-world music-making applications, such as music inpainting, conditional music generation, music variation generation, etc. This involves exploring various approaches and techniques used in the field, including integrating control mechanisms during model training, enhancing control post-training, and employing sophisticated editing methods.

\subsection{Approaches to Enhancing Controllability and Editability}

Previous approaches to control in small generative models often rely on disentangling representations in latent space~\citep{luo2019learning, polydis, xai}. For large-scale music generative models, control mechanisms are typically implemented through various strategies:

\begin{enumerate}
    \item \textbf{Integrating Trainable Control Modules into Pretrained Models:} This category focuses on augmenting pretrained large-scale models with additional control mechanisms, enabling the generation of music that adheres to specific parameters without retraining the entire model. Examples include Music ControlNet~\citep{musiccontrolnet}, Diff-A-Riff~\citep{diffariff}, and JASCO~\citep{jasco}, which introduce control modules or conditioning methods to adjust dynamics, melody, rhythm, and other musical attributes during the generation process.

    \item \textbf{Specialised Editing Models:} These models are designed explicitly for editing existing music content based on user instructions, often utilising text guidance to modify or enhance music. For instance, AUDIT~\citep{audit} and InstructME~\citep{instructme} perform various audio editing tasks such as adding, dropping, replacing, inpainting, and super-resolution, guided by human text instructions. They provide fine-grained control over the editing process, allowing users to adjust specific aspects of the music while preserving others.

    \item \textbf{Agent-Based Methods for Compositional Music Generation and Editing:} This approach employs agent-based systems or multi-modal AI frameworks to enable interactive and compositional music generation and editing, integrating large language models with other modalities. Examples include AudioGPT~\citep{audiogpt}, which incorporates audio foundation models to enhance large language models' capabilities in processing complex audio information and engaging in spoken dialogues, and M$^2$UGen~\citep{m2ugen}, which leverages large language models for multi-modal music understanding and generation from diverse sources such as music, images, and videos.

    \item \textbf{Inference-Time Optimisation Methods:} These methods focus on controlling and editing music generation during the inference stage of pretrained models without requiring additional training or fine-tuning. Techniques like DITTO~\citep{ditto}, SMITIN~\citep{smitin}, ZETA~\citep{zeta}, and the approach by \citet{simon2024} utilise inference-time optimisation, classifier probes, or zero-shot editing to steer the output of generative models toward desired musical characteristics. This allows for dynamic adjustments and fine-tuning of the generated music to better match user intentions.
    
\end{enumerate}

\subsection{Integrating Trainable Control Modules into Pretrained Models}

\subsubsection{Music ControlNet}

Music ControlNet~\citep{musiccontrolnet} presents a novel approach to enhancing the controllability of music generation models. It is pioneering work that adapts the ControlNet architecture, commonly used in image generation, for the domain of music. The primary goal is to introduce dynamic, melody, and rhythm controls that allow users to fine-tune the output of music generation models more precisely. By doing so, Music ControlNet aims to bridge the gap between automated music generation and real-world music production requirements.

Music ControlNet integrates control signals into the generative process of large-scale transformer-based models. The input consists of control signals and initial music input. These control signals, which include dynamic markings, melody lines, and rhythmic patterns, are processed through a series of transformations and then incorporated into the self-attention and cross-attention layers of the transformer. This integration allows the model to adjust various aspects of the music generation process based on the provided control signals, ensuring that the generated music adheres to user specifications.

The model architecture can be formally described as follows:

\begin{itemize}
    \item \textbf{Input Representation}: The input consists of control signals \( c \) and initial music input \( x \). Control signals can include dynamic markings, melody lines, or rhythmic patterns.
    \item \textbf{Control Module Integration}: The control module is integrated into the transformer-based generative model. The control signals \( c \) are processed through a series of transformations:
    \begin{equation}
    c' = W_c \cdot c + b_c,
    \end{equation}
    \noindent where \( W_c \) and \( b_c \) are weight and bias parameters for the control signals.
    \item \textbf{Attention Mechanisms}: The modified control signals \( c' \) are incorporated into the self-attention and cross-attention layers of the transformer. For self-attention, the attention weights \( w \) are adjusted based on \( c' \):
    \begin{equation}
    w' = A \cdot (w + c'),
    \end{equation}
    \noindent where \( A \) is a low-rank adaptation matrix. Similarly, for cross-attention, the process is:
    \begin{equation}
    w'' = B \cdot (w' + c').
    \end{equation}
    \item \textbf{Output Generation}: The adjusted attention weights are then used to generate the output music sequence \( y \):
    \begin{equation}
    y = \text{Decoder}(w'').
    \end{equation}
    This process ensures that the generated music adheres to the control signals provided by the user.
\end{itemize}

Music ControlNet is evaluated on a diverse dataset comprising various music genres and styles, including both MIDI files and audio recordings annotated with dynamic, melody, and rhythm information. The experiments focus on measuring the controllability and quality of the generated music. Results indicate significant improvements in controllability compared to baseline models, allowing users to effectively adjust dynamics, melody, and rhythm. The quality of the generated music, assessed through both objective metrics like pitch accuracy and rhythmic consistency and subjective human evaluations, demonstrates that Music ControlNet produces high-quality music that closely aligns with user specifications.

\subsubsection{Diff-A-Riff} 

\citet{diffariff} introduce a novel method for generating high-quality instrumental accompaniments designed to be adaptable to various musical contexts. The model is based on Latent Diffusion Models (LDMs), offering control through audio references, text prompts, or a combination of both, while producing 48kHz stereo audio. 

At the heart of the Diff-A-Riff system are two critical technological elements: a Consistency Autoencoder (CAE) with a high compression rate, which improves inference time and memory usage, and the expressive power of Elucidated Diffusion Models (EDMs). EDMs are known for their robust handling of complex data distributions and enhanced efficiency in model parameterisation and inference. The model is trained to reconstruct accompaniments given a context and a CLAP (Cross-modal Latent Audio Processing) embedding derived from a random sub-segment of the target accompaniment itself. 

\subsubsection{JASCO} 

\citet{jasco} present a model named JASCO, which aims to enhance the controllability and editability of text-to-music generation models. JASCO is a temporally controlled text-to-music generation model that utilises both symbolic and audio-based conditions. This approach allows for the generation of high-quality music samples that are conditioned on global text descriptions as well as fine-grained local controls. The model is based on the Flow Matching modelling paradigm combined with a novel conditioning method, enabling local and global control over music generation.

The components of JASCO include a source separation network for drum extraction, an F0 saliency detector model for melody extraction, and a chord progression extraction model for harmonic conditioning. The model leverages a continuous latent representation of audio, obtained from a compression model, and incorporates discrete token sequences for audio conditioning. JASCO applies information bottleneck layers in conjunction with temporal blurring to extract relevant information with respect to specific controls. This allows the incorporation of both symbolic and audio-based conditions in the same text-to-music model.

In terms of algorithmic formulas, the paper introduces the Conditional Flow Matching (CFM) objective function, which is defined as:
\begin{equation} L_\text{CFM}(\theta; z_0, z_1, t|Y) = | v_\theta(z, t|Y) - (z_1 - (1 - \sigma_{min}) \cdot z_0) |^2. \end{equation}
where \( z_0 \sim N(0, I) \) is sampled noise, \( z_1 \sim S \) is the latent representation of a data sample, and \( z = (1 - (1 - \sigma_{min}) \cdot t) \cdot z_0 + t \cdot z_1 \) is an interpolation between the noise and the data sample. The model is trained to predict the vector field of the continuous latent audio variable \( z \), given \( t \) and a set of conditions \( Y \). Additionally, the paper discusses the use of a weighted loss function during training, which is given by:
\begin{equation} L_\text{WeightedCFM} = \mathbb{E}_{t \sim U(0,1), z_0 \sim N(0,1), z_1 \sim S} \left[ (1 + t) \cdot L_{CFM}(\theta; z_0, z_1, t|Y) \right]. \end{equation}

\subsection{Specialised Editing Models}

\subsubsection{AUDIT}\label{sec:audit}

\citet{audit} presents AUDIT, a specialised text-guided audio editing model that operates based on latent diffusion models. This model is designed to perform various audio editing tasks such as adding, dropping, replacement, inpainting, and super-resolution, guided by human text instructions. Unlike previous methods that rely on pretrained text-to-audio models and require complete descriptions of the output audio, AUDIT utilises simple edit instructions, enhancing its flexibility and suitability for real-world applications. The model demonstrates a novel approach to integrating control mechanisms within audio editing tasks, focusing on the editability and controllability of the audio output based on textual guidance. This work pioneers the field of specialised text-guided audio editing models, providing a foundation for subsequent research in this domain.

AUDIT employs a control mechanism that leverages triplet training data consisting of instructions, input audio, and output audio. The model is trained in a supervised manner, utilising the input audio and text instructions as conditions to generate the edited audio output. This approach allows the model to automatically learn which segments of the audio need modification, thereby enhancing control over the editing process. The control mechanism in AUDIT is distinct as it does not require a full description of the target audio, aligning with the practical scenarios where detailed descriptions may not be available. This method of control is innovative as it directly uses the input audio as conditional input, forcing the model to ensure consistency of unedited segments before and after the editing process.

AUDIT comprises several components, including a variational autoencoder (VAE), a T5 text encoder, a diffusion network, and a vocoder. The VAE model projects the input mel-spectrogram into a latent space and reconstructs it, while the T5 encoder converts text instructions into embeddings. The diffusion network operates in the latent space, guided by the text embeddings and the latent representation of the input audio, to generate the edited audio. The vocoder then reconstructs the waveform from the output mel-spectrogram. The training loss of the autoencoder is expressed as:
\begin{equation} L_\text{VAE} = \lambda_1 L_1 + \lambda_2 L_2 + \lambda_\text{KL} L_\text{KL} + \lambda_\text{GAN} L_\text{GAN}, \end{equation}
where \( L_1 \) and \( L_2 \) are the L1 and L2 reconstruction losses, \( L_\text{KL} \) is the Kullback-Leibler loss, and \( L_\text{GAN} \) is the GAN loss. The latent diffusion model is trained to learn the distribution \( p(z_\text{out} | z_\text{in}, c_\text{text}) \), where \( z_\text{out} \) is the latent representation of the edited mel-spectrogram, \( z_\text{in} \) is the latent representation of the input mel-spectrogram, and \( c_\text{text} \) is the embedding of the editing instruction. The training loss for the latent diffusion model is given by:
\begin{equation} L_{LDM} = \mathbb{E}_{z_\text{in}, z_\text{out}, \text{instruction}} \left[ \mathbb{E}_{\epsilon \sim \mathcal{N}(0, I)} \left[ || \epsilon_{\theta}(z_t, t, z_\text{in}, c_\text{text}) - \epsilon ||^2 \right] \right]. \end{equation}

The model is trained and evaluated on several datasets, including AudioCaps, AudioSet, FSD50K, and ESC50. These datasets consist of audio clips with corresponding labels or captions, providing a rich source of data for training the triplet data required for the various editing tasks. The training process involves generating approximately 0.6 million triplet data points, covering the five editing tasks.

\subsubsection{InstructME}\label{sec:instructme}

\citet{instructme} introduce InstructME, a novel framework for instruction-guided music editing and remixing based on latent diffusion models. InstructME leverages natural language instructions to perform music editing tasks such as adding, removing, extracting, replacing, and remixing instrument tracks. The framework is designed to maintain the intrinsic harmony and coherence of music, which is often compromised by direct applications of image and audio modification techniques. InstructME employs a multi-scale aggregation strategy and incorporates chord progression matrices to enhance harmonic consistency during editing. Additionally, it utilises a chunk transformer to handle extended musical pieces by discerning long-term temporal dependencies. The framework's effectiveness is demonstrated through both subjective and objective evaluations, outperforming previous systems in music quality, text relevance, and harmony.

InstructME is composed of several key components, including a variational autoencoder (VAE), a T5 text encoder, a diffusion network, and a chunk transformer. The VAE model compresses the input audio into a latent representation, which is then reconstructed by the decoder. The T5 encoder converts text instructions into embeddings that guide the diffusion process. The diffusion model, based on a time-conditional U-Net, generates new audio samples from noisy embeddings, conditioned on text and source music embeddings. The chunk transformer addresses the computational inefficiency of self-attention in long music sequences by processing the data in chunks, thereby modelling long-term dependencies with reduced computational cost. The control mechanism in InstructME is achieved through a combination of classifier-free guidance and classifier guidance, allowing for both fine-grained semantic control and flexibility in generation.

The objective function for model optimisation is the reweighted bound:
\begin{equation} L_\text{DM} = \mathbb{E}_{\epsilon, t, z_0} \left[ | \epsilon - \epsilon_{\theta}(t, T(y), z_s, z_t) |_2^2 \right]. \end{equation}
In addition, the framework incorporates multi-scale aggregation and chord conditioning to improve consistency and harmony:
\begin{equation} L_\text{CDM} = \mathbb{E}_{\epsilon, t, z_0} \left[ | \epsilon - \epsilon_{\theta}(t, p_s, z_s, z_t) |_2^2 \right], \end{equation}
\noindent where \(p_s\) denotes the chord progression matrix of the source music.

InstructME is trained on a collection of 417 hours of music audio, resampled to a 24kHz sample rate and divided into non-overlapping 10-second clips. The training data includes triplets of text instructions, source music, and target music, covering remixing, adding, and replacement tasks.

\subsection{Agent-based Methods for Compositional Music Generation and Editing}

\subsubsection{AudioGPT}

\citet{audiogpt} introduces AudioGPT, a multi-modal AI system designed to enhance the capabilities of large language models (LLMs) in processing complex audio information and engaging in spoken dialogues. Unlike regular LLMs that primarily focus on text, AudioGPT incorporates audio foundation models to enable understanding and generation tasks across speech, music, sound, and talking head modalities. The system aims to improve the controllability and editability of pretrained text-to-music models by integrating these audio capabilities. The control mechanisms employed by AudioGPT include the use of speech recognition, text-to-speech, and audio-to-text models to facilitate dialogue and audio processing, thereby enhancing the system's ability to interact with users in a more natural and intuitive manner.

AudioGPT is composed of several components that work in tandem to achieve its objectives. The model can be divided into four main stages: 
1. Modality Transformation: This stage uses input/output interfaces such as automatic speech recognition (ASR) and text-to-speech (TTS) to transform between speech and text, bridging the gap between spoken language models and text-based LLMs.
2. Task Analysis: The dialogue engine and prompt manager in this stage help the LLM understand the user's intention to process audio information.
3. Model Assignment: Here, the LLM assigns audio foundation models based on structured arguments for prosody, timbre, and language control.
4. Response Generation: After the execution of foundation models, responses are generated and returned to the users.

These components collectively enable AudioGPT to perform a wide range of tasks, including generating audio from text, synthesising speech and singing voices, and processing audio for various applications.

\subsubsection{M$^2$UGen}\label{sec:m2ugen}

\citet{m2ugen} introduces a framework that leverages large language models (LLMs) for multi-modal music understanding and generation. The authors aim to bridge the gap in research that combines both understanding and generation using LLMs. The M$^2$UGen framework integrates the capabilities of LLMs to comprehend and generate music from diverse sources of inspiration, including music, images, and videos. This is achieved through the use of pretrained models such as MERT, ViT, ViViT, and LLaMA 2, and music generation models like AudioLDM 2 and MusicGen. The framework also employs the MULLaMA model to generate extensive datasets that support text/image/video-to-music generation, facilitating the training of the M$^2$UGen framework.

The M$^2$UGen framework employs a systematic approach to controllability and editability in text-to-music models. It incorporates multiple modal encoders to represent image, video, and music inputs, utilising ViT and ViViT for image and video modalities, and MERT as the music encoder. The feature representations from these encoders are fed into understanding adaptors, which are then comprehended by the LLaMA 2 model. For music generation, the framework explores two music decoders, AudioLDM 2 and MusicGen. The control mechanism in M2UGen is achieved through the integration of these components, allowing for the generation of music based on multi-modal inputs. The framework also supports music editing based on natural language prompts, enhancing its editability.

The M2UGen framework is evaluated on a range of tasks, including music question answering, text-to-music generation, image-to-music generation, video-to-music generation, and music editing. The authors have generated four datasets to train the model: MUCaps, MUImage, MUVideo, and MUEdit. These datasets consist of text-music pairs, image-music pairs, video-music pairs, and music editing instructions, respectively. The model is trained in a phased manner, initially focusing on comprehending diverse modalities, followed by refining the LLaMA 2 model's capability to condition music generation based on input captions. The final training stage employs a LoRA training strategy to fine-tune the LLaMA 2 model, multi-modal understanding adapters, and output projection layer.

\subsection{Inference-time Optimisation Methods}

\subsubsection{DITTO}

\citet{ditto} introduce a framework called Diffusion Inference-Time T-Optimisation (DITTO), which aims to enhance the controllability and editability of pretrained text-to-music diffusion models at inference time. Unlike methods that focus on integrated control during pre-training or injecting modules like adapters for control, DITTO operates by optimising the initial noise latents to achieve a target output. This approach allows for fine-grained control over various musical features without the need for further training or fine-tuning of the underlying model.

The control mechanism in DITTO is based on optimising an arbitrary differentiable feature matching loss. This enables the model to adjust its output to match a desired feature, such as intensity, melody, or musical structure. The framework leverages gradient checkpointing to ensure memory efficiency during the optimisation process. This method stands out as it provides a training-free way to control pretrained models, offering flexibility and efficiency in music generation tasks.

The core components of the DITTO model include the pretrained diffusion model, a feature extractor function \(f(\cdot)\), a loss function \(L\), and an optimiser. The algorithm operates by initialising the noise latents \(x_T\), running the diffusion sampling process to generate a music spectrogram \(x_0\), extracting features from the generated content, and then optimising the initial noise latents to fit any differentiable loss. The pseudo-code for the DITTO algorithm is as follows:

\begin{algorithm}
\caption{Pseudocode}
\begin{algorithmic}[1]
\State \textbf{Input:} $\epsilon_\theta$, Sampler, $T$, $f$, $L$, $y$, $x_T$, $c$, $K$, $g$
\For{$i = 1$ \textbf{to} $K$}
    \State $x_t \leftarrow x_T$ 
    \For{$t = T$ \textbf{to} $1$}
        \State $x_{t-1} = \text{Checkpoint}(\text{Sampler}, \epsilon_\theta, x_t, t, c)$
    \EndFor
    \State $\hat{y} = f(x_0)$
    \State $x_T \leftarrow x_T - g(\nabla_{x_T} L(\hat{y}, y))$
\EndFor
\State \textbf{Output:} $x_0$
\end{algorithmic}
\end{algorithm}

In this algorithm, \(\epsilon_\theta\) represents the diffusion model, \(T\) is the number of sampling steps, \(f\) is the feature extractor, \(L\) is the loss function, \(y\) is the target feature, \(x_T\) is the initial noise latent, \(c\) is the text conditioning, \(K\) is the number of optimisation steps, and \(g\) is the optimiser.

\subsubsection{SMITIN}

   \citet{smitin} present SMITIN (Self-Monitored Inference-Time INtervention), an innovative approach for controlling autoregressive generative music transformers. Unlike previous methods that focus on integrated control during pre-training or injecting modules for control, SMITIN employs classifier probes to steer the model's attention heads towards capturing specific musical traits. These probes are trained on a small dataset of audio examples and are used to dynamically adjust the intervention strength at each generation step. This ensures that the output incorporates the desired musical characteristics while maintaining temporal coherence. The method is validated for both audio continuation and text-to-music applications, demonstrating its effectiveness in adding controls to large generative models without the need for retraining or fine-tuning.

   The core components of the SMITIN model are the pretrained autoregressive music transformer, classifier probes, and a self-monitoring intervention mechanism. The transformer generates music by predicting a sequence of audio frames, with each layer consisting of multiple attention heads. The classifier probes are trained to recognise specific musical traits and are used to modify the attention heads' outputs. The self-monitoring intervention mechanism dynamically adjusts the intervention strength based on the probe output, preventing excessive intervention that could disrupt the music's temporal coherence. The model also incorporates a soft-weighting approach to determine the weights of the attention heads, eliminating the need for empirical tuning.

\subsubsection{ZETA}

\citet{zeta} explore two novel zero-shot editing techniques for audio signals, leveraging pretrained diffusion models through DDPM inversion. The first technique, named ZEro-shot Text-based Audio (ZETA) editing, adapts a method from the image domain, allowing for text-guided manipulations of audio signals. The second technique, called ZEro-shot UnSupervised (ZEUS) editing, introduces an unsupervised approach to discover semantically meaningful editing directions. These methods aim to enhance the controllability and editability of text-to-music models without the need for training or test-time optimisation, thereby providing a more flexible and efficient means of audio editing.

The control mechanism employed in this paper is primarily through text guidance in the ZETA technique and semantic perturbation in the ZEUS technique. In ZETA, the authors utilise classifier-free guidance (CFG) to steer the generative process towards the desired text prompt, enabling edits such as changing the style or genre of a musical piece. In contrast, ZEUS relies on perturbing the output of the denoiser in the directions of the top principal components (PCs) of the posterior. This unsupervised approach allows for the generation of variations that adhere to the original key, rhythm, and style, but are not limited by text descriptions.

The components of the model include the pretrained AudioLDM2 model, which operates in a latent space and generates mel-spectograms conditioned on text. These mel-spectograms are then decoded into waveforms using HiFi-GAN. The DDPM inversion is used to extract latent noise vectors corresponding to the source signal. For the ZETA technique, the generative process is controlled by changing the text prompt supplied to the denoiser model. For the ZEUS technique, the output of the denoiser is perturbed along the top PCs of the posterior covariance.

The algorithms presented in the paper involve the DDPM inversion process and the application of PCs for editing.  For the ZEUS technique, the perturbation of the denoiser's output is given by:
\begin{equation} x_{t-1} = \mu_t(x_t) + \gamma c_t \lambda_i^{1/2} v_{i|t'} + \sigma_t z_t, \quad t = T, \ldots, 1 \end{equation}
where \( c_t \) is a correction factor, \( \lambda_i^{1/2} \) and \( v_{i|t'} \) are the eigenvalues and eigenvectors of the posterior covariance, and \( \gamma \) is a user-chosen parameter controlling the strength of the modification.

\subsubsection{\citet{simon2024}} 

\citet{simon2024} propose two distinct strategies for conditioning a language model-based music generation system with audio input. The first strategy, known as textual inversion, utilises a pretrained text-to-music model to map audio input to corresponding ``pseudowords" in the textual embedding space. The second strategy involves training a music language model from scratch, jointly with a text conditioner and a quantised audio feature extractor. This approach allows for the combination of textual and audio conditioning during inference, balanced through a novel double classifier-free guidance method. The paper contributes to the field by adapting the textual inversion method to a pretrained text-to-music model and introducing a style conditioner method based on a frozen audio feature extractor, a transformer encoder, a Residual Vector Quantizer (RVQ), and temporal downsampling.

The model's components include a style conditioner that is jointly trained with the language model. During training, a 30-second music excerpt paired with a textual description is input to the model. The textual description is processed through a frozen T5 tokenizer and transformer encoder, while the style encoder takes a random subsample of the input audio and encodes it. The text and style latent representations are projected to match the dimension of the transformer language model and provided as a prefix to the sequence to model. The input audio is encoded by a pretrained EnCodec model, and the language model is trained autoregressively with a cross-entropy loss. Additionally, the tokens corresponding to the random subsample fed into the style encoder are masked in the loss to reduce the model's tendency to copy the style audio input.

The textual inversion method involves optimising the textual embedding by taking successive gradient steps on the cross-entropy loss of the music language model. The style conditioning method uses a style conditioner with bottlenecks (RVQ and downsampling) to prevent transmitting all the information of the conditioning audio excerpt to the model. The double classifier-free guidance formula is given by:
\begin{equation} l_{\text{double CFG}} = l_{\emptyset} + \alpha \left[ l_{\text{style}} + \beta (l_{\text{text},\text{style}} - l_{\text{style}}) - l_{\emptyset} \right],\end{equation}
\noindent where \( l_{\text{style}} \) and \( l_{\text{text},\text{style}} \) are the logits of the model conditioned on style and textual description, respectively, and \( \alpha \) and \( \beta \) are parameters used to balance the importance of the style and text conditioning.

\section{\highlight{Interactive Interfaces for Text-to-Music Systems}}

\highlight{Interactive music creation interfaces have evolved significantly with the integration of artificial intelligence (AI) and machine learning technologies. These interfaces aim to assist users in the music creation process by providing tools that are both powerful and intuitive. \citet{development} discussed the development and practice of AI technology for the production of contemporary popular music, highlighting the importance of integrating AI tools into the creative workflow of artists. }

\highlight{Before text-based interfaces were introduced, a number of interactive music creation interfaces already existed.} Some of these interfaces are built upon AI models~\citep{cococo, visualisation, iteratta}, while others extend traditional music software with AI capabilities, such as Band-in-a-Box.\footnote{\url{https://www.pgmusic.com/}}CoCoCo~\citep{cococo} is an interactive interface based on the CoCoNet model~\citep{coconet}, trained on Bach's chorales to assist users in composing four-part harmonies. The system allows users to steer the AI generation process, enabling a collaborative composition experience. \citet{visualisation} developed a front-end interface for the MelodyRNN model, where the system provides multiple candidate melodies for users to choose from and allows editing at different levels of granularity. These AI-based interfaces offer varying degrees of control over the music creation process; however, their functionality is typically tied to a single backend model, which limits their adaptability and the range of tasks they can support.

\highlight{Text-to-music generation can be regarded as a form of text-based interactive music creation and itself considered a new interface. While these models represent a significant advancement, their interfaces are often limited to simple text prompts and lack the interactive refinement capabilities that musicians may require.}

\begin{figure}[htbp]
    \centering
    \includegraphics[width=\linewidth]{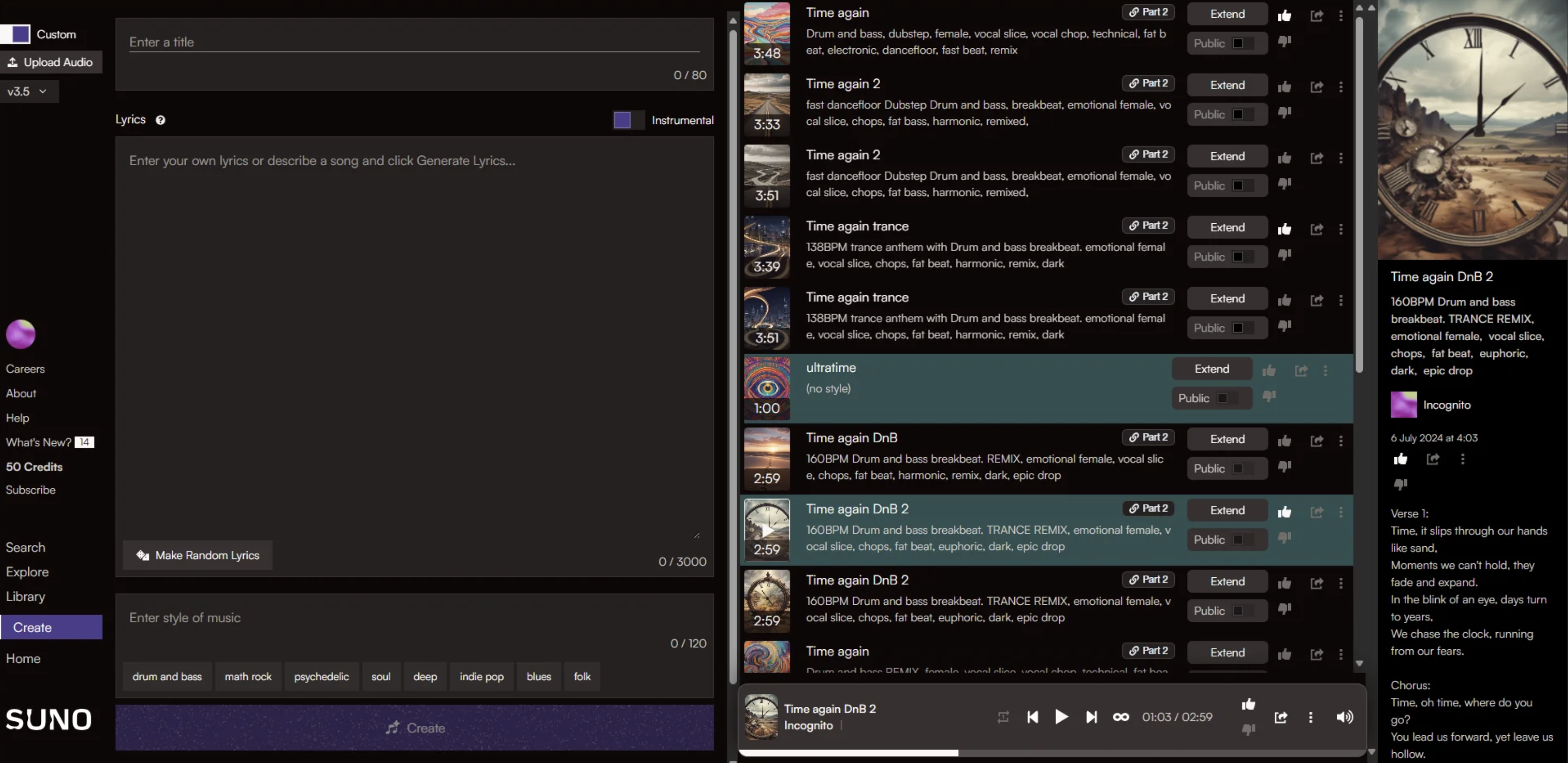}
    \caption{Screenshot of Suno. }
    \label{fig:suno}
\end{figure}

\begin{figure}[htbp]
    \centering
    \includegraphics[width=\linewidth]{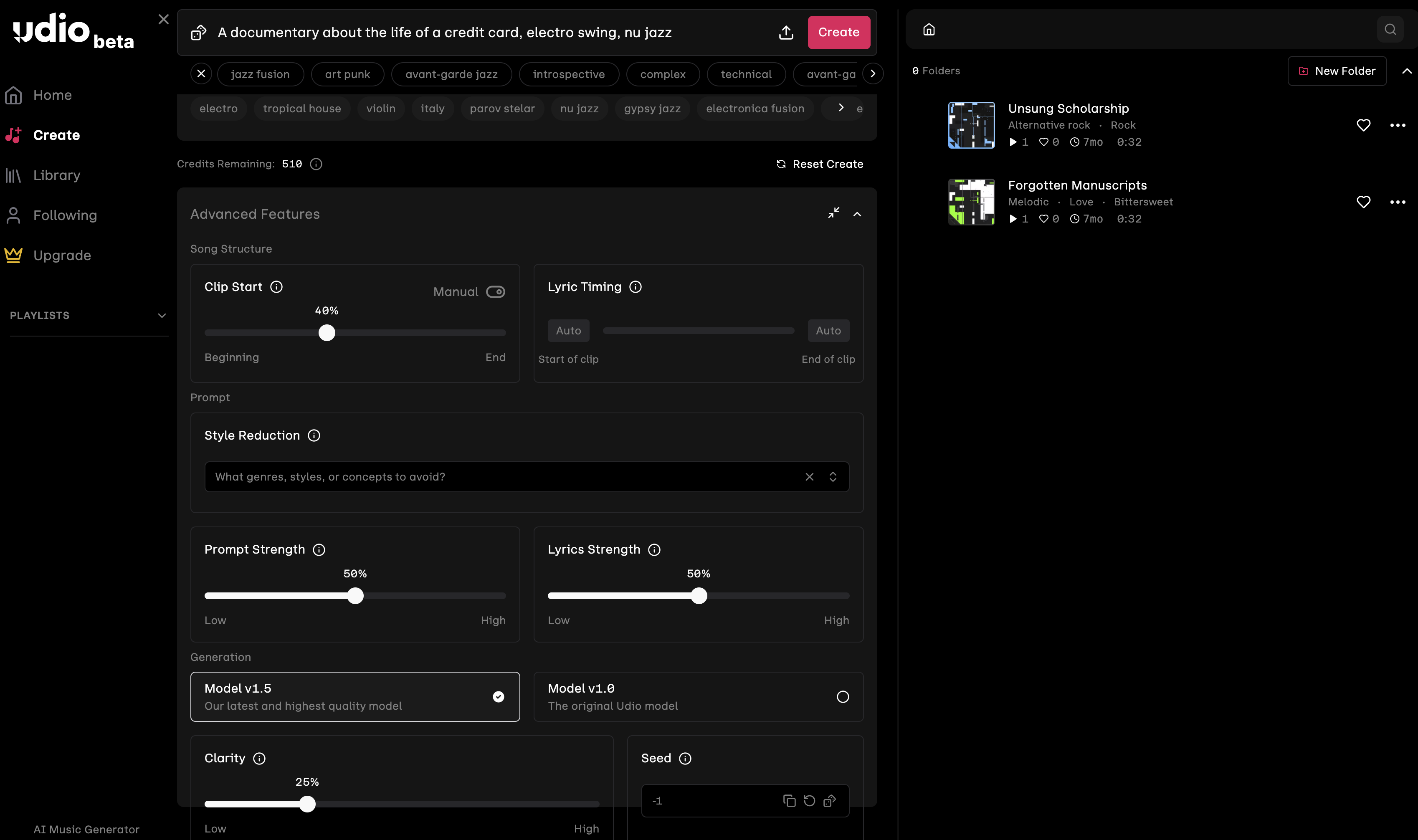}
    \caption{Screenshot of Udio. }
    \label{fig:udio}
\end{figure}

\highlight{Given that text-to-music research is still a relatively new field, there is limited HCI research focused on discovering how users should best interact with these models using improved interfaces. However, there are a few key developments in this area. One promising direction is the development of multiple control features, which can be enhanced through graphical user interfaces (GUIs). With the appearance of libraries like Gradio~\footnote{\url{https://www.gradio.app/}}, it has become easier to create standard interfaces that facilitate model interaction. Recent text-to-music models such as Suno~\footnote{\url{https://suno.com/}} and Udio~\footnote{\url{https://www.udio.com/}} have introduced user-friendly interfaces for tasks like music inpainting, music continuation, and lyric modification. These platforms aim to provide users with greater control over their music generation processes and incorporate a range of AIGC functionalities. For example, the interfaces of Suno (Figure~\ref{fig:suno}\footnote{Credit from \url{https://www.reddit.com/r/SunoAI/comments/1e1tumo/customized_create_screen_for_suno/}}) and Udio (Figure~\ref{fig:udio}) adopt the layouts similar to traditional music or audio editing software, often divided into three sections: the left panel for navigation or track management, the central panel for interaction with control modules (such as adjusting musical parameters or selecting generation options), and the right panel displaying a list of generated music outputs. This design, which closely resembles the structure of popular audio editing tools, appears to be becoming the standard in the field. On the product level, it is evident that many developers are converging on this layout, likely because it offers a familiar and intuitive user experience, allowing users to easily manage and refine their generated content.}

\highlight{Another development includes tools like WavTools~\footnote{\url{https://wavtool.com/}}, which integrates text prompts into existing Digital Audio Workstations (DAWs) to manipulate music elements. While such integrations are not yet mainstream, they represent a significant step toward combining the flexibility of traditional music production with the power of AI. However, there remains a gap in the literature regarding the effectiveness of these interfaces, particularly in terms of user experience and usability.} 

\highlight{Although these tools present promising new directions for text-to-music generation, there is a lack of research quantifying how useful or intuitive these interfaces are for users. \citet{paguri} conducted a user experience study with PAGURI, investigating how musicians interact with text-to-music models and incorporating personalisation techniques to better meet users' creative needs.}

Recent studies have also explored the use of text-to-audio models in interactive interfaces. \citet{iteratta} introduced IteraTTA, an interface that allows users to explore both text prompts and audio priors in generating music with text-to-audio models. 

These works highlight the potential and challenges of interactive music creation interfaces powered by AI. While significant progress has been made, there remains a need for interfaces that are both flexible and intuitive, supporting a wide range of creative tasks and allowing for natural interaction between the user and the system. Text-to-music generation, as a form of interactive interface, opens new possibilities for next-generation music creation tools, although it is not the only path forward. Our work seeks to demonstrate the potential of such interfaces while acknowledging the diversity of approaches in this evolving field.

\section{Datasets and Metrics}

\subsection{Text-to-Music Datasets}

Datasets play a crucial role in training and evaluating text-to-music generation models and interactive music creation interfaces. However, the availability of large-scale, publicly accessible datasets for music generation is limited due to licensing restrictions. Many models are trained on proprietary datasets comprising vast amounts of music data that are not publicly released.

For evaluation purposes, smaller datasets with text-audio pairs have been developed. {\textbf{MusicCaps}} \citep{musiclm} is one such dataset, consisting of 5.5k music-text pairs with rich textual descriptions provided by human experts. Each example in MusicCaps includes a 10-second music clip from the AudioSet dataset \citep{audioset} and a detailed caption focusing on how the music sounds, rather than metadata like artist names.

Another dataset is the {\textbf{Song Describer Dataset}} (SDD) \citep{mulab}, which contains approximately 1.1k captions for 706 music recordings. SDD is designed for evaluating models that address music-and-language tasks such as music captioning, text-to-music generation, and music-language retrieval. The dataset provides comprehensive descriptions that can be used to assess the quality of generated music in terms of adherence to textual prompts.

\highlight{The two datasets mentioned above are commonly used as standard evaluation benchmarks. Additionally, several other datasets have the potential to contribute to the training of text-to-music datasets. \citet{sid} introduced the \textbf{Song Interpretation Dataset}, which provides triplets of audio, lyrics, and lyric interpretations; Music captioning models \citet{lpmusiccaps,mullama,midicaps, jmla} can be used to generate synthetic text descriptions for music audio and to improve the performance of text-to-audio \citep{kong2024improving} and text-to-music \citep{xu2024generating} models as augmented data.}

\subsection{Constructing Instruction Datasets}

For text-based music editing and controllable music generation, standard datasets are scarce. Researchers often generate their own datasets by processing existing music data. 

In the context of controllable music generation, datasets with metadata or extracted audio features are commonly used to provide control signals. For example, models may use tempo, pitch, or instrumentation information extracted from music data to condition the generation process. However, there is no standardised dataset for controllable music generation, and researchers often rely on custom datasets tailored to their specific needs.

For instance, \citet{instructme} created the {InstructME} dataset by collecting 417 hours of music and generating triplet data (instruction, source music, target music) for various editing tasks such as remixing, adding, removing, extracting, and replacing musical elements. Similarly, \citet{audit} generated triplet training data for their model {AUDIT}, focusing on editing tasks like adding, dropping, replacement, inpainting, and super-resolution.

The workflow for triplet data generation in the InstructME dataset is as follows:

\begin{itemize}
    \item \textbf{Remix}: Randomly select a clip and mix all its instrument tracks to create the source music. Obtain rhythm and timing information from the source music, then retrieve another clip with the same rhythm to serve as the target music. Align the source and target music using time-step information and generate the instruction using a template like \textit{``Remix with \{instrument/genre\}''}.
    
    \item \textbf{Add}: Choose a clip and select \( i \) instrument tracks (where \( i \in [1, 2, 3, 4] \)) to mix into the source music. Create the target music by adding another instrument track to the source music. The instruction follows a template such as \textit{``Add \{instrument\}''}.
    
    \item \textbf{Remove}: Reverse the adding operation by using the target music from the adding task as the source and the source music as the target. The instruction template is \textit{``Remove \{instrument\}''}.
    
    \item \textbf{Extract}: Select one instrument track from a clip to serve as the target music. Mix this track with additional instruments to create the source music. The instruction is \textit{``Extract \{instrument\}''}.
    
    \item \textbf{Replace}: Choose two different instrument tracks and mix them with other instruments to form the source and target music. The instruction template is \textit{``Replace \{instrument A\} with \{instrument B\}''}.
\end{itemize}

They provided specific text command templates and examples for each task to facilitate the model's learning of diverse editing instructions. Examples include \textit{``Add distorted electric guitar,'' ``Remove accordion,'' ``Extract viola,'' ``Replace flute with accordion,''} and \textit{``Remix with drums, bass, guitar, piano.''} This structured approach to dataset construction enabled the model to learn various editing operations in a controlled and scalable manner.

\begin{table}[h!]
\centering
\begin{tabular}{ll}
\toprule
\textbf{Task} & \textbf{Text Command Template} \\
\midrule
Remix   & Remix with \{instrument/genre\} \\
Add     & Add \{instrument\} \\
Remove  & Remove \{instrument\} \\
Extract & Extract \{instrument\} \\
Replace & Replace \{instrument A\} with \{instrument B\} \\
\bottomrule
\end{tabular}
\caption{Text command templates used for generating instructions in the InstructME dataset~\cite{instructme}.}
\end{table}

\subsection{Evaluation Metrics}

Evaluating the performance of text-to-music generation models involves both \highlight{objective} and subjective metrics. \highlight{Objective} metrics provide objective measures of the quality and relevance of the generated music. \highlight{For instance, metrics like Fréchet Distance (FD) can reflect perceptual audio quality by comparing statistical distributions of audio features, while metrics like KL divergence and CLAP score can assess semantic similarity and alignment with the input prompt. Subjective metrics, on the other hand, are designed to evaluate human perceptions of attributes such as musicality, creativity, and adherence to input prompts.}

\subsubsection{\highlight{Objective} Metrics}

Several quantitative metrics have been proposed to evaluate different aspects of generated music:

\begin{itemize}
    \item \textbf{Fréchet Distance (FD)}: The Fréchet Distance measures the similarity between the statistical distributions of the generated audio and a reference dataset. Traditionally, FD is computed using features extracted by models like VGGish \citep{fad}, which operates at 16~kHz. However, for high-fidelity audio at higher sampling rates, \textbf{FD\textsubscript{OpenL3}} has been introduced. OpenL3 \citep{openl3} accepts audio signals up to 48~kHz, allowing for evaluation of full-bandwidth audio. FD\textsubscript{OpenL3} computes the Fréchet Distance in the OpenL3 feature space, capturing perceptual aspects of audio quality and diversity in generated samples.

    \item \textbf{Kullback-Leibler Divergence (KL)}: The KL divergence measures the difference between two probability distributions. In this context, \textbf{KL\textsubscript{PaSST}} uses PaSST \citep{passt}, a state-of-the-art audio tagging model trained on AudioSet \citep{audioset}, to compute the KL divergence over the predicted class probabilities between the generated and reference audio. A lower KL\textsubscript{PaSST} indicates that the generated audio shares similar semantic content with the reference audio, reflecting the model's ability to generate relevant content.

    \item \textbf{CLAP Score}: The CLAP (Contrastive Language-Audio Pretraining) score \citep{clap} computes the cosine similarity between the embeddings of the input text prompt and the generated audio, as obtained from a CLAP model. A higher CLAP score indicates better alignment between the textual description and the generated audio, measuring how well the model captures the semantics of the input prompt.

\end{itemize}

These metrics have been adapted to handle variable-length and high-resolution audio. For instance, to evaluate long-form audio, KL\textsubscript{PaSST} can be computed by segmenting the audio into overlapping windows and averaging the predictions \citep{musicgen}. Similarly, the CLAP score can be calculated using a feature fusion approach, combining embeddings from different segments of the audio to handle durations longer than the model's default input length \citep{musicgen}. These adaptations ensure that the metrics accurately reflect the quality and relevance of longer and higher-fidelity audio samples generated by modern models.

\subsubsection{Subjective Metrics}

Subjective evaluation remains crucial in assessing the quality of generated music, as it captures human perceptions that are difficult to quantify objectively. Human listeners are typically asked to rate samples based on criteria such as:

\begin{itemize}
    \item \textbf{Overall Quality (OVL)}: Listeners rate the perceptual quality of the audio, considering factors like sound fidelity, absence of artifacts, and general pleasantness.

    \item \textbf{Relevance to Text Input (REL)}: Listeners assess how well the generated music matches the given text prompt, evaluating the semantic alignment between the description and the audio.

\end{itemize}

Studies like those by \citet{musicgen} and \citet{noise2music} have conducted human evaluations using platforms such as Amazon Mechanical Turk. These studies ensure the reliability of the evaluations by filtering out outliers and ensuring that listeners fully engage with the samples. For instance, \citet{musicgen} use the CrowdMOS package to filter noisy annotations and remove unreliable raters. Subjective evaluations provide insights into the musicality and creative aspects of the generated music that are not captured by quantitative metrics.

\subsubsection{Metrics for Music Editing}

For music editing tasks, additional metrics are essential for evaluating the effectiveness of editing operations. 

Metrics such as the Scale-Invariant Signal-to-Distortion Ratio (SI-SDR) \citep{si-sdr} are commonly used to assess the quality of source separation or the removal of specific tracks from a mix. SI-SDR evaluates the fidelity of the reconstructed signal compared to the original source while disregarding scale differences.

Furthermore, classifiers can be used to verify whether certain attributes have been successfully modified, such as confirming whether an instrument has been replaced or a genre has been changed. Pretrained classifiers or evaluators provide a quantitative measure of whether the desired editing operations have been performed as expected.

Several additional metrics can also be employed to address specific aspects of music editing. For instance, in \textbf{InstructME}, the instruction accuracy (IA) metric measures the relevance of the text-music pair by calculating the accuracy of edited music tags, such as instrument, mood, and genre, compared to the input command.

Similarly, in \textbf{Music ControlNet}, various metrics are used to evaluate the editing process. Melody accuracy checks whether the pitch classes (C, C\#, etc.) align between the input melody control and the generated output. Dynamics correlation, calculated using Pearson’s correlation, assesses the relationship between the input and generated dynamics values, both on a micro and macro scale. Micro correlation focuses on individual generations, while macro correlation evaluates across multiple generations. Finally, Rhythm F1, based on beat/downbeat detection, quantifies the alignment of timestamps between the input rhythm control and the generated output, considering alignment within 70 milliseconds.

\section{Conclusion}

In conclusion, this chapter has provided a comprehensive overview of the key concepts, models, and techniques that form the foundation of text-to-music generation and editing systems. I have explored various representations of music and text, examined the architecture and functionality of backbone models such as diffusion models and audio language models, and discussed the importance of controllability and editability in music generation.

The background presented here sets the stage for the novel contributions of this thesis, which directly address some of the most pressing challenges in the field of AI-driven music creation and manipulation. Our work builds upon and extends the concepts discussed in this chapter in several key ways.

\begin{enumerate}
    \item \textbf{MusicMagus: Inference-Time Optimisation Model for Music Editing. }This contribution leverages the understanding of backbone models and control mechanisms to develop a novel approach that allows for precise editing of generated music at inference time. This work aims to provide a more flexible and responsive tool for the music editing task.
    \item \textbf{Instruct-MusicGen: Instruct tuning for music editing.} Building on the Parameter-Efficient Finetuning (PEFT) techniques discussed, such as LoRA and Llama-Adapter, our instruct tuning approach demonstrates how pretrained models can be efficiently adapted for specific music editing tasks. This work showcases the practical application of the theoretical concepts of model adaptation and specialised editing models outlined in this chapter.
    \item \textbf{Loop Copilot: Agent-based music editing model.} This work combines the power of large language models with multiple music generation models. It builds upon the agent-based methods for compositional music generation and editing discussed in this chapter, extending these concepts to create a more interactive and intuitive music editing experience.
\end{enumerate}

These contributions collectively advance the state-of-the-art in controllable and editable music generation, which are two key challenges identified in our overview of current models and techniques. By grounding our work in the fundamental concepts and recent advancements outlined in this chapter, we position our research at the forefront of AI-driven music creation and manipulation.

The following chapters will delve deeper into the methodologies, implementations, and results of these novel approaches, demonstrating how they push the boundaries of what is possible in AI-assisted music composition and editing.

\chapter{Loop Copilot: Conducting AI Ensembles for Music Generation and Iterative Editing}
\label{ch:loop}

This chapter represents the initial exploration into the challenge of text-based music editing within the context of AI-assisted music creation. The work presented here builds on two contributions: \textbf{COSMIC}, an early-stage proof of concept, and \textbf{Loop Copilot}, an extended system that addresses some of the limitations observed in COSMIC. \textbf{COSMIC} proposed the innovative idea of using a combination of multiple specialised music models, coordinated by a state machine, to facilitate music composition through a simple text-based interface. However, the system's reliance on a manually designed state machine and its limited interaction capabilities highlighted the need for a more adaptable and user-friendly approach. \textbf{Loop Copilot} extends this concept by integrating a large language model (LLM) as an AI agent, thereby improving the system’s usability and making the interaction more natural for users. This chapter also includes a user study to evaluate the effectiveness and limitations of Loop Copilot in real-world scenarios.

The core contribution of this chapter lies in the introduction and detailed examination of Loop Copilot. The system is designed to enable users to generate and iteratively refine music through a multi-round dialogue interface, powered by an LLM that interprets user intentions and selects appropriate AI models for specific tasks. These models, each specialised in different aspects of music creation, work together to achieve the user's goals, with their outputs being aggregated to maintain coherence in the musical piece. A key innovation in Loop Copilot is the Global Attribute Table (GAT), a dynamic data structure that records and preserves essential musical attributes throughout the iterative process. This ensures that the integrity of the music is maintained as modifications are made. The chapter concludes with a user study, conducted through semi-structured interviews and questionnaires, which provides valuable insights into the system’s practical utility, as well as its strengths and areas for improvement.

Positioned as the foundational work in this thesis, this chapter sets the stage for the subsequent exploration of more advanced text-guided music editing techniques. While Loop Copilot demonstrates the feasibility and potential of a multi-model approach to music creation, it also exposes limitations, particularly in its ability to perform precise and detailed edits. These limitations serve as a catalyst for the development of the more sophisticated methods discussed in the following chapters, particularly MusicMagus and Instruct-MusicGen, which aim to address these challenges and further advance the state of the art in AI-assisted music production.\footnote{Code available at \url{https://github.com/ldzhangyx/loop-copilot}.}

\section{Introduction}

Music creation is an art that has traditionally been the domain of expert human musicians. Recently, with the advent of artificial intelligence (AI) music models~\citep{survey}, the music creation process is becoming more democratised. However, in the real world, there are two major challenges in the human music creation process: first, music creation involves multiple phased tasks, from harmony and melody crafting, to arrangement and mixing; second, music creation is an inherently iterative process that cannot be achieved in one step. It usually undergoes multiple refinements before reaching its final form. Most current AI models, including interactive music interfaces and dedicated generative models, fall short in at least one of these two challenges.

\begin{figure}[tbp]
\centering
  \includegraphics[width=\linewidth]{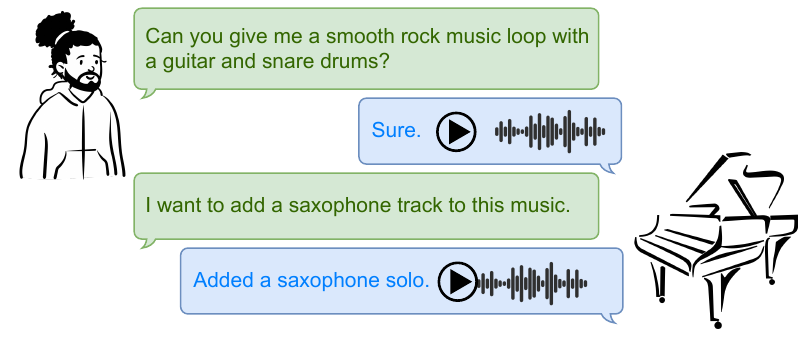}
  \caption{A conceptual illustration of interaction with Loop Copilot. The diagram depicts a two-round conversation: initially, a user requests music generation and the AI provides a loop. In the subsequent round, the user seeks modifications, and the AI offers a refined loop, emphasising Loop Copilot's iterative feedback-driven music creation process.}
  \label{fig:teaser}
\end{figure}

Interactive music interfaces excel in melody inpainting ~\citep{cococo, visualisation} but often lack adaptability for diverse music creation. Current popular interactive music interfaces, such as Magenta Studio~\citep{magentastudio} and Flow Machines~\footnote{\url{https://www.flow-machines.com/}}, are powerful and user-friendly, but they predominantly focus on a singular type of musical modification: melody inpainting—filling in gaps based on an existing melody. These models, with their intuitive human-in-the-loop interactions, have undoubtedly lowered the entry barrier for users. However, these AI-based interfaces for music creation, although recognising the importance of iterative generation and refinement, often rely on a single task throughout the process. This reliance not only hampers their flexibility but also restricts their adaptability to diverse music creation needs. 

On the other hand, dedicated music models offer broad capabilities but tend to have a narrow focus, limiting their application. Existing dedicated music generative models have demonstrated significant capabilities across a myriad of tasks in music creation, such as controlled music generation using chord progressions~\citep{polyffusion, inpainting}, text prompts~\citep{musicgen,musiclm}, images \citep{vis2mus}, and emotion~\citep{fadernets}. They also span a spectrum of music style transfer tasks at the score~\citep{polydis,accomontage}, performance~\citep{mididdsp}, and timbre~\citep{timbretransfer} levels. However, a prevalent issue with these models is their `one-off' design approach. They often treat music generation as a singular process, either focusing strictly on music generation or specific editing tasks, like style transfer. As a result, users looking to engage in a comprehensive music creation process find themselves scouting for various models to cater to different aspects of their musical needs.

This chapter introduces Loop Copilot, a system designed to address these challenges. It allows users to generate a music loop and iteratively refine it through a multi-round dialogue with the system. By leveraging a large language model~\citep{llmsurvey}, Loop Copilot seamlessly integrates various specialised models catering to different phases of music creation. It harnesses the power of individual models to provide a rich set of generation and editing tools. The intuitive and unified interaction is facilitated through a conversational interface, reminiscent of the benefits of the first type of above models, while applying the strengths of the second type. 

Loop Copilot is built on three key components: a large language model (LLM) controller, which interprets user intentions, selects suitable AI models for task execution, and gathers the outputs of these models; a set of backend AI models, which carry out specific tasks; and a Global Attribute Table (GAT), which records necessary music attribute information to ensure continuity throughout the creation process. Intuitively, users can utilise the LLM to `conduct' the AI ensemble, guiding the music creation process through conversation. 

In summary, the main contributions are:

\begin{enumerate}
    \item This chapter introduces Loop Copilot, a novel system that integrates LLMs with specialised AI music models. This enables a conversational interface for collaborative human-AI creation of music loops. 
    \item This chapter develops the Global Attribute Table that serves as a dynamic state recorder for the music loop under construction, thereby ensuring that the musical attributes remain consistent in the iterative editing process.
    \item This chapter conducts an interview-based comprehensive evaluation, which not only measures the performance of our system but also sheds light on the advantages and limitations of using an LLM-driven iterative editing interface in music co-creation.
\end{enumerate}

\section{Preliminary Work: COSMIC}

The COSMIC system serves as the foundation for Loop Copilot, sharing the underlying concept of using a form-filling approach to maintain consistency in music creation. 

In COSMIC, the design focuses on two main aspects: building a dialogue system and enabling controlled music generation. The dialogue system is adaptable to the needs of music composition through the use of a dialogue tracker, which creates a form instance for each music creation task. This form records the status of the music creation and translates the user's inputs into form-filling actions. Once the form is completed, COSMIC passes the information to the backend models responsible for music and lyric generation. In the current version, COSMIC employs the BUTTER model~\citep{butter} for controlled music generation, which utilises natural language as a condition, along with a GPT-2 based model for controlled lyric generation. 

\begin{figure}[htbp]
	\centering
		\includegraphics[width=1\textwidth]{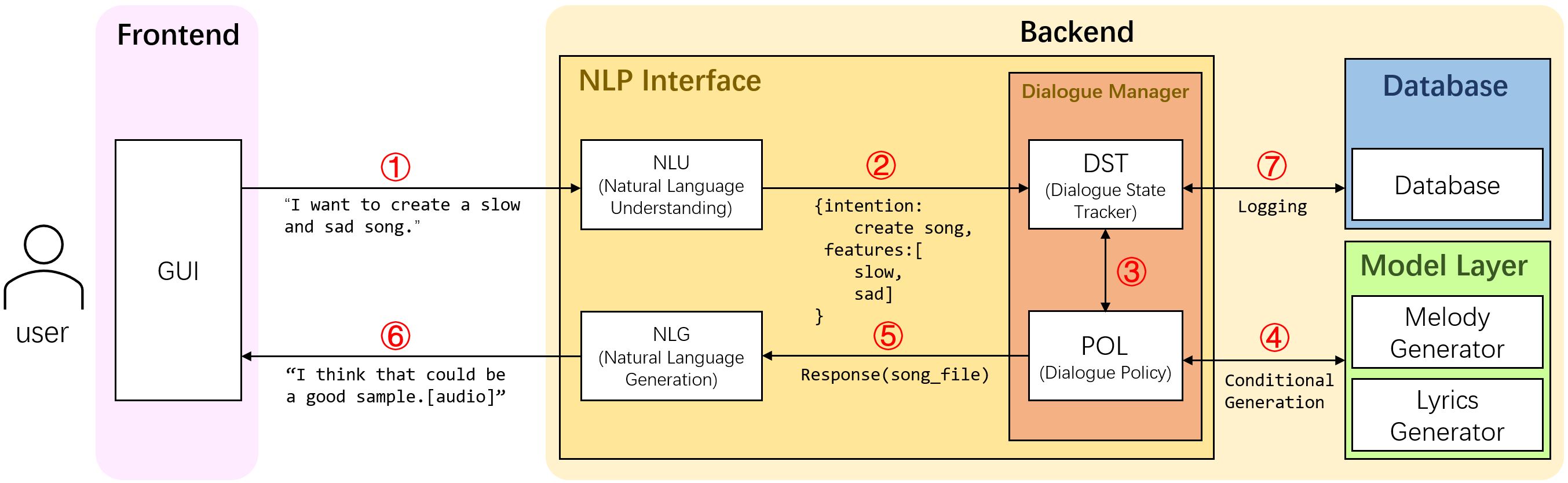}
	\caption{A diagram of COSMIC's handling of a single round of dialogue.}
	\label{fig:steps}
\end{figure}

Figure \ref{fig:steps} shows the working mechanism of COSMIC during a single-round dialogue with an example. The user says to COSMIC, "I want to create a slow and sad piece of music".

\begin{enumerate}
    \item This sentence is first transferred from COSMIC's front-end to the NLP interface in RESTful format and received by the Natural Language Understanding (NLU) module, as shown in step 1. 
    \item The NLU module parses the user's requirements and keywords from text input and transfers them to the Dialogue State Tracker (DST) module in Dialogue Manager in json format, as shown in step 2. 
    \item After receiving the data, the DST module fills in the creation form and determines the current dialogue status. After determining the status, the DST transmits the creation form and the current status to the POL module as shown in step 3. 
    \item The Dialogue Policy (POL) module is responsible for converting the creation table into a condition and sending conditional generation commands to the model layer as shown in step 4. 
    \item After POL gets the generated music data from the model layer, it sends a response command to the NLG module as shown in step 5. 
    \item The Natural Language Generation (NLG) module translates the command into natural language that humans can understand and sends it back to the COSMIC front-end along with the music data as shown in step 6; 
    \item Finally, COSMIC logs this session to the database as shown in step 7.
\end{enumerate}

\textbf{Loop Copilot} extends this concept by integrating a large language model (LLM) as an AI agent, thereby improving the system’s usability and making the interaction more natural for users. 

\section{System Design of Loop Copilot}

\subsubsection{Model Formulation}

I begin with an example of a typical interaction process, shown in Figure \ref{fig:teaser}.  It comprises two key steps: the user initially:

\begin{enumerate}
    \item \textbf{drafts a music loop} (\textit{``Can you give me a smooth rock music loop with a guitar and snare drums?"});
    \item \textbf{iteratively refines it} through multiple rounds of dialogue (\textit{``I want to add a saxophone track to this music."}).
\end{enumerate}

After completing the 2-round dialogue, the current status can be represented by a sequence $[(Q_1, A_1), (Q_2, A_2)]$, where $Q$ means the user's question and $A$ means the answer of Loop Copilot.

To formally define the interaction process, let us consider a sequence $H_T = [(Q_1, A_1), ..., (Q_T, A_T)]$, where each $(Q_t, A_t)$ pair denotes a user query and the corresponding system response in the $t$-th round dialogue. At each step $t$, the system generates a response $A_t$ using the Loop Copilot function: 

$$A_t = \text{LoopCopilot}(Q_t, H_{t-1}).$$

\begin{figure}[htbp]
    \centering
\includegraphics[width=\linewidth]{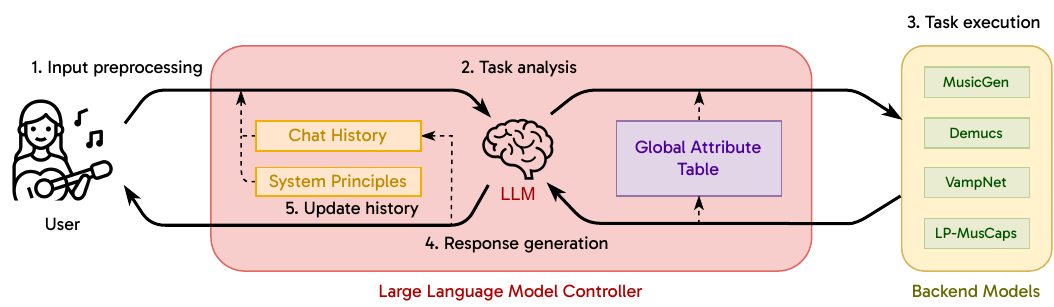}
    \caption{Loop Copilot's workflow. Once the user inputs the request, firstly, Loop Copilot \textbf{preprocesses the input} and converts it to textual modality; secondly, the LLM, based on the input, the system principles, and the chat history, performs the \textbf{task analysis} and calls the corresponding models; after that, the backend models \textbf{execute the task} and output the result; finally, the LLM does the \textbf{final processing} of the output and returns it.}
    \label{fig:diagram}
\end{figure}

Figure \ref{fig:diagram} shows the workflow of our proposed system. Loop Copilot comprises 5 key components: (1) The \textbf{large language model ($\mathcal{M}$)} for understanding and reasoning; (2) The \textbf{system principles ($\mathcal{P}$)} that provide basic rules to guide the large language model; (3) A list of \textbf{backend models ($\mathcal{F}$)} responsible for executing specific tasks; (4) A \textbf{global attribute table ($\mathcal{T}$)} that maintains crucial information to ensure continuity throughout the creative process; and (5) A \textbf{framework handler ($\mathcal{D}$)} that orchestrates the interactions between these components.

%

The workflow of Loop Copilot involves several steps. 

\begin{enumerate}
    \item \textbf{Input preprocessing.} the system processes the input by unifying the modality of the input. The framework handler $\mathcal{D}$ utilises a music captioning model to describe the input music, while textual inputs are kept as they are.
    \item \textbf{Task analysis.} the framework handler performs task analysis if the text input contains an explicit demand. It calls the large language model $\mathcal{M}$ to analyse the task, resulting in a sequence of steps, which may involve a call to a single model or multiple chained calls to models, as the large language model may need to handle the task step by step. Section \ref{sec:supported_tasks} demonstrates the details.
    \item \textbf{Task execution.} After task analysis, the framework handler records all the steps and proceeds to execute the tasks. It calls the backend models in the specified order, providing them with the necessary parameters obtained from the large language model. If it requires a chained call of multiple models, the intermediate results generated by the previous model will be used in the next model.
    \item \textbf{Response generation. } Once the task execution is complete, the handler $\mathcal{D}$ collects the final result and sends it to the large language model for the final output. 
\end{enumerate}

Throughout this process, all operations are tracked and recorded in the global attribute table $\mathcal{T}$, ensuring consistency and continuity in the generation process. Algorithm \ref{alg:workflow} illustrates the process during a T-round dialogue. 

\begin{algorithm}[htbp]
\caption{The workflow of Loop Copilot}\label{alg:workflow}
\renewcommand{\algorithmicrequire}{\textbf{Input:}}
\renewcommand{\algorithmicensure}{\textbf{Output:}}
\begin{algorithmic}
\Require user queries $Q = \{Q_1, ..., Q_T\}$
\Ensure responses $A = \{A_1, ..., A_T\}$
\State Initialise components: $\mathcal{M}, \mathcal{P}, \mathcal{F}, \mathcal{T}, \mathcal{D}$ 
\State Initialise chat history $H_0$ 
\State Define $A_0$ as initial music state or silence
\highlight{\State At timestep $T$:}
    \State $Q'_t \gets \mathcal{D}(Q_t)$ \Comment{\textbf{Input preprocessing}}
    \State $\mathcal{F}_{1:N} \gets \mathcal{M}(Q'_t, H_{t-1})$ \Comment{\textbf{Task analysis}}
    \State $A'_{t,0} \gets A_{t-1}$ \Comment{initialise the chain}
    \For{$n$ in $[1, N]$}
        \State $A'_{t, n} \gets \mathcal{F}_n(A'_{t,n-1})$ \Comment{\textbf{Task execution}}
    \EndFor
    \State $A_t \gets \mathcal{M}(A'_{t,N})$ \Comment{\textbf{Response generation}}
    \State $H_t \gets$ Append$(H_{t-1}, (Q_t, A_t))$  \Comment{Update chat history}
    \State Update $\mathcal{T}$ with key attributes from $A_t$ 

\end{algorithmic}
\end{algorithm}

\subsubsection{Supported Tasks} \label{sec:supported_tasks}

The interaction process within Loop Copilot is essentially a multi-stage workflow, as illustrated in Figures \ref{fig:teaser} and \ref{fig:diagram}. The first stage involves the user drafting a music loop, while the second stage is dedicated to iterative refinement through dialogue. Each stage necessitates different tasks. In the initial stage, the focus is on creating music from an ambiguous demand, essentially with a global description. The later stages shift the focus to music editing, where fine-grained localised revisions are made. These revisions can include regenerating specific areas, adding or removing particular instruments, and incorporating sound effects. A comprehensive list of all supported tasks is presented in Table \ref{tab:tasks}.

Each task in Table \ref{tab:tasks} corresponds to one or more specific backend models, which are sequentially called as needed. For instance, consider the task ``Style Imitation". Here, a user can reference the title of a real-world music track. Loop Copilot first invokes ChatGPT to generate a description based on the given music title, which is then forwarded to MusicGen to generate the music audio. This ability to chain multiple models opens up a wealth of opportunities to accomplish new tasks that have scarcely been explored before, although the results may not be as good as for models trained for specific tasks. 

Specifically, I explore new methods for the tasks below:

\begin{enumerate}
    \item \textit{Imitate rhythmic pattern.} I utilise MusicGen's continuation feature to use the input drum pattern as a prefix while guiding the model with a target text description for generation.
    \item \textit{Style Imitation.} For the `imitation' descriptions that are not musical features but a reference to existing recordings, such as bands and titles, I first use ChatGPT to convert them into descriptions of musical features, and then call MusicGen to generate music audio. During the generation process, Loop Copilot does not directly copy music pieces from the original recording to avoid direct plagiarism of copyright works.
    \item \textit{Add a track.} There are still no publicly available models supporting this feature. I instead utilise MusicGen's continuation feature to take the original audio as a prefix and use the new track text description to guide model generation. To ensure stability, I use the CLAP model to verify that the similarity between the generated result and the new text description is above a threshold.
\end{enumerate}

Note that Loop Copilot can comprehend complex demands that necessitate the combination of existing tasks. For instance, if a user wishes to ``generate jazz music and add medium level background noise, like in a pub", the large language model will dissect this demand into a series of tasks: ``text-to-music" and ``add sound effects". Within each task, if necessary, backend models are chained accordingly. Thus, the sequential invocation can occur at both the task and model levels. However, the final output presented to the user is the seamlessly integrated ``jazz music with background noise".

\begin{sidewaystable}[htbp]
    \centering
    \begin{tabular}{lcll}
    \toprule
    Task & Stage & Examples of text input & Backend models  \\
    \midrule
    Text-to-music & 1 & Generate rock music with guitar and drums.  & MusicGen~\citep{musicgen} \\
    Drum pattern to music$^{\dag}$ & 1 & Generate rock music with guitar based on this drum pattern.  & MusicGen \\
    Style Imitation$^{*}$ & 1 & Generate a music loop that feels like ``Hey Jude".  & ChatGPT, MusicGen \\
    Stylistic rearrangement & 1 & Rearrange this music to jazz with sax solo.  & MusicGen \\
    Music variation generation & 1 & Generate a music loop that sounds like this music.  & VampNet~\citep{vampnet} \\
    \midrule
    Add a track$^{*}$ & 2+ & Add a saxophone solo to this music loop. & MusicGen, CLAP~\citep{clap} \\
    Remove a track & 2+ & Remove the guitar from this music loop.  & Demucs~\citep{demucs} \\
    Re-generation/inpainting & 2+ & Re-generate the 3-5s part of the music loop. & VampNet \\
    Add sound effects & 2+ & Add some reverb to the guitar solo. & pedalboard~\tablefootnote{\url{https://doi.org/10.5281/zenodo.7817838}} \\ 
    Pitch shifting & 2+ & Transpose this music to G major. & pedalboard \\ 
    Tempo changing & 2+ & Make the music a bit slower. & torchaudio~\citep{torchaudio} \\ 
    \midrule
    Music captioning & N/A & Describe the current music loop.  & LP-MusicCaps \\
    \bottomrule
    \end{tabular}
    \caption{The list of all supported tasks in Loop Copilot at stage 1 (generation) and later stages (editing). I explore new training-free methods for those tasks with $*$ marks, as described in Section \ref{sec:supported_tasks}.}
    \label{tab:tasks}
\end{sidewaystable}

\subsection{Global Attribute Table}

The Global Attribute Table (GAT) is an integral component of the Loop Copilot system, designed to encapsulate and manage the dynamic state of music being generated and refined during the interaction process. Its role is to offer a centralised repository for the various attributes that define the musical piece at any given moment. This centralisation is pivotal for Loop Copilot's ability to provide continuity, facilitate task execution, and maintain musical coherence. 

The design philosophy behind the GAT draws inspiration from  ``blackboard" architectures~\citep{blackboard}. In this paradigm, the GAT can be likened to a blackboard---a shared workspace where different components of the system can access and contribute information. 

Table \ref{tab:gat} provides an example, showing the GAT state in the scenario of Figure \ref{fig:teaser}.

\begin{table}[htbp]
    \centering
    \begin{tabular}{lp{2.5cm}ll}
    \toprule
     \textbf{bpm}         & 90 & \textbf{key}  & E$\flat$ major \\
     \midrule
     \textbf{genre}       & rock & \textbf{mood} & smooth \\
     \midrule
     \textbf{instruments} & \multicolumn{3}{l}{saxophone, guitar, snare drum} \\
     \midrule
     \textbf{description} & \multicolumn{3}{l}{\parbox{5.5cm}{smooth rock music loop with saxophone, a guitar arrangement and snare drum}} \\
     \midrule
     \multirow{2}{*}{\textbf{tracks}}   & \textbf{mix}  & \multicolumn{2}{l}{c540d5a6.wav} \\
     \cmidrule{2-4}
         & \textbf{stems} & \multicolumn{2}{l}{N/A} \\
    \bottomrule
    \end{tabular}
    \caption{An example of the Global Attribute Table in the scenario of Figure \ref{fig:teaser}.}
    \label{tab:gat}
\end{table}

GAT's significance can be further expounded upon through its multifaceted functionalities:

\begin{enumerate}
    \item \textbf{State Continuity:} The GAT ensures that users experience a seamless dialogue with the Loop Copilot by persistently tracking musical attributes and evolving based on both user input and system output.
    
    \item \textbf{Task Execution:} During the task execution phase, backend models often require contextual information. The GAT provides this context, thereby enhancing the models' performance.
    
    \item \textbf{Musical Coherence:} For any music creation tool, maintaining musical coherence is paramount. By storing key attributes like musical key and tempo, the GAT ensures the harmonious and consistent evolution of music throughout the creative process.
\end{enumerate}

In practice, the GAT information is attached to the system prompt, served as an additional document to be retrieved. \highlight{Since GAT is represented as a table, and LLMs do not support table-based inputs, the GAT is converted into a plain text format before being provided to the model. It is important to clarify that the implementation of GAT imposes no vocabulary constraints; it is populated with relevant keywords in natural language form.}

\section{Experiments}

\begin{figure}
    \centering
    \includegraphics[width=\linewidth]{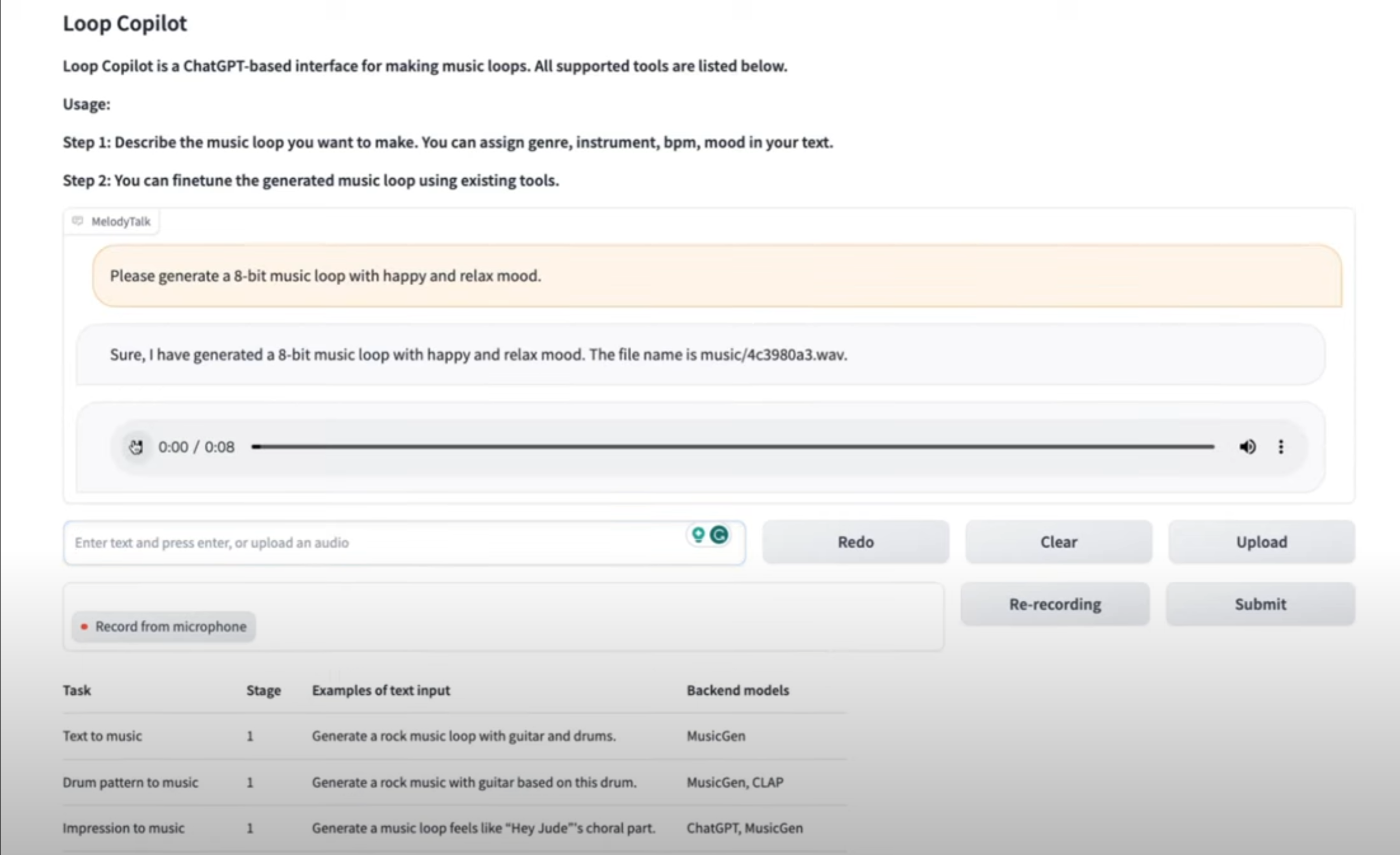}
    \caption{Caption}
    \label{fig:loopcopilot-interface}
\end{figure}

To evaluate the efficacy and usability of Loop Copilot, a mixed-methods experimental design was adopted. This design aligns with the triangular research framework~\citep{triangle}. A screenshot of the demo system is shown in Figure~\ref{fig:loopcopilot-interface}. \footnote{Demo available at \url{https://sites.google.com/view/loop-copilot}.}

\subsection{Participants}

I recruited 8 volunteers (N=8) who were interested in AI-based music production, and work in the field of music and audio technology or production, though not necessarily professional-level musicians.
Participants provided informed consent, and data anonymisation protocols were strictly followed to maintain ethical standards ~\footnote{The experiment was approved by Yamaha ethics committee.}. The distribution of the participants was as follows:

\begin{enumerate}
    \item \textit{Experience in Music Production}: 3 starters (0-2 years), 3 intermediate (2-5 years), 2 experts (\textgreater 5 years).
    \item \textit{Experience in Music Performance}: 2 starters (0-2 years), 2 intermediate (2-5 years), 4 experts (\textgreater 5 years).
    \item \textit{Age}: 4 (18-35 years), 2 (35-45 years), 2 (\textgreater 45 years).
\end{enumerate}

\subsection{Measures}

I measure the following constructs: 

\begin{enumerate}
    \item \textit{Usability.} Usability serves as a critical metric for assessing the ease with which users can interact with Loop Copilot. It measures not only the system's efficiency but also gauges the intuitive nature of the user interface. I adopted the System Usability Scale (SUS)~\citep{sus} (5-point Likert scale, see Appendix A) as a validated tool for this aspect of the evaluation. After linear projection, SUS scores have a range from 0 to 100, where a value over 68 is considered acceptable.
    \item \textit{Acceptance.} Understanding user acceptance is crucial for assessing whether Loop Copilot would be willingly incorporated into existing workflows. This encompasses factors like the perceived ease of use and the perceived usefulness of the system. The Technology Acceptance Model (TAM)~\citep{tam} served as the theoretical framework for evaluating these dimensions. Our TAM questionnaire (5-point Likert scale, see Appendix B) consists of 11 questions categorised into perceived ease of use  (Q1-4), perceived usefulness (Q5-8), and overall impressions (Q9-11).
    \item \textit{User experience.} Beyond usability and acceptance, the qualitative aspect of user experience provides a more nuanced understanding of the system's impact. This involves exploring the emotional and cognitive perceptions that users have when using Loop Copilot, such as the joys and frustrations they experience. Open-ended questions were designed to capture these subjective aspects in detail.
\end{enumerate}

\subsection{Procedure}

Experiments were conducted in a quiet, controlled environment to ensure consistency and minimize distractions. The experimental session for each participant consisted of three phases:

\begin{enumerate}
    \item \textit{Orientation Phase (10 minutes)}: 
    During this phase, participants were acquainted with the functionalities and features of Loop Copilot. This briefing aimed to standardize the initial level of understanding across participants.
    Specifically, the system was shown to the subjects with a brief explanation of how to use the interface.  Furthermore, the participants were presented with the example inputs in Table~\ref{tab:tasks} as examples of possible prompts supported by the system. 

    \item \textit{Interactive Usage Phase (20 minutes)}: 
    Participants were allowed to freely interact with Loop Copilot for music composition. Observational notes were made in real-time to capture immediate insights and identify areas for potential system improvement.

    \item \textit{Feedback and Evaluation Phase (15 minutes)}: 
    Upon completion of the interaction, participants were asked to fill out the SUS and TAM questionnaires. Additionally, a semi-structured interview based on the responses from the questionnaires was conducted to obtain qualitative feedback on their experience.

\end{enumerate}

Both quantitative (SUS, TAM scores) and qualitative (interview notes) data were collected. Data during the interview section were collected primarily through observational notes. These notes were aimed at capturing immediate insights, identifying potential areas for system improvement, and gathering qualitative feedback on the user experience. 

\highlight{In addition to recording the questionnaire results, we took textual notes to capture the key points from the interviews. The decision to use note-taking instead of audio recording was made to respect participants' preferences, as some individuals did not want their voices to be recorded. This approach was chosen to protect participant privacy. The notes are further analysed in Section \ref{sec:individual_feedback}.}

\subsection{Quantitive Results}

\subsubsection{System Usability Scale (SUS)}

\begin{figure}[htbp]
    \centering
    \includegraphics[trim={0.5cm 0 0.3cm 0}, width=\linewidth, clip]{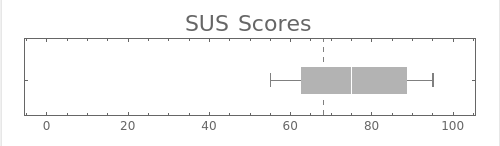}
    \caption{The box plot depicting SUS score results with an average of 75.31$\pm$15.32. The dotted line marks the threshold for effectiveness.}
    \label{fig:sus}
\end{figure}

The SUS was used to measure the overall usability of Loop Copilot. The mean SUS score was \(75.31\) with a standard deviation of \(15.32\). According to the conventional SUS scale, a score above \(68\) is considered above average, suggesting that the participants found the system to be generally usable. A visualisation is shown in Figure \ref{fig:sus}, and Table~\ref{tab:sus_table} shows the detailed results.

\begin{sidewaystable}[htbp]
\centering
\begin{tabular}{c|l|c}
\toprule
\textbf{Question No.} & \textbf{Question} & \textbf{Score (std)} \\
\midrule
1 & I think that I would like to use this system frequently. & 4.13 (0.83) \\
2 & I found the system unnecessarily complex. & 1.50 (0.76) \\
3 & I thought the system was easy to use. & 4.13 (0.83) \\
4 & I think that I would need the support of a technical person to be able to use this system. & 2.63 (1.41) \\
5 & I found the various functions in this system were well integrated. & 3.88 (0.99) \\
6 & I thought there was too much inconsistency in this system. & 2.00 (0.93) \\
7 & I would imagine that most people would learn to use this system very quickly. & 3.88 (1.36) \\
8 & I found the system very cumbersome to use. & 1.38 (0.52) \\
9 & I felt very confident using the system. & 3.38 (0.74) \\
10 & I needed to learn a lot of things before I could get going with this system. & 1.75 (1.04) \\
\bottomrule
\end{tabular}
\caption{SUS survey questions with scores and standard deviations.}
\label{tab:sus_table}
\end{sidewaystable}

The SUS scores revealed a generally favourable perception of the system's usability~\footnote{For SUS scores, higher scores are better for odd-numbered problems and lower scores are better for even-numbered problems. To transform the results: for the odd-numbered questions, subtract 1 from the given score; for the even-numbered questions, subtract the score from 5. After adjusting the values for all questions, sum the resulting scores. Finally, multiply the total by 2.5 to obtain the final score out of 100 points.}. Users indicated a willingness to use the system frequently (Q1, 4.13$\pm$0.83), highlighting its perceived ease of use (Q3, 4.13$\pm$0.83) and quick learnability (Q7, 3.88$\pm$1.36).

However, some concerns were raised about the need for technical support (Q4, 2.63±1.41), suggesting that while the system is user-friendly, it may involve layers of complexity that require expert guidance or better onboarding processes. Although the system’s features were generally regarded as well-integrated (Q5, 3.88±0.99), the moderate scores for system consistency (Q6, 2.00±0.93) indicate potential areas for improvement in unifying the system's functionalities. This inconsistency may be linked to the varying responsiveness of different AI models, which could explain why some dialogues experience significantly longer response times than others.

\subsubsection{Technology Acceptance Model (TAM)}

\begin{enumerate}
    \item \textit{Perceived Usefulness (PU).} The average score for Perceived Usefulness was \(3.58\) with a standard deviation of \(1.13\). This indicates a moderate level of agreement among the participants that the system is useful. 
    \item \textit{Perceived Ease of Use (PEOU).} The average score for Perceived Ease of Use was \(3.89\) with a standard deviation of \(0.80\). This suggests that participants generally found the system easy to use.
    \item \textit{Overall TAM Scores.} The overall average TAM score was \(4.09\) with a standard deviation of \(1.09\), which suggests a favourable perception towards both the ease of use and usefulness of the system.
\end{enumerate}

\begin{figure}[htbp]
    \centering
    \includegraphics[trim={0 0 0.3cm 0}, width=\linewidth, clip]{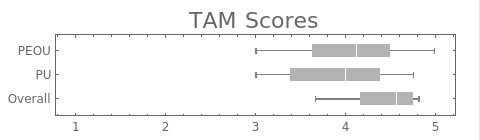}
    \caption{Box plot of the TAM score results. Perceived Usefulness (PU) with an average of 3.58$\pm$1.13; Perceived Ease of Use (PEOU) averaging 3.89$\pm$0.80; Overall TAM score of 4.09$\pm$1.09. These scores reflect participants' favourable perceptions of the system's utility and usability.}
    \label{fig:tam}
\end{figure}

Table ~\ref{tab:tam_scores} and Figure~\ref{fig:tam} show the results of TAM scores. 

\begin{sidewaystable}[htbp]
\centering
\begin{tabular}{c|l|c}
\toprule
\textbf{Question No.} & \textbf{Question} & \textbf{Score (std)} \\
\midrule
1 & I find Loop Copilot useful in live music performance. & 4.25 (0.89) \\
2 & Using Loop Copilot improves my experience in music performance. & 3.25 (1.67) \\
3 & Loop Copilot enables me to accomplish tasks more quickly. & 4.13 (0.64) \\
4 & I find that Loop Copilot increases my productivity in music performance. & 4.00 (0.93) \\
5 & I find Loop Copilot easy to use. & 4.13 (0.83) \\
6 & Learning to operate Loop Copilot is easy for me. & 4.63 (0.52) \\
7 & I find it easy to get Loop Copilot to do what I want it to do. & 2.88 (1.13) \\
8 & I find the interface of Loop Copilot to be clear and understandable. & 4.63 (0.74) \\
9 & Given the chance, I intend to use Loop Copilot. & 4.88 (0.35) \\
10 & I predict that I would use Loop Copilot in the future. & 4.75 (0.46) \\
11 & I plan to use Loop Copilot frequently. & 4.00 (0.76) \\
\bottomrule
\end{tabular}
\caption{TAM survey questions with scores and standard deviations.}
\label{tab:tam_scores}
\end{sidewaystable}

The positive impact of the system on music creation is evident in both its perceived usefulness and ease of use. The questions related to usefulness (Q1-Q4) received favourable responses (Q1: ``I find Loop Copilot useful in music creation" – 4.25±0.89, Q2: ``Using Loop Copilot improves my experience in music creation" – 3.25±1.67, Q3: ``Loop Copilot enables me to accomplish tasks more quickly" – 4.13±0.64, Q4: ``I find that Loop Copilot increases my productivity in music creation" – 4.00±0.93). The user-friendly nature of the interface was similarly rated positively (Q5-Q8) with most responses reflecting ease of use (Q5: ``I find Loop Copilot easy to use" – 4.13±0.83, Q6: ``Learning to operate Loop Copilot is easy for me" – 4.63±0.52, Q7: ``I find it easy to get Loop Copilot to do what I want it to do" – 2.88±1.13, Q8: ``I find the interface of Loop Copilot to be clear and understandable" – 4.63±0.74).

It is notable that Q7, which is related to control over the system, shows a lower score, suggesting potential areas for improvement in making the system more responsive to user intent. Finally, there was a strong indication that users are inclined to incorporate the system into their future workflows (Q9-Q11), with particularly high responses for Q9: ``Given the chance, I intend to use Loop Copilot" (4.88±0.35), Q10: ``I predict that I would use Loop Copilot in the future" (4.75±0.46), and Q11: ``I plan to use Loop Copilot frequently" (4.00±0.76). These findings underscore the perceived value of the system and the ease of integration into creative processes of users.

\subsection{Individual Feedback}\label{sec:individual_feedback}

\highlight{We collected and analysed individual feedback from the user study. Given the small sample size (N=8), we conducted a manual analysis based on the interview notes. For each participant, we carefully reviewed the feedback provided and identified recurring themes mentioned by at least two individuals. This process allowed us to categorise the feedback into four main areas, as outlined below. The analysis aimed to capture key patterns in user experience, which were then triangulated with the quantitative findings to provide a more comprehensive understanding of the system’s performance and user reception.}

\subsubsection{Overall Impressions} 

Participants generally found value in Loop Copilot as a tool for music generation. Users widely considered it a promising starting point for creative inspiration.

\begin{enumerate}

\item Some participants found text-to-music conversion not fully meeting their specific musical visions, indicating a gap between user expectations and system output.

\item Participants thought that Loop Copilot was useful for getting creative inspiration.

\end{enumerate}
    
\subsubsection{Positive Feedback} 

\begin{enumerate}
    \item \textit{Ease of Use.} Most participants, especially beginners and intermediates, appreciated the intuitive nature of the interface. Most users found the system to be straightforward and easy to understand.

    \item \textit{Design and Interaction} Users lauded the design potential and interactive methods, suggesting that they represent a fertile ground for future development.
\end{enumerate}

\subsubsection{Areas of Concern}

\begin{enumerate}
    \item \textit{Limited Control and Precision.} Participants commonly mentioned the limited control they had over the musical attributes. Some cited specific instances where text prompts like \textit{``Add a rhythmic guitar"} or \textit{``Remove reverb"} were not adequately reflected in the output.
    \item \textit{Integration with Existing Workflows.} 
 Some users thought the system's current specifications were limited as a stand-alone music production system, and wanted it instead as a part of existing music creation systems, like a digital audio workstation.   
\end{enumerate}

\subsubsection{Future Expectations}
\begin{enumerate}
    \item \textit{Feature Extensions.} Many users called for additional features like volume control, the ability to upload their own melody lines, and options for chord conditioning. Users also highlighted the need for multiple output options to choose from, rather than a single output.
    \item \textit{Improved Responsiveness.} Given that some participants found the system occasionally unresponsive to specific prompts, they hoped future versions could offer improved interpretation and execution of user commands.
\end{enumerate}

\section{Conclusion}

This chapter presented Loop Copilot, a system designed to integrate Large Language Models (LLMs) with specialised AI music models to support human-AI collaborative creation of music loops. The system utilises a conversational interface that allows users to engage in an interactive and iterative music creation process. A key innovation of Loop Copilot is the introduction of the Global Attribute Table, which tracks the evolving state of the music to ensure that modifications are coherent and consistent throughout the editing process. Additionally, the system employs a chaining mechanism that enables training-free music editing by leveraging existing AI music models. Our evaluation, supported by interview-based insights, demonstrates the potential of conversational interfaces for facilitating iterative music production.

While the current system demonstrates the feasibility of controlling non-text-to-music generation models, such as VampNet, there are limitations in the precision and scope of music editing tasks that can be achieved. The input and output remain constrained by the capabilities of the individual models. Although chaining multiple models offers some degree of control, it is not sufficient for more complex tasks. For instance, there is currently no model that can precisely add a new track to existing music audio based on a natural language description of the track—a limitation that has inspired the development of Instruct-MusicGen (Chapter~\ref{ch:instruct}). Similarly, there are other potential text-based controls that require deeper exploration into text-to-music foundation models, motivating the research behind MusicMagus (Chapter~\ref{ch:musicmagus}).

In summary, while Loop Copilot provides a usable interface for AI-assisted music creation, its limitations of the performance of specific components 
highlight the need for more advanced models and techniques. These insights have directly informed the subsequent research presented in this thesis, where I seek to overcome these challenges and further advance the capabilities of text-based music production.

\chapter{MusicMagus: Zero-Shot Text-to-Music Editing via Diffusion Models}
\label{ch:musicmagus}

This chapter introduces MusicMagus, a system developed to explore the challenges associated with text-guided music editing. Building on insights gained from Loop Copilot, MusicMagus presents a novel approach that leverages pre-trained diffusion models to enable zero-shot music editing. The system is designed to modify specific musical attributes, such as genre, mood, and instrumentation, while preserving the overall structural integrity of the original piece. This capability is particularly significant given the iterative nature of music production, where nuanced and flexible edits are often required to achieve the desired artistic outcome.

MusicMagus operates by manipulating the latent space of diffusion models, transforming the text-based editing process into a task of latent space exploration. This method incorporates an additional constraint that enforces consistency, ensuring that the edits do not disrupt non-targeted musical elements. By adopting this approach, MusicMagus demonstrates that stylistically coherent edits can be achieved without additional training, positioning it as a versatile tool for real-world music editing scenarios.

However, MusicMagus has its limitations. The system primarily focuses on intra-stem editing, i.e. making changes within individual musical stems, but encountering challenges with more complex operations, such as adding or removing instruments. Additionally, the diffusion-based approach is inherently constrained by the length of the generated audio, which is typically shorter than what can be produced by music language models. These limitations highlight opportunities for further advancements in text-guided music editing, motivating the development of the more sophisticated system, Instruct-MusicGen, discussed in the following chapter.

\section{Introduction}

Recent advances in text-to-music generation have opened up new possibilities in musical creativity~\citep{butter,accomontage,musecoco,polyffusion,musiclm,mousai,musicgen,musicldm}. However, a significant challenge persists in how to \textit{edit} the generated results as music production usually involves iterative refinements. Building on this momentum, I regard `text-to-music editing' as the process of using text queries to edit music, and I see two major types of operations: \textit{inter-stem editing}, such as adding or removing instruments (e.g., ``add a saxophone" or ``remove the drums"), and \textit{intra-stem editing}, which involves modifications within the same stem, such as adding effects or changing instruments (e.g., ``add reverb to this stem" or ``transfer the timbre of the specified notes"). In this context, a ``stem'' refers to an individual track or component within a music piece, such as a specific instrument or vocal part. The primary focus of this paper is on the latter, \textit{intra-stem editing}. 

One of the fundamental challenges of text-to-music editing is the difficulty of accommodating flexible text operations in both dataset construction and model training. This is not only a matter of data pair scarcity, but also of the complexity inherent in the vast array of possible text-based edits that can be applied to music. Existing research~\citep{audit,instructme,m2ugen} has primarily focused on manually constructing datasets. However, these models are constrained to a few predefined operations, which undermines their effectiveness in text-to-music editing, which requires flexibility and variety. This highlights the need for a new approach that moves away from traditional supervised learning reliant on specific data pairs and toward a more adaptable, unsupervised, or zero-shot approach.

\begin{figure}[tb]
    \centering
\includegraphics[width=\linewidth]{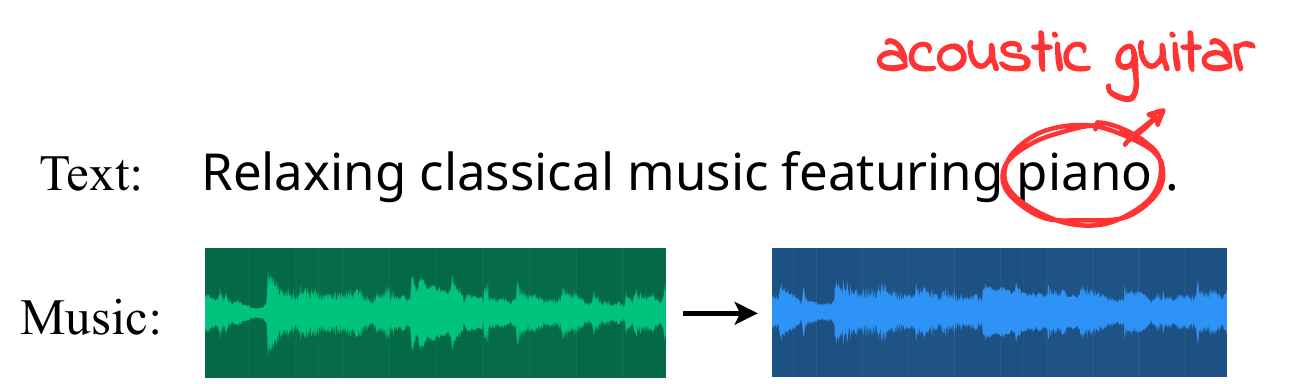}
    \caption{Text-to-music editing with MusicMagus. The edit from ``piano" to ``acoustic guitar" in the text prompt directly alters the corresponding musical attribute, while leaving others unchanged.}
    \label{fig:musicmagus-teaser}
\end{figure}

In this work, I introduce MusicMagus, which focuses on text-based \textit{intra-stem} music editing. Leveraging the inherent capabilities of pre-trained diffusion models, MusicMagus is able to perform zero-shot editing without requiring additional training pairs. As illustrated in Figure~\ref{fig:musicmagus-teaser}, I utilise word swapping to direct the editing process. This operation is implemented as a manipulation within the diffusion model's semantic space. Recognising the sensitivity of the diffusion process, where minor alterations can propagate significant changes, I employ an additional constraint to ensure that the resultant music maintains the structural integrity and stylistic coherence of the original music. 

Although I mainly focus on the editing of music clips generated from diffusion models, I also discuss how to edit real-world music audio using the Denoising Diffusion Implicit Model (DDIM) inversion technique ~\citep{ddim}.

In summary, the main contributions are as follows:

\begin{enumerate}[itemsep=0pt, parsep=0pt]
    \item I propose a flexible and user-friendly text-to-music editing method using word swapping.
    \item I contribute MusicMagus, a system capable of zero-shot music editing on diverse tasks without any dependence on \textit{paired} training data.
    \item Comparative experiments validate that MusicMagus outperforms existing zero-shot methods and some supervised approaches in tasks such as style and timbre transformation. 
\end{enumerate}

\section{Method}

To illustrate the idea, I refer to the example in Figure~\ref{fig:musicmagus-teaser}. Initially, a music clip, denoted as $x$, is generated from the text prompt ``Relaxing classical music featuring piano", which I refer to as $y$. The next step involves altering this text prompt by substituting ``piano" with ``acoustic guitar", thereby creating a new prompt $y'$. The aim is to produce a revised music piece $x'$, where only the specified attribute is changed, while maintaining all other aspects.

The explanation of the idea is two-fold. In Section~\ref{sec:method_1}, I detail the method for altering the text prompt in the semantic domain. Subsequently, in Section~\ref{sec:method_2}, I discuss the approach to enforce suitable constraints over the cross-attention map during diffusion to preserve the integrity of the remaining elements of the music.


\subsection{Finding Editing Direction}\label{sec:method_1}

In this section, I introduce a strategy to calculate a difference ($\Delta$) vector in the latent space to guide the editing direction. This method is chosen over direct word swapping as it better preserves semantic coherence and contextual relevance, especially in cases of varying phrase lengths and complex content alterations. I will further explain it in Section~\ref{sec:method_2}; besides, previous research finds that similar operations can facilitate a more robust edit, especially when the keywords subject to modification are sparsely represented in the training dataset~\citep{pix2pix}.


I first introduce the text embedding method in Audio\-LDM 2. AudioLDM 2 uses a two-branch text encoder to embed the text prompt $y$ to two embeddings: $E = \{E_\text{T5}, E_\text{GPT}\}$, where $E_\text{T5}$ encodes the sentence-level representation, and $E_\text{GPT}$ captures the more fine-grained semantic information inside $y$.

First, the FLAN-T5~\citep{flant5} encoder, utilising a T5 model~\citep{t5}, encodes \( y \) into a feature vector \( E_\text{T5} \in \mathbb{R}^{L \times 1024} \), where $L$ represents the prompt length. In parallel, the CLAP~\citep{clap} text encoder leverages a RoBERTa~\citep{roberta} model to transform \( y \) into a flattened vector \( E_\text{CLAP} \in \mathbb{R}^{1 \times 512} \): 

\begin{equation}
\left\{
\begin{aligned}
E_\text{T5} &= \text{T5}(y), \\
E_\text{CLAP} &= \text{CLAP}(y).
\end{aligned}
\right.
\end{equation}

Then, $E_\text{T5}$ and $E_\text{CLAP}$ are linearly projected to \( P \in \mathbb{R}^{768} \). A GPT-2 model, pre-trained on an AudioMAE~\citep{audiomae}, is then employed to auto-regressively generate 8 new tokens \( E_\text{GPT} \in \mathbb{R}^{8 \times 768} \):

\begin{equation}
E_\text{GPT} = \text{GPT-2}(\text{Proj}(E_\text{T5}, E_\text{CLAP})).
\end{equation}

The LDM takes both $E_\text{T5}$ and $E_\text{GPT}$ as input in the diffusion process:

\begin{align}
\epsilon_\theta &= \epsilon_\theta(z_t, E, t), \\
z_{t-1} &= \text{Denoise}(z_t, \epsilon_\theta, E, t).
\end{align}

Similarly, the new prompt $y'$ can be encoded to $E' = \{E'_\text{T5}, E'_\text{GPT}\}$. The goal is to find $E^\text{edit} = \{E^\text{edit}_\text{T5}, E^\text{edit}_\text{GPT}\}$.

\begin{figure}[tb]
    \centering
    \includegraphics[width=\linewidth]{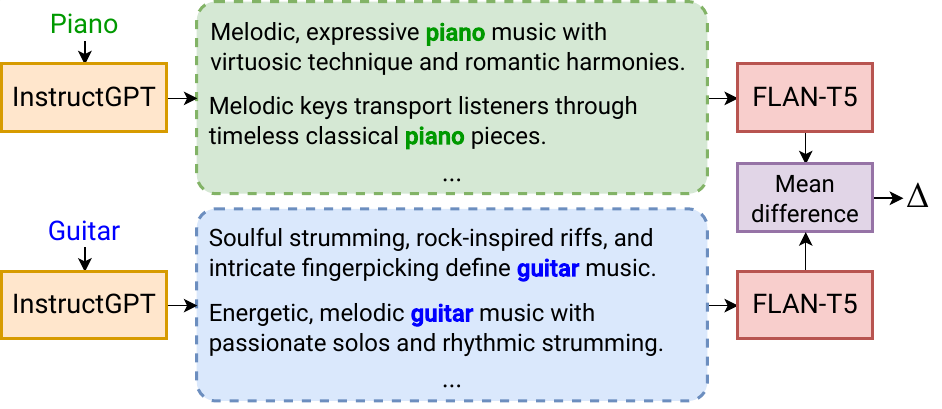}
    \caption{The pipeline of finding the editing direction $\Delta$. I first use InstructGPT to generate a large number of captions and then calculate the mean difference between the two embedding sets.}
    \label{fig:delta}
\end{figure}

I use the following method to find the editing vector $\Delta$, as shown in Figure~\ref{fig:delta}:

\begin{enumerate}[itemsep=0pt, parsep=0pt]
    \item I first generate a multitude of music-related captions using a pretrained InstructGPT model~\citep{instructGPT}. These captions are designed to contain the original and new keywords.
    \item Subsequently, I input these two sets of captions into the FLAN-T5 encoder and compute the mean embeddings for each set of encoded vectors.
    \item The final step is calculating the difference between these two mean embeddings, which is then employed as the vector for the editing direction $\Delta$.
\end{enumerate}

I employ different strategies to edit $E_\text{T5}$ and $E_\text{GPT}$. For $E_\text{T5}$, the edited embedding is:

\begin{equation}
    E^\text{edit}_\text{T5} = E_\text{T5} + \Delta.
\end{equation}

The aforementioned editing method faces challenges when applying $\Delta$ to \( E_{\text{GPT}} \). The core issue is that \( E_{\text{GPT}} \) is obtained through the GPT-2 model, where the addition of a \( \Delta \) to the embedding may not constitute a semantically valid operation. Consequently, in practical applications, I resort to using \( E^\text{edit}_{\text{GPT}} = E'_{\text{GPT}}\), which is derived directly from the encoding of the new prompt.

Finally, I have the edited embeddings:

\begin{equation}
    E^\text{edit} = \{E_\text{T5} + \Delta, E'_{\text{GPT}}\}.
\end{equation}

\subsection{Adding Constraints over Cross-Attention}\label{sec:method_2}

\begin{figure}[htbp]
    \centering
    \includegraphics[width=\linewidth]{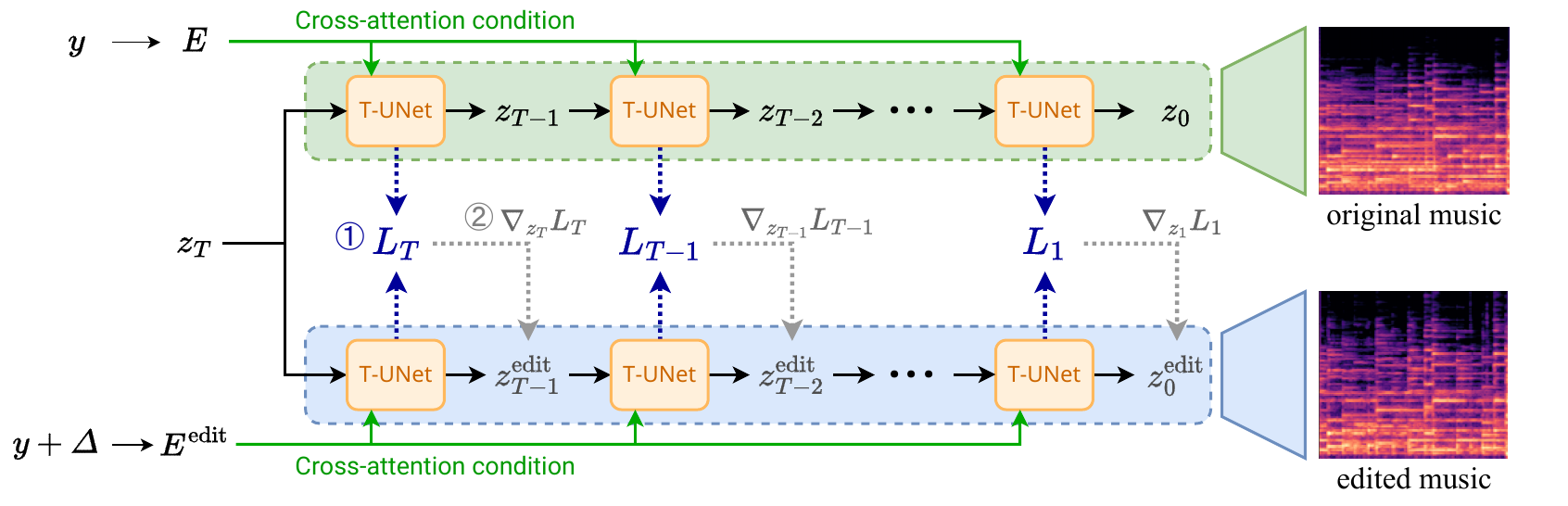}
    \caption{The workflow of the MusicMagus model. To constrain the diffusion model at timestep $t$, I need to: (1) calculate the L2 loss $L_t$ between the cross-attention map $M^\text{edit}_t$ and $M^\text{origin}_t$; (2) compute the gradient of $L_t$ with respect to $z_t$, and then perform a single-step optimisation to update $\epsilon_\theta^\text{edit}$ of the diffusion model.}
    \label{fig:musicmagus-diagram}
\end{figure}

Diffusion models exhibit randomness in their generation output. By setting a fixed random seed and using the same text prompts, the model can generate the same musical output. However, even minor variations in the text prompt can result in significantly different music clips. Previous studies have demonstrated that imposing external constraints on the cross-attention map between the text condition and the diffusion latent space enhances the consistency of the music generation, particularly for the remaining attributes that need to remain unchanged~\citep{prompt2prompt,pix2pix,plugandplay}. Building on this concept, I introduce a method designed to constrain the text-to-music diffusion model specifically for editing purposes.

To begin, I examine the acquisition of the cross-attention map. During the denoising process at the time step $t$, the model computes the cross-attention score between the encoded text $\{E_\text{T5}, E_\text{GPT}\}$ and the intermediate features of LDM $\epsilon_\theta$:

\begin{equation}
\begin{split}
\text{Attention}(Q, K, V) &= M \cdot V, \\
\text{where } M &= \text{Softmax}\left(\frac{QK^T}{\sqrt{d}}\right).
\end{split}
\end{equation}

In this context, $Q = W_Q\phi(z_t),~ K = W_kE,~ V = W_vE$ are defined, where $W = \{W_Q, W_K, W_V\}$ represents projection layers, and $E = \{E_\text{T5}, E_\text{GPT}\}$ represents the text embeddings. AudioLDM 2 proposes the T-UNet architecture, which is distinct from the UNet architecture, to extract intermediate spatial features $\phi(x_t)$. T-UNet incorporates a transformer block after each encoder and decoder block's convolution operation, and the cross-attention occurs in the transformer block's final layer. The term $d$ denotes the dimension of the projected keys and queries.

As illustrated in Figure~\ref{fig:musicmagus-diagram}, to apply the editing, I first reconstruct the music $x$ with the original text embeddings $E$. I record cross-attention maps for each timestep $t \in [1, T]$:

\begin{equation}
    M^\text{origin} = \{M^\text{origin}_1, ..., M^\text{origin}_T\}.
\end{equation}

\noindent Then I use the edited text embeddings $E^\text{edit}$ to generate an edited music clip. Similarly, at timestep $t$, I have a cross-attention map $M^\text{edit}_t$.

At each time step $t$, I apply the constraint by calculating the $L_2$ loss between $M^\text{origin}_t$ and $M^\text{edit}_t$:

\begin{equation}
    L_t = | M^\text{edit}_t - M^\text{origin}_t|_2.
\end{equation}

\noindent I then compute the gradient $\nabla_{z_t}L_t$ and perform a single-step optimisation with the step length $\alpha$:

\begin{equation}
    \epsilon^\text{edit}_\theta = \epsilon_\theta(z_t - \alpha\nabla_{z_t}L_t, E^\text{edit}, t).
\end{equation}

Subsequently, I execute the $t$-step denoising process using the updated $\epsilon^\text{edit}$:

\begin{equation}
    z_{t-1} = \text{Denoise}(z_t, \epsilon^\text{edit}_\theta, E^\text{edit}, t).
\end{equation}

\noindent This optimisation is applied at every step until the denoising process is completed to enhance structural consistency during denoising.

To effectively utilise the constraint of cross-attention, it is essential to employ $\Delta$ for editing. This method is crucial, especially when dealing with cases that involve substituting text of varying lengths, exemplified by replacing a shorter expression with a longer one (such as ``piano" $\rightarrow$ ``acoustic guitar"). Utilising $\Delta$ maintains the uniformity of embedding lengths during the editing process. In contrast, techniques such as word swapping can alter these lengths, leading to discrepancies between $M^\text{edit}$ and $M^\text{origin}$, and consequently errors in calculating $L_t$. Furthermore, $\Delta$ facilitates the insertion of words at different sentence positions without disrupting the position-related cross-attention maps, ensuring that the attention mechanism remains focused on the correct semantic context.

\section{Experiments}

In the domain of text-to-music editing, comprehensive model evaluation is inherently challenging due to the countless number of possible editing schemes. To address this, I focus on two key aspects: timbre transfer and style transfer, and compare the model's performance against established baselines in these areas. This comparison is conducted through objective and subjective testing methodologies. 

\subsection{Baselines}

I benchmark the model against three distinct models in the field: AudioLDM 2~\citep{audioldm2}, Transplayer~\citep{transplayer}, and MusicGen~\citep{musicgen}. Although the approach uses AudioLDM 2 as its backbone, AudioLDM 2 independently offers methods for both timbre and style transfer tasks, making it a relevant baseline.

\noindent\textbf{AudioLDM 2.} AudioLDM 2 is a diffusion-based model supporting unified speech, audio, and music generation at 16kHz. AudioLDM 2 is able to perform audio style transfer through the interpolation of audio latents and subsequent denoising with a new prompt.

\noindent\textbf{Transplayer.} This state-of-the-art, diffusion-based model is trained on the POP909~\citep{pop909} and the MAESTRO~\citep{maestro} datasets, specialising in timbre transfer at 16kHz. Unlike typical timbre transfer models that require training for each instrument pair, Transplayer is trained on multiple pairs, enabling versatile many-to-many timbre transfers.

\noindent\textbf{MusicGen.} A leading text-to-music generation model, MusicGen is a supervised model trained on a dataset of over 20,000 high-quality music pieces, generating 32kHz music. It uniquely allows for the inclusion of an extra melody condition, facilitating the style transfer task within the text-to-music generation process by generating music with new text prompt conditioning on the original music audio.

\subsection{Metrics}

I employ different metrics for subjective and objective experiments. For subjective evaluation, I incorporate the following metrics, where OVL and REL are as follows~\citep{audiogen}:

\textbf{Overall Quality (OVL).} This metric is used to assess the overall music quality, encompassing aspects like sound clarity and musicality. It primarily evaluates whether the editing process enhances or diminishes the quality of the original music audio. The scoring for this metric ranges from 0 to 100.

\textbf{Relevance (REL).} REL measures the perceived semantic closeness between the edited music and the new text prompt. It is a subjective score, also ranging from 0 to 100.

\textbf{Structural Consistency (CON).} I define a new metric CON to evaluate the consistency of the pitch contour and structural aspects in the subjective test. Similar to the others, its scoring range is from 0 to 100.

The objective experiments utilise the following metrics:

\textbf{CLAP Similarity (CLAP).}\citep{clap} This metric measures the semantic relevance between edited music and a new text prompt by leveraging the CLAP model, which is pretrained on multimodal data to capture cross-domain relationships between audio and text. The metric is particularly suitable for assessing semantic similarity in text-to-music generation tasks, as it is designed to understand both musical content and linguistic descriptions. \highlight{CLAP is also flexible and can be applied to style and timbre transfer tasks. For instance, by editing the text prompt through word swapping, we can expect the model to generate audio that aligns with the edited text prompt. In this case, the resulting audio will have a higher CLAP similarity score with the new text prompt, while the original audio will retain a higher similarity score with the original text prompt, reflecting the semantic alignment.} A higher CLAP score indicates a stronger alignment between the music and the text prompt, with values ranging from 0 (no similarity) to 1 (perfect similarity). This metric is implemented with the MuLaB library\citep{mulab}, which provides an efficient interface for computing CLAP Similarity.

\textbf{Chromagram Similarity (Chroma).} I use this new metric to gauge the preservation of pitch contours and rhythm patterns in the music. It involves computing the cosine similarity between the chromagrams of the original and edited music. A higher score suggests better retention of the structure and pitch contour, with values ranging from 0 to 1. This metric is implemented with the librosa library~\citep{librosa}.

\subsection{Data Preparation}

\subsubsection{Objective Experiments}

To evaluate our model's performance on the timbre transfer task, I began by generating a set of music audio samples using AudioLDM 2. I crafted text prompts based on the template: "A {\textit{mood}} {\textit{genre}} music with {\textit{timbre}} performance," where {\textit{mood}} is randomly selected from the set \{``upbeat", ``relaxing", ``peaceful"\}. This approach ensured a diverse range of initial audio samples in terms of mood, genre, and timbre.

I focused on three specific timbre swapping pairs supported by the Transplayer model: (piano $\rightarrow$ organ), (viola $\rightarrow$ piano), and (piano $\rightarrow$ acoustic guitar). \highlight{Note that MusicMagus can perform arbitrary word swapping, whereas Transplayer has been trained on a fixed set of pairs. To ensure a fair comparison, we restricted the swapping to specific pairs.}

For each pair, I generated audio samples and implemented a quality-based filtering process. Samples that fell below a predefined quality threshold—such as those with significant artifacts or misalignment with the text prompt—were excluded. I continued this process until I accumulated 60 suitable samples, aiming to represent the typical outputs of AudioLDM 2.

For the music style transfer task, I applied a similar methodology. Using the same text prompt template and random mood selection, I generated audio samples for various style conversions, including (jazz $\rightarrow$ classical), (country $\rightarrow$ metal), (jazz $\rightarrow$ metal), and (jazz $\rightarrow$ rock). After applying the quality-based filtering criteria, I curated a dataset comprising 50 samples.

It is important to acknowledge that our proposed model relies on audio generated by AudioLDM 2. This dependency introduces a potential bias when comparing editing results between models that also utilize AudioLDM 2 and those that do not. Models sharing the same underlying generation process might exhibit compatibility advantages, potentially influencing the evaluation outcomes. I recognize this inherent bias and have taken it into consideration during our analysis.

\subsubsection{Subjective Experiments}

For the subjective test, I randomly selected a subset of data points from the objective test dataset. Specifically, 8 data points were chosen for the timbre transfer task and 5 data points for the style transfer task. Each data point included results from both the baseline models and the ablation studies. The results are shown in Tables~\ref{tab:timbre_sub} and~\ref{tab:style_sub}. 

\subsection{Experimental Setup}

I choose the \textit{AudioLDM2-base} model~\footnote{\url{https://huggingface.co/cvssp/audioldm2}} as our backbone model. During inference, I configure the DDIM steps to 100, and generate 5-second audio clips at a sampling rate of 16kHz. A uniform gradient step length ($\alpha = 0.04$) is applied for both timbre transfer and style transfer tasks. All inference is performed on a single NVIDIA A100 GPU.

For the Transplayer model, I use the official pre-trained checkpoint~\footnote{\url{https://github.com/Irislucent/TransPlayer}} without any modifications to its weights or code. For MusicGen, I opt for the \textit{MusicGen-melody} checkpoint~\footnote{\url{https://huggingface.co/facebook/musicgen-melody}}, which has 1.5B parameters. To maintain consistency, all generated samples from these models are subsequently downsampled to 16kHz resolution.

\subsection{Results}

\subsubsection{Subjective Experiments}

I conducted a subjective listening test for both the timbre transfer and style transfer tasks. This test involved disseminating an online survey within the Music Information Retrieval (MIR) community and our broader research network, which resulted in the collection of 26 complete responses. The gender distribution of the participants was 19 males (73\%), 6 females (23\%) and 1 others (4\%). Regarding musical experience, 5 participants (19\%) had less than 1 year of experience, 5 (19\%) had between 1 and 5 years, and the majority, 16 participants (62\%), had more than 5 years of experience. This subjective test was approved by the ethics committee of both Sony AI and Queen Mary University of London (QMERC20.565.DSEECS23.129).

The data presented in Table~\ref{tab:timbre_sub} reveals that the proposed model exhibits superior performance in the timbre transfer task when compared to two baseline models. Specifically, AudioLDM 2 demonstrates a notable limitation in transferring to novel semantics, resulting in edited samples that closely resemble the original ones. This is evident from its low Relevance (REL) score and high Consistency (CON) score. Contrary to expectations, the performance of Transplayer is consistently inferior, suggesting that its generalisation capability may be inadequate for complex tasks such as many-to-many instrument timbre transfer in practical applications. The model is the best on the average of altering semantic content and maintaining structural integrity. 

\begin{table}[tb]
\small
    \centering
    \begin{tabular}{l|l|c@{\enspace}c@{\enspace}c|c}
    \toprule
    \textbf{Model name} & \textbf{Type}    &  \textbf{REL}  & \textbf{OVL} & \textbf{CON} & \textbf{Avg.} \\
    \midrule
    AudioLDM 2 &   Zero-shot & 15.7 & 49.9 & \textbf{80.6} & 48.7\\
    Transplayer & Supervised & 28.3 & 28.9 & 34.6 & 30.6\\
    \midrule
    MusicMagus \textit{w/o L2 \& $\Delta$} & Zero-shot & 78.0 & 61.6 & 50.4 & 63.3\\
    MusicMagus \textit{w/o L2} & Zero-shot &\textbf{78.8}&\textbf{62.4}&51.3 & 64.2\\
    \midrule
    \textbf{MusicMagus (final)} & Zero-shot &76.2&62.1&66.6 & \textbf{68.3}\\
    \bottomrule
    \end{tabular}
    \caption{The subjective evaluation results on the timbre transfer task.}
    \label{tab:timbre_sub}
\end{table}

Insights gleaned from the ablation study further elucidate these findings. The inclusion of the additional constraint significantly enhances performance in terms of Structure Consistency (CON), highlighting its role in bolstering structural coherence. However, subjective experiments do not indicate marked differences in Relevance (REL) scores between methods. This observation aligns with expectations, since the primary objective of $\Delta$ usage is to ensure the consistency of the cross-attention maps, particularly during complex editing operations or in scenarios involving underrepresented words stated in Section~\ref{sec:method_1}, which may not be fully reflected by the current subjective test settings.

I also evaluated the model performance in the style transfer task, as detailed in Table~\ref{tab:style_sub}. Similar to the previous findings, the model demonstrates superior performance over the baseline models in this task as Ill.

\begin{table}[tb]
\small
    \centering
    \begin{tabular}{l|l|c@{\enspace}c@{\enspace}c|c}
    \toprule
    \textbf{Model name} & \textbf{Type}    &  \textbf{REL}  & \textbf{OVL} & \textbf{CON} & \textbf{Avg.} \\ 
    \midrule
    AudioLDM 2 &   Zero-shot & 19.8 & 53.2 & \textbf{84.2} & 52.4\\
    MusicGen & Supervised & 63.3 & \textbf{66.0} & 48.2 & 59.1\\
    \midrule
    MusicMagus \textit{w/o L2 \& $\Delta$} & Zero-shot &69.2&56.9&58.9 & 61.7\\
    MusicMagus \textit{w/o L2} & Zero-shot &\textbf{71.3}&53.8&55.0 & 60.0\\
    \midrule
    \textbf{MusicMagus (final)} & Zero-shot & 65.7&57.8&65.6 & \textbf{63.1}\\
    \bottomrule
    \end{tabular}
    \caption{The subjective evaluation results on the style transfer task.}
    \label{tab:style_sub}
\end{table}

AudioLDM 2 exhibits notable limitations in style transfer, with its performance generally unstable; MusicGen, despite its downsampled audio quality from 32KHz to 16kHz, retains a high level of audio quality, as indicated by its high Overall Quality (OVL) score. However, MusicGen struggles with precisely preserving the original melody in the style transfer process, particularly in maintaining polyphonic melodies, which introduces some instability in its output.

In contrast, the method not only changes the semantics but also maintains the overall quality, resulting in the best average score; it also maintains the structural integrity and pitch consistency, which are critical in music style transfer.

\subsubsection{Objective Experiments}

I compare the performance of the model and the zero-shot and supervised baselines. The results for the timbre transfer and style transfer tasks are shown in Tables~\ref{tab:timbre_obj} and~\ref{tab:style_obj}.

In the timbre transfer task (Table~\ref{tab:timbre_obj}), the model demonstrated better performance in semantic transfer. The incorporation of a constraint on the cross-attention mechanism greatly improved pitch and rhythm accuracy, reinforcing the insights obtained from subjective experiments. These results substantiate the efficacy of the model in maintaining semantic integrity while facilitating timbre transfer results.

\begin{table}[tb]
\small
    \centering
    \begin{tabular}{l|l|cc|c}
    \toprule
    \textbf{Model name} & \textbf{Type}    &  \textbf{CLAP}  & \textbf{Chroma} & \textbf{Avg.}  \\
    \midrule
    AudioLDM 2 &   Zero-shot & 0.16 & 0.72 & 0.44 \\
    Transplayer & Supervised & 0.18 & 0.56 & 0.37 \\
    \midrule
    MusicMagus \textit{w/o L2 \& $\Delta$} & Zero-shot  & 0.33 & 0.68 & 0.51\\
    MusicMagus \textit{w/o L2} & Zero-shot & \textbf{0.34} & 0.69 & 0.52\\
    \midrule
    \textbf{MusicMagus (final)} & Zero-shot & 0.33 & \textbf{0.76} & \textbf{0.55}\\
    \bottomrule
    \end{tabular}
    \caption{The objective evaluation results on the timbre transfer task.}
    \label{tab:timbre_obj}
\end{table}

Table~\ref{tab:style_obj} presents the findings for the style transfer task. Here, the model outperformed the baselines in terms of structural and pitch consistency. However, in terms of semantic transfer, the differences between the model and the baselines were less pronounced. This suggests that while the model excels in maintaining the structural and pitch elements during style transfer, the semantic changes are comparable to those achieved by the baseline models.

\begin{table}[tb]
\small
    \centering
    \begin{tabular}{l|l|cc|c}
    \toprule
    \textbf{Model name} & \textbf{Type}    &  \textbf{CLAP} & \textbf{Chroma} & \textbf{Avg.} \\
    \midrule
    AudioLDM 2 &   Zero-shot & 0.18 & \textbf{0.80} & \textbf{0.49}\\
    MusicGen & Supervised & \textbf{0.24} & 0.66 & 0.45\\
    \midrule
    MusicMagus \textit{w/o L2 \& $\Delta$} & Zero-shot & 0.22 & 0.65 & 0.44\\
    MusicMagus \textit{w/o L2} & Zero-shot & 0.22 & 0.67 & 0.45\\
    \midrule
    \textbf{MusicMagus (final)} & Zero-shot & 0.21 & 0.77 & \textbf{0.49}\\
    \bottomrule
    \end{tabular}
    \caption{The objective evaluation results on the style transfer task.}
    \label{tab:style_obj}
\end{table}

\section{Discussion}

\subsection{Real Music Audio Editing}\label{sec:real}

\begin{figure}[htbp]
    \centering
    \includegraphics[width=\linewidth]{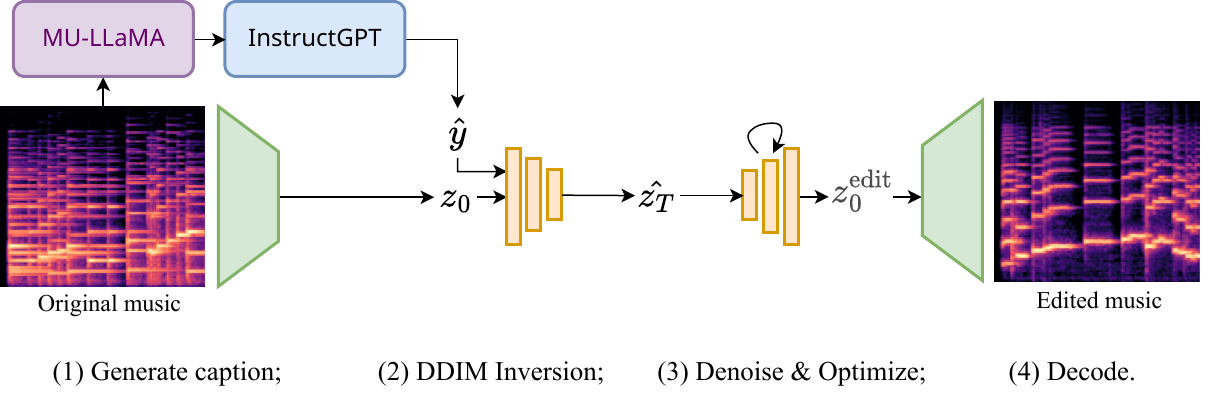}
    \caption{The diagram of the real music audio editing pipeline using MusicMagus with DDIM inversion and diffusion model editing.}
    \label{fig:inversion}
\end{figure}

MusicMagus offers capabilities for editing real-world music audio, although it is noted that the performance may not match the editing of synthesised music audio generated from diffusion models. I begin with the DDIM inversion~\citep{ddim} to estimate the latent representation $\hat{z_T}$ of a given real music audio $x$. This step is crucial to facilitate editing with the diffusion model, as depicted in Figure~\ref{fig:inversion}.

Inversion requires a corresponding text prompt $\hat{y}$, which is initially generated by a pre-trained music captioning model, MU-LLaMA~\citep{mullama}. Due to the discrepancy between the text prompt distributions of AudioLDM 2 and MU-LLaMA, the InstructGPT model is employed to refine the generated captions, aligning them more closely with AudioLDM 2's distribution. This refinement includes condensing the caption into a single concise sentence and emphasising essential characteristics such as the key instruments, mood, and genre.

DDIM inversion, while effective, is not a perfect reconstruction method. It faces a trade-off between the editability of the estimated latent $\hat{z_T}$ and its reconstruction fidelity~\citep{prompt2prompt}. A balance is sought by selecting an intermediate value for classifier-free guidance, set to 1. Additionally, the diffusion latent is typically modeled as Gaussian noise. To mitigate the autocorrelation that may arise during inversion, I adopt a strategy from Parmar et al.~\citep{pix2pix}, introducing autocorrelation regularisation to diminish its impact, thus enhancing the estimation of $\hat{z_T}$.

Subsequent to obtaining the estimated latent $\hat{z_T}$, the caption $\hat{y}$ is edited, and the MusicMagus editing algorithm is applied within the diffusion model framework to produce the edited music audio~\footnote{I provide listening samples at \url{https://bit.ly/musicmagus-demo}.}.

\subsection{Limitations}

The current implementation of MusicMagus, while effective, is based on the AudioLDM 2 model, which is not without its constraints. One significant limitation is the model's challenge in generating multi-instrument music when such complexity is specified. This inherently restricts the scope of creative expression and diversity that the model can offer. The performance of AudioLDM 2 was not enhanced in the approach, which is an aspect I aim to address moving forward.

Moreover, the zero-shot method exhibits instability, as evidenced by a notable number of failure cases. These failures are often due to unsuccessful application of the delta and word-swapping techniques, highlighting an area ripe for improvement. Currently, the scope of alterations I can apply to the music is somewhat modest; the system struggles to introduce substantial changes, such as adding or removing an instrument, adding sound effects, etc., without compromising the overall structure and quality of the audio.


Another factor that confines the system is the inherent limitations of the base model itself. For instance, the diffusion process struggles to generate very long sequences, which in turn limits the practical applications of the model. Addressing this limitation could potentially open up new domains where longer sequence generation is essential.

Lastly, the quality of generated audio, is another significant limitation, often resulting in artifacts that can detract from the listener's experience. Enhancing the fidelity of the audio is an important step that will bring us closer to a model that can produce professional-grade audio, which is crucial for both consumer applications and artistic endeavors. The pursuit of higher audio quality and the reduction of artifacts are critical goals for the future work.

\section{Ethical Statement}

Subjective tests are approved by the ethics committees of both Sony AI and Queen Mary University of London (QMERC20.565.DSEECS23.129).

\section{Conclusion}

In this chapter, I explored MusicMagus, a system developed to investigate the challenges and possibilities of text-guided music editing through latent space manipulation within pre-trained diffusion models. MusicMagus represents a significant step in AI-assisted music creation, offering a method to modify specific musical attributes while preserving the structural integrity of the original music. This approach, which does not require additional training, demonstrates the potential for stylistically coherent edits in real-world music production scenarios.

However, MusicMagus also has its limitations. The system’s focus on latent space manipulation constrains its ability to perform more complex editing tasks, such as adding or removing entire musical elements. Additionally, the diffusion models used are limited by the length of the generated audio, making them less suitable for longer music sequences compared to music language models. These limitations suggest areas for further research and innovation in text-guided music editing, leading to the development of Instruct-MusicGen. By examining these challenges, Instruct-MusicGen aims to expand the scope of music editing operations, offering greater flexibility and precision in handling more complex and dynamic music production scenarios.

In general, MusicMagus contributes to the evolving landscape of AI-assisted music creation, highlighting both the potential and the challenges of zero-shot text-to-music editing. It sets the stage for future explorations that aim to push the boundaries of what is possible in this emerging field.

\chapter{Instruct-MusicGen: Unlocking Text-to-Music Editing for Music Language Models via Instruction Tuning}
\label{ch:instruct}

This chapter introduces Instruct-MusicGen, a system designed to address the limitations of existing text-to-music editing models by enhancing both the precision and flexibility of music generation. Building on the foundational work of Coco-mulla and the advancements realised in MusicMagus, Instruct-MusicGen offers a novel approach that integrates instruction tuning with a dual-modality fusion process. This integration allows the model to interpret complex text-based editing instructions and apply them directly to musical content, thereby enabling more nuanced and dynamic modifications.

Instruct-MusicGen operates by concurrently processing textual instructions and audio inputs, a capability made possible through the introduction of text and audio fusion modules. These modules allow for precise control over the editing process, whether it involves adding, removing, or altering specific musical stems. This system is particularly significant in the context of music production, where the ability to make detailed, text-guided edits, improves AI models' usability for achieving specific artistic goals.

While Instruct-MusicGen builds on the strengths of earlier systems like MusicMagus by broadening the scope of text-to-music editing, it also addresses some of the limitations observed in previous approaches. Specifically, it overcomes challenges related to intra-stem editing and the constraints of diffusion models by leveraging the capabilities of music language models. However, Instruct-MusicGen is not without its own challenges, such as issues related to signal-level precision and the reliance on paired data for fine-tuning. These aspects highlight the areas where further research is needed to fully realise the potential of text-to-music editing.

By carefully examining both its capabilities and limitations, this chapter aims to position Instruct-MusicGen within the broader context of this thesis, contributing to the ongoing exploration of text-guided music generation. Through this analysis, I will explore how Instruct-MusicGen advances the field while also acknowledging the opportunities for future work that remain. \footnote{Code, model weights and demo are available at: \url{https://bit.ly/instruct-musicgen}.}

\section{Introduction}\label{sec:intro}

As discussed in Chapter~\ref{ch:background}, there is a growing interest in the development of models that offer greater controllability~\citep{cocomulla, musiccontrolnet, mustango, airgen} and editability~\citep{instructme, musicmagus, m2ugen} over the music generation process. In music production, a stem, a mixed group of tracks often related by instrument type (like drums or lead vocals), is essential for mixing and mastering because it allows producers to isolate, adjust, and manipulate individual elements of a song. Following the definition in Chapter~\ref{ch:musicmagus}, ``text-to-music editing" involves using textual queries to modify various aspects of a music recording, which can be categorised into two main types: intra-stem editing, which focuses on modifying a single stem (e.g., changing the instrument, timbre, or performance style) and inter-stem editing, which involves altering the relationships among stems (e.g., adding, removing, or separating stems). The work mainly focuses on the problem of inter-stem editing. 

Previous attempts to develop text-based music editing models have encountered several challenges. Some approaches~\citep{audit,instructme} have focused on the training of specialised editing models from scratch, which is resource-intensive and may not yield results comparable to state-of-the-art music generation models. Other work~\citep{loopcopilot, m2ugen, uniaudio} has sought to leverage existing language models (LLMs) and MusicGen~\citep{musicgen}, allowing the LLM to interpret editing instructions without further training the music model. Although this approach offers flexibility, it often lacks the ability to precisely reconstruct the conditional audio, leading to unreliable results. To address these limitations, an ideal solution should harness the knowledge embedded in pretrained models to ensure high-quality audio output while adapting the architecture to accommodate the specific requirements of music editing tasks.

\begin{figure}[tbp]
    \centering
    \hspace*{-1cm}    \includegraphics[width=\linewidth, trim=0cm 24cm 7cm 0cm]{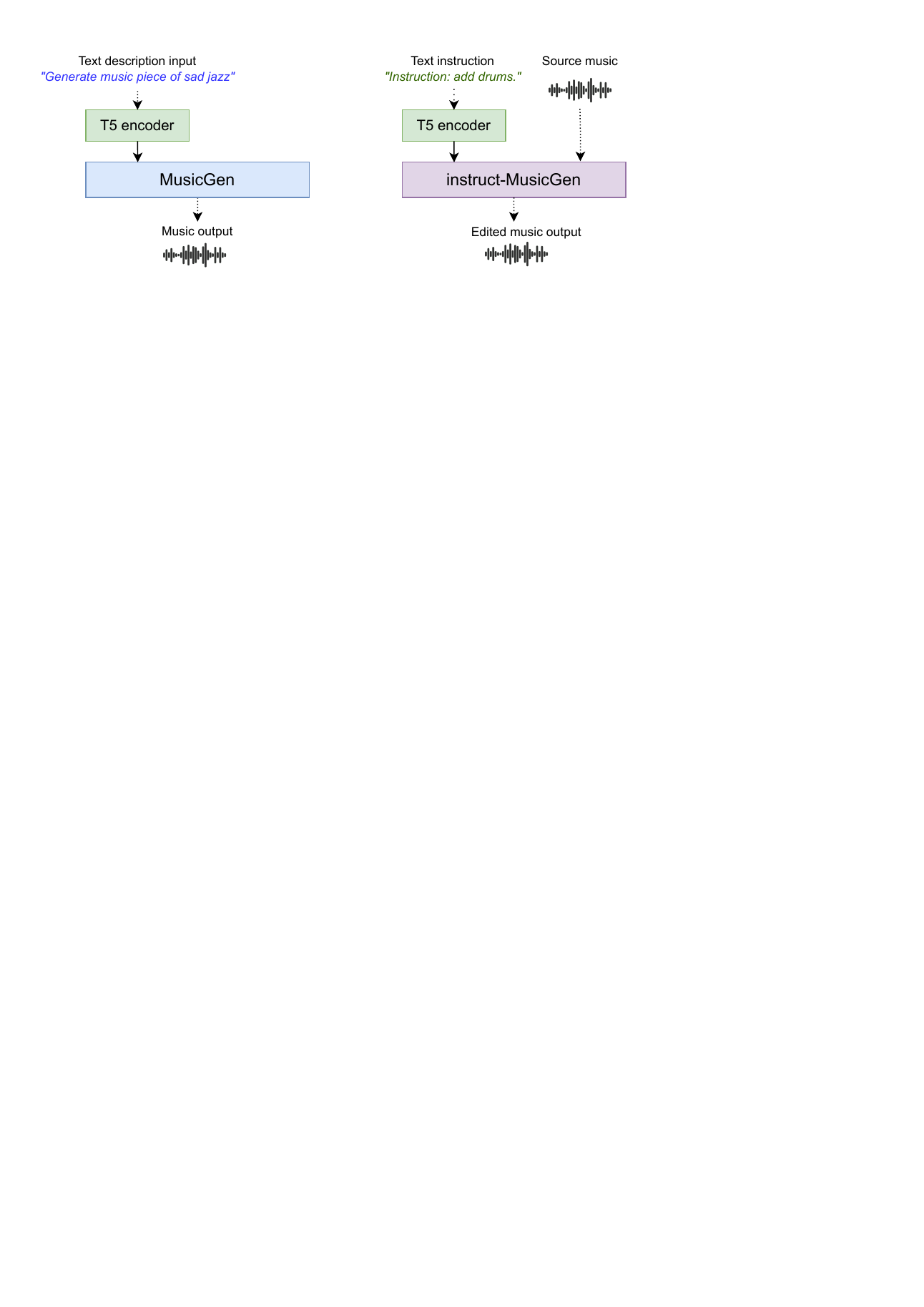}
    \caption{Comparison between MusicGen and Instruct-MusicGen. Instruct-MusicGen accepts both audio input and editing instruction text as conditions.}
    \label{fig:preview}
\end{figure}

As shown in Figure \ref{fig:preview}, Instruct-MusicGen is a novel approach that applies an instruction-following tuning strategy to the pretrained MusicGen model, enhancing its ability to follow editing instructions effectively without finetuning all its parameters. By incorporating an audio fusion module based on LLaMA-adapter~\citep{llamaadapter, cocomulla} and a text fusion module based on LoRA~\citep{lora} into the original MusicGen architecture, I allow the model to process both precise audio conditions and text-based instructions simultaneously, while the original MusicGen does not. This enables Instruct-MusicGen to perform a wide range of editing tasks, such as adding, separating, and removing stems. To train Instruct-MusicGen, I synthesise an instructional dataset using the Slakh2100 dataset~\citep{slakh}, introducing only ~8\% additional parameters compared to the original model, and finetune the model for only 5K steps, which is less than 1\% of training a music editing model from scratch.

I evaluate Instruct-MusicGen on two datasets: the Slakh test set and the out-of-domain MoisesDB dataset~\citep{moisesdb}. The model outperforms existing baselines and achieves performance comparable to models specifically trained for individual tasks. This demonstrates the effectiveness of the approach in leveraging pretrained models for text-to-music editing while maintaining high-quality results. The main contributions of this work are threefold:

\begin{itemize}
   
 \item \textbf{I propose Instruct-MusicGen}, a novel approach that finetunes pretrained large music language models to follow editing instructions effectively. This method addresses the limitations of previous approaches by leveraging pretrained models to ensure high-quality audio output and precise reconstruction, combining both music and text multi-modal control.

 \item \textbf{I enhance the capabilities of a pretrained text-to-music model} to perform multiple tasks, including adding, separating, and extracting stems from music audio, all within a single training process. This approach significantly reduces the computing resources cost compared to training specialised editing models from scratch.

 \item \textbf{I extend the benchmarking for text-to-music editing tasks}, providing a comprehensive comparison across various models. The model outperforms existing baselines and achieves performance comparable to models specifically trained for individual tasks, demonstrating the effectiveness and versatility of the approach.
\end{itemize}

\section{Background: Coco-mulla}

Before introducing Instruct-MusicGen, it is important to first understand its precursor, Coco-mulla. I contributed to this work as a co-author, and through expanding its concepts within the field of music editing, we developed Instruct-MusicGen. However, Coco-mulla is not presented as a primary contribution in my thesis.

Coco-mulla is a framework developed to explore the integration of content-based controls within music generation, specifically addressing the limitations of models that rely solely on text-based descriptions. Previous text-based controls, while functional, often fall short in capturing intricate musical elements such as chord progressions, rhythm patterns, and melodic structures. Coco-mulla seeks to address these challenges by introducing a joint embedding approach that combines both symbolic (e.g., chord and MIDI representations) and acoustic (e.g., drum patterns) features within a unified generative model.

\subsection{Implementation of Coco-mulla}

Coco-mulla is implemented using a parameter-efficient fine-tuning (PEFT) framework designed to enhance pre-trained music language models, such as MusicGen, without necessitating extensive retraining. The model introduces two primary components: a \textit{joint symbolic and acoustic embedding encoder} and a \textit{condition adaptor}.

\subsubsection{Joint Symbolic and Acoustic Embedding}

The joint embedding encoder in Coco-mulla is designed to integrate symbolic music data and acoustic signals into a coherent representation. Symbolic music data is encoded using MIDI and chord structures. For instance, a chord is represented as a combination of its root pitch and bass pitches, as well as chroma, forming a multi-hot vector $\mathbf{c}_i$ for each frame $i$:

\[
\mathbf{c}_i = 
\begin{cases} 
[e(\text{root}); e(\text{bass}); \mathbf{m}; 0], & \text{if chord exists in frame } i \\
[0;0;0; 1], & \text{otherwise} 
\end{cases}
\]

\noindent where $e(\text{root})$ and $e(\text{bass})$ are basis vectors representing the root and bass pitches, and $\mathbf{m}$ is the chroma vector. Acoustic features, particularly drum tracks, are encoded using pre-trained embeddings derived from EnCodec tokens. These embeddings are then mapped to a lower-dimensional space via a trainable matrix, facilitating the model's ability to manage continuous representations effectively.

\subsubsection{Condition Adaptor}

The condition adaptor is critical for integrating the joint embeddings into the generative process. This module modifies the self-attention layers of the pre-trained MusicGen model by introducing a \textit{condition prefix} that captures the temporal evolution of symbolic and acoustic conditions. The condition prefix is inserted into the final layers of the MusicGen decoder, where it interacts with the hidden states derived from the original text-based prompts. The condition information is processed through a series of self-attention and cross-attention mechanisms, ensuring that the generated music adheres to both the symbolic and acoustic conditions provided.

This foundation, while robust in its approach, highlighted several areas where further advancements could be made, particularly in enhancing the flexibility and precision of music editing tasks. These insights directly informed the development of Instruct-MusicGen, which aimed to build upon and extend the capabilities introduced by Coco-mulla.

\subsection{Evolution into Instruct-MusicGen}

Building upon the framework established by Coco-mulla, Instruct-MusicGen was developed to address some of the limitations and expand the potential of content-based controls in music generation. By incorporating a more sophisticated instruction-following mechanism, Instruct-MusicGen enables dynamic and precise text-based editing of music, thus offering greater flexibility and control over the generated output.

While Coco-mulla utilises a joint embedding strategy to merge symbolic and acoustic controls, Instruct-MusicGen introduces an additional \textit{text fusion module}. This module enables the model to interpret and execute complex editing instructions alongside processing audio inputs. The architecture of Instruct-MusicGen integrates the LLaMA-adapter framework into the condition adaptor, enhancing its capacity to manage both symbolic and text-based controls.

Moreover, Instruct-MusicGen employs a dual-modality fusion process, where the \textit{audio fusion module} combines the original music input with the desired edits specified by text instructions. This process allows Instruct-MusicGen to perform tasks such as adding or removing specific instruments, changing musical styles, or separating stems with a higher degree of precision, which Coco-mulla’s original implementation was not designed to achieve to the same extent.

The primary distinction between Coco-mulla and Instruct-MusicGen is the latter's ability to handle complex editing tasks using text instructions without compromising on the quality or fidelity of the generated music. While Coco-mulla provides a framework for generating music based on combined symbolic and acoustic conditions, it does not offer the same level of flexibility required for detailed music editing—an area where Instruct-MusicGen shows marked improvements.

Instruct-MusicGen also benefits from an optimised training process. Despite introducing only $\sim$8\% additional parameters compared to the original MusicGen model, it achieves enhanced performance across a range of editing tasks as seen in Table~\ref{tab:editing_4_stems}. The use of instruction tuning, combined with the advanced fusion modules, allows Instruct-MusicGen to adapt and edit existing music more effectively than its predecessor.

The following sections detail the architecture, training strategy, and inference methods implemented in Instruct-MusicGen, highlighting how these innovations build upon the groundwork laid by Coco-mulla.

\section{Method}\label{sec:method}

\subsection{Instruct-MusicGen}

\begin{figure}[htbp]
    \centering
\includegraphics[width=\linewidth]{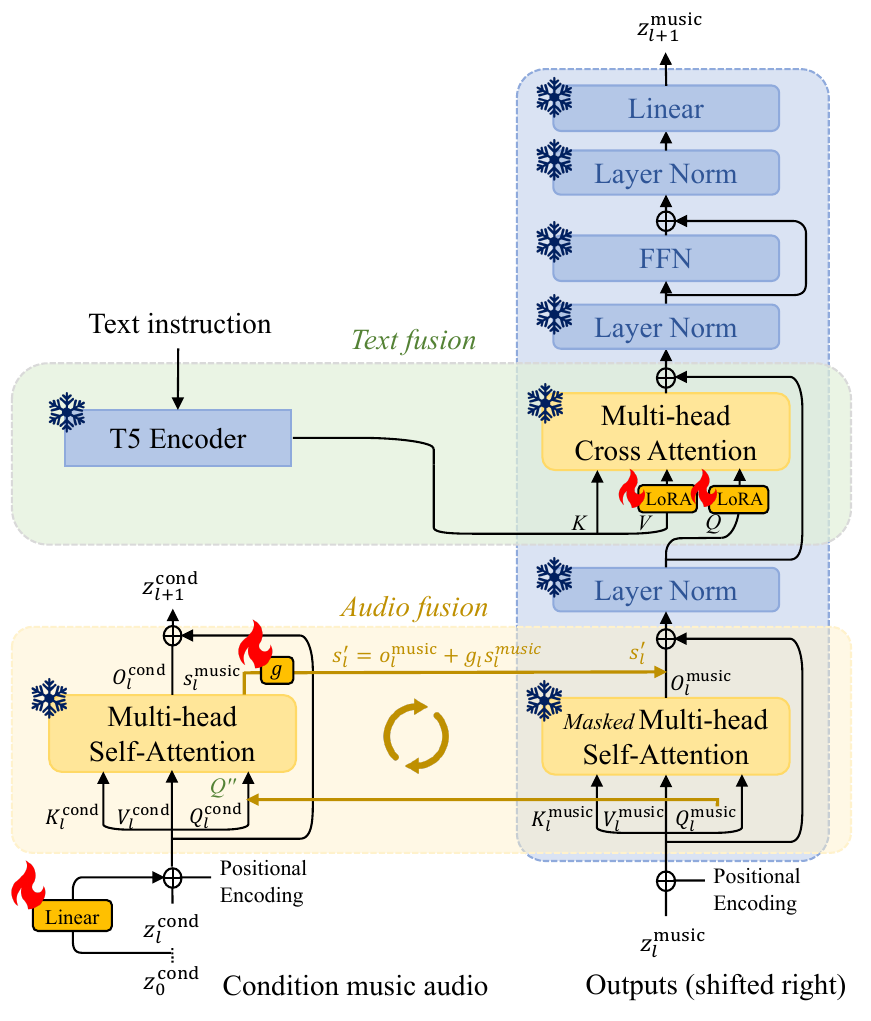}
    \caption{The fusion mechanism inside the Transformer module of Instruct-MusicGen. The audio fusion module transforms the conditional music audio into embeddings using a duplicated encoder and integrates these embeddings into the MusicGen decoder. The text fusion module modifies the cross-attention mechanism to handle text instructions by finetuning specific layers (marked by \textit{Flame}) while keeping the text encoder parameters frozen (marked by \textit{Snowflake}). }
    \label{fig:2}
\end{figure}

Instruct-MusicGen takes a music audio input $X^{\text{cond}}$ and a text instruction $X^{\text{instruct}}$ (e.g., ``Add guitar") as inputs. The model then edits the music audio $X^{\text{cond}}$ according to the instruction $X^{\text{instruct}}$ and generates the desired edited music $X^{\text{music}}$. As illustrated in Figure~\ref{fig:2}, Instruct-MusicGen incorporates two additional modules into the vanilla MusicGen: an audio fusion module and a text fusion module. 

Note that we no longer use the symbolic conditioning in Coco-mulla, which is because audio conditioning has already contained all information of the input stems. 

\subsubsection{Audio Fusion Module}

The audio fusion module enables Instruct-MusicGen to accept external audio inputs, which is inspired by \citet{llamaadapter} and \citet{cocomulla}. Figure~\ref{fig:2} presents the diagram of the audio fusion module. Initially, I convert $X^{\text{cond}}$ into EnCodec tokens, followed by re-encoding these tokens into the embedding $z^{\text{cond}}$ through the pre-trained embedding layers of MusicGen. Similarly, I transform $X^{\text{music}}$ into the pretrained embedding $z^{\text{music}}$.

The module begins by duplicating self-attention modules of the pretrained MusicGen model to extract latent representations of $z^{\text{cond}}$. Given that MusicGen consists of $M$ layers, I denote 

\begin{equation}
    Z^{\text{cond}} = \{z^{\text{cond}}_0, z^{\text{cond}}_1, \ldots, z^{\text{cond}}_M\},
\end{equation}

\begin{equation}
    Z^{\text{music}} = \{z^{\text{music}}_0, z^{\text{music}}_1, \ldots, z^{\text{music}}_M\},
\end{equation}

which represent the hidden states of $X^{\text{cond}}$ and $X^{\text{music}}$ respectively. Note that I use a learnable input embedding as $z^{\text{cond}}_0$ and initialise $z^{\text{music}}_0$ with $z^{\text{music}}$. 

I compute the vanilla self attention for $X^{\text{music}}$ as follows:
\begin{equation}
    Q^{\text{music}}_l, K^{\text{music}}_l, V^{\text{music}}_l = \text{QKV-projector}(z^{\text{music}}_l),
\end{equation}
\begin{equation}
    o^{\text{music}}_l = \text{SelfAttn}(Q^{\text{music}}_l, K^{\text{music}}_l, V^{\text{music}}_l).
\end{equation}

I project $z^{\text{cond}}$ to a high-dimension representation $h$ via a linear layer $f_l$ and learnable positional encoding $e_l$,
\begin{equation}
h=f_l(z^{\text{cond}}) + e_l.
\end{equation}
Then, I compute the $(l+1)$-th layer hidden states of $X^{\text{cond}}$ as follows:
\begin{equation}
    Q^{\text{cond}}_l, K^{\text{cond}}_l, V^{\text{cond}}_l = \text{QKV-projector}(z^{\text{cond}}_l + h),
\end{equation}
\begin{equation}
   z^{\text{cond}}_{l+1} = {\rm SelfAttn}(Q^{\text{cond}}_l, K^{\text{cond}}_l, V^{\text{cond}}_l).
\end{equation}


To fuse information of $X^{\text{cond}}$ into $X^{\text{music}}$, I compute the cross attention between them,
\begin{equation}
    s^{\text{music}}_l = \text{CrossAttn}(Q^{\text{music}}_l+Q^{\text{cond}}_l, K^{\text{cond}}_l, V^{\text{cond}}_l).
\end{equation}

Finally, the attention output of $X^{\text{music}}$ is updated as follows,
\begin{equation}
s'_{l} = o^{\text{music}}_l + g_l \cdot s^{\text{music}}_l,
\end{equation}
\begin{equation}
z^{\text{music}}_{l+1} = \text{TextFusion}(s'_{l}, X^{\text{instruct}}),
\label{eq:textfusion}
\end{equation}
where $g$ is a zero-initialised learnable gating factor.

Thus, the total trainable parameters in Instruct-MusicGen include the input embedding $z^{\rm cond}_0$, linear layers $f_l$, learnable position embeddings $e_l$, learnable gating factors $g$, and learnable parameters in the text fusion module.

\subsubsection{Text Fusion Module}

To replace the text description input with instruction input, I modify the behavior of the current text encoder.
I achieve this by finetuning only the cross-attention module between the text embedding and the music representations while keeping the text encoder's parameters frozen.

The instruction is embedded and encoded by the T5 text encoder as $z^{\text{instruct}} = \text{T5}(X^{\text{instruct}})$. 
For efficient finetuning of the cross-attention module, I apply LoRA to the query and value projection layers. Thus, I expand Equation~\ref{eq:textfusion} as follows,
\begin{equation}
    Q_l, K^{\text{instruct}}_l, V^{\text{instruct}}_l = \text{QKV-Lora}(s'_l, z^{\text{instruct}}),
\end{equation}
\begin{equation}
    z^{\text{music}}_{l+1} = \text{CrossAttn}(Q_l, K^{\text{instruct}}_l, V^{\text{instruct}}_l).
\end{equation}
During fine-tuning, only query and value projection layers are trainable in the text fusion module.






\section{Experiments}\label{sec:experiment}

I conduct both subjective experiments and objective experiments for evaluation, and also provide audio samples in Figure~\ref{fig:subfigures} and a demo page~\footnote{\url{https://bit.ly/instruct-musicgen}}.

\subsection{Objective Experiments}

\subsubsection{Dataset}

For the objective evaluations, I utilise two distinct datasets, each serving a specific purpose in assessing both in-domain and out-of-domain performance capabilities of various models.

\begin{enumerate}
    \item \textbf{Slakh2100 dataset}~\citep{slakh}. The Synthesised Lakh (Slakh) Dataset, originally derived from the Lakh MIDI Dataset v0.1, comprises audio tracks synthesised using sample-based virtual instruments. Known as Slakh2100, this dataset features 2100 tracks complete with corresponding MIDI files. These tracks are methodically divided into subsets designated for training (1500 tracks), validation (375 tracks), and testing (225 tracks), collectively amounting to approximately 145 hours of audio mixtures. This dataset primarily supports audio source separation tasks within a controlled, synthesised environment.
    \item \textbf{MoisesDB dataset}~\citep{moisesdb}. In contrast to Slakh2100, the MoisesDB dataset caters to source separation with a broader scope. It includes 240 real audio tracks sourced from 45 diverse artists spanning twelve musical genres. Uniquely, MoisesDB organises its tracks into a detailed two-level hierarchical taxonomy of stems, offering a varied number of stems per audio track, each annotated with textual descriptions. 
\end{enumerate}

The rationale for selecting two datasets lies in their diverse configurations and common applications. While the Slakh dataset is typically utilised for training models tailored to a four-stem arrangement \citep[e.g.,][]{GMSDI}, the model, Instruct-MusicGen, although initially trained on this dataset, is designed to generalise to various stem configurations. Conversely, models such as InstructME and AUDIT are trained on private or larger, more diverse datasets. By employing both Slakh2100 and MoisesDB, I ensure a comprehensive evaluation, allowing us to fairly compare the adaptability and performance of different models under varying conditions of data familiarity and complexity.

\subsubsection{Data Preprocessing}

I utilised the Slakh2100 dataset to construct an instruction-based dataset for training and testing, employing the following pipeline:

\begin{itemize}
    
\item A data point was randomly selected from the Slakh training dataset.
\item \highlight{An instruction was randomly selected from a predefined set \textit{{add, remove, extract}}, along with a target stem. The target stem represents the difference between the input and output, where the input consists of $n$ stems, and the output consists of $1$, $n-1$ or $n+1$ stems \footnote{The number of both input stems and output stems should range from $[1, N]$, where $N$ is the total number of audio stems}. For example, if the chosen instruction is``add guitar", the target stem would be the distinction between the input and output, with the input containing $n$ stems and the output containing $n+1$ stems. Then, $n$ other stems were randomly chosen from the remaining stems to complete the dataset.}
\item An offset was randomly determined to cut a 5-second audio clip. If the target stem contained more than 50\% silence, a different offset was selected.
\item The stems were mixed according to the specified instructions to create a triplet consisting of \textit{\{instruction text, condition audio input, audio ground truth\}}.
\end{itemize}

\subsubsection{Experimental Setup}

For the finetuning of MusicGen, I jointly trained the audio fusion module and the text fusion module. The optimisation process utilised the AdamW optimiser, with a learning rate set at $5e^{-3}$. I use L2 loss over latent token embeddings as the training objective. Training incorporated a Cosine Annealing scheduler with an initial warmup of 100 steps. The training regimen extended over 5,000 steps with an accumulated batch size of 32, achieved through setting the batch size to 8 and using gradient accumulation over 4 iterations. The finetuning process was executed on a single NVIDIA A100 GPU and was completed within a span of two days.

\subsubsection{Baselines}
In this section, I explore two baseline models, each distinguished by their unique methodologies for handling audio data. 

\begin{enumerate}[itemsep=0pt, parsep=0.6pt]
    \item \textbf{AUDIT}~\citep{audit} (Chapter \ref{sec:audit}): AUDIT is an instruction-guided audio editing model, consisting of a variational autoencoder (VAE) for converting input audio into a latent space representation, a T5 text encoder for processing edit instructions, and a diffusion network that performs the actual audio editing in the latent space. The system accepts mel-spectrograms of input audio and edit instructions, and generates the edited audio as output. 
    
    \item \textbf{M$^2$UGen}~\citep{m2ugen} (Chapter \ref{sec:m2ugen}): The M$^2$UGen framework leverages large language models to comprehend and generate music across various modalities, integrating abilities from external models such as MusicGen~\citep{musicgen} and AudioLDM 2~\citep{audioldm2}. 

\end{enumerate}

InstructME (Chapter \ref{sec:instructme}) is a novel framework for instruction-guided music editing and remixing that leverages latent diffusion models. However, InstructME's model weights and evaluation protocol have note been publicly released, so it is only possible to record the evaluation results which were reported in the original paper.


Table~\ref{tab:baselines} shows a comparison among the models. 

\begin{table}[htbp]
\small
    \centering
    \begin{tabular}{ccccc}
    \toprule
        \textbf{Model}  & \textbf{Param count} & \textbf{Dataset} & \textbf{Hours (h)} & \textbf{Steps}\\
    \midrule
        AUDIT  & 942M (1.5B) & Multiple & $\sim$6500 & 0.5M\\
        InstructME & 967M (1.7B) & Multiple & 417 & 2M \\
        M$^2$UGen  & 637M ($\sim$9B) & MUEdit & 60.22 & - \\
    \midrule
        \textbf{Instruct-MusicGen} & \textbf{264M} (3.5B) & Slakh & 145 & \textbf{5K} \\
     \bottomrule
    \end{tabular}
    \caption{Comparison of different models, where the param count numbers are the number of trainable parameters and total parameters respectively. The proposed model has the lowest parameter size, and only requires 5K training steps.}
    \label{tab:baselines}
\end{table}

\subsubsection{Metrics}

Having been introduced in Chapter~\ref{ch:background}, the metrics to evaluate model performance are listed below.

\begin{enumerate}[itemsep=0pt, parsep=0.6pt]

\item \textbf{Fr\'{e}chet Audio Distance (FAD)}~\citep{fad}\footnote{\url{https://github.com/gudgud96/frechet-audio-distance}.} measures the similarity between two sets of audio files by comparing multivariate Gaussian distributions fitted to feature embeddings from the audio data. I use the FAD score to evaluate the overall audio quality of the predicted music.

\item \textbf{CLAP score (CLAP)}~\citep{clap}\footnote{\url{https://github.com/LAION-AI/CLAP}.} evaluates the alignment and relevance of audio content with corresponding textual descriptions, using models pretrained on both audio and language tasks to assess semantic coherence between audio and text. A high CLAP score indicates the predicted music aligns with the text description well.

\item \textbf{Kullback-Leibler Divergence (KL)}\footnote{\url{https://github.com/haoheliu/audioldm_eval}.} assesses the difference between the probability distributions of audio features from two sources, indicating information loss when approximating one distribution with another. A low KL score indicates the predicted music shares similar features with the ground truth.

\item \textbf{Structural Similarity (SSIM)}~\citep{ssim} is an image quality metric that I adapt to evaluate structural similarity between predicted music and ground truth.

\item \textbf{Scale-Invariant Signal-to-Distortion Ratio (SI-SDR)} \citep{si-sdr} quantifies audio quality, especially in source separation tasks. It is scale-invariant, useful for varying audio volumes, and measures distortion relative to a reference signal. I use SI-SDR score to evaluate the signal loss of the predicted audio.

\item \textbf{Scale-Invariant Signal-to-Distortion Ratio improvement (SI-SDRi)} \citep{si-sdri} extends SI-SDR, measuring the improvement in signal-to-distortion ratio after processing. It is commonly used in audio enhancement and separation contexts.

\end{enumerate}

To further investigate if the model can successfully add, remove or extract the instrument, I propose the \textbf{P-Demucs score} to evaluate model performance. This metric specifically focuses on detecting the presence of a newly added instrument in the generated audio. It leverages the Demucs model, a source separation model, to isolate the target instrument from the audio. After separation, the root-mean-square energy (RMSE) of the isolated track is used for judging~\footnote{\url{https://librosa.org/doc/main/generated/librosa.feature.rms.html}}. For example, if the instruction is to ``add guitar," the success of the model is indicated by the presence of a non-silent guitar track. P-Demucs (Precision over Demucs) measures the precision of this detection, focusing exclusively on the new instrument's presence without considering changes to existing instruments.

\subsubsection{Results}

The evaluation of Instruct-MusicGen demonstrates its superior performance across various tasks compared to existing text-to-music editing baselines (AUDIT, InstructME, M$^2$UGen). On the Slakh dataset (Table~\ref{tab:editing_4_stems}), Instruct-MusicGen performs well in adding, removing, and extracting stems, achieving the lowest Fréchet Audio Distance (FAD) and the highest CLAP and SSIM scores in the addition task. It also significantly improved the signal-to-noise ratio (SI-SDR) in the removal task, showing balanced performance across all metrics and proving its robustness in various editing scenarios. 

\begin{sidewaystable}[htbp]
    \centering
    \begin{tabular}{c|l|ccccccc}
    \toprule
    \toprule
    \textbf{Task} & \textbf{Models}  & \textbf{FAD}$\downarrow$ & \textbf{CLAP}$\uparrow$ & \textbf{KL}$\downarrow$ & \textbf{SSIM}$\uparrow$ &\textbf{P-Demucs}$\uparrow$& \textbf{SI-SDR}$\uparrow$ & \textbf{SI-SDRi}$\uparrow$ \\
    \midrule
    \midrule
    \multirow{5}{*}{\textbf{Add}} & AUDIT & 6.88 & 0.12  & 1.02   & 0.21 &0.53& - & - \\
    & M$^2$UGen  & 7.24 & 0.22 & 0.99  & 0.20 &0.43 & - & - \\
    \cmidrule{2-9}
    \cmidrule{2-9}
    & \textbf{Instruct-MusicGen}  & \textbf{3.75} & \textbf{0.23}  & \textbf{0.67}   & \textbf{0.26}  & \textbf{0.80} & - & - \\
    \midrule
    \midrule
    \multirow{3}{*}{\textbf{Remove}} & AUDIT  & 15.48 & 0.07  & 2.75  & 0.35 & 0.33  & -45.60  & -47.28  \\
    & M$^2$UGen  & 8.26 & 0.09 & 1.59 & 0.23 & 0.70  & -44.20 &  -46.13 \\
    \cmidrule{2-9}
    \cmidrule{2-9}
    & \textbf{Instruct-MusicGen}  & \textbf{3.35} & \textbf{0.12}  & \textbf{0.66}  & \textbf{0.45} & \textbf{0.76}  & \textbf{-2.09}  & \textbf{-3.77}   \\
    \midrule
    \midrule
    \multirow{3}{*}{\textbf{Extract}} & AUDIT  & 15.08 & 0.06  & 2.38   & 0.42 & 0.61 & -52.90  & -50.16  \\
    & M$^2$UGen  & 8.14 & 0.11  & 2.15  & 0.31 & 0.60 & -46.38 & -43.53 \\
    \cmidrule{2-9}
    \cmidrule{2-9}
    & \textbf{Instruct-MusicGen} & \textbf{3.24} & \textbf{0.12}  & \textbf{0.54}  & \textbf{0.52} & \textbf{0.75}  & \textbf{-9.00}  & \textbf{-6.15}  \\    
    \bottomrule
    \bottomrule
    \end{tabular}
    \caption{Comparison of text-based music editing models on the Slakh dataset (4 stems). }
    \label{tab:editing_4_stems}
\end{sidewaystable}

Similarly, in the MoisesDB dataset evaluations (Table~\ref{tab:moisesDB}), Instruct-MusicGen continued to demonstrate strong performance. It achieved competitive FAD scores and showed improvements in CLAP and SSIM metrics for both addition and removal tasks. Our model consistently outperformed baseline models, highlighting its efficiency and effectiveness in text-to-music editing applications.

\begin{sidewaystable}[htbp]
    \centering
    \begin{tabular}{c|l|cccccccc}
    \toprule
    \toprule
    \textbf{Task} & \textbf{Models} & \textbf{FAD}$\downarrow$ & \textbf{CLAP}$\uparrow$ & \textbf{KL}$\downarrow$ & \textbf{SSIM}$\uparrow$ &\textbf{P-Demucs}$\uparrow$& \textbf{SI-SDR}$\uparrow$ & \textbf{SI-SDRi}$\uparrow$ \\
    \midrule
    \midrule
    \multirow{3}{*}{\textbf{Add}} & AUDIT  & 4.06 & 0.12  & 0.84   & 0.21 &0.50 & - & - \\
    & M$^2$UGen  & 5.00 & \textbf{0.18} & 0.83  & 0.20 &0.45& - & - \\
    \cmidrule{2-9}
    & \textbf{Instruct-MusicGen}  & \textbf{3.79} & \textbf{0.18} & \textbf{0.35}  & \textbf{0.35} &\textbf{0.77}& - & - \\
    \midrule
    \midrule
    \multirow{3}{*}{\textbf{Remove}} & AUDIT  & 10.72 & 0.10 & 2.46 & \textbf{0.34} &0.41 & -44.32 & -57.10 \\
    & M$^2$UGen  & \textbf{3.75} & \textbf{0.13} & 1.27 & 0.19 &0.72& -43.94 & -56.73 \\
    \cmidrule{2-9}
    & \textbf{Instruct-MusicGen}  & 5.05 & 0.10 & \textbf{0.84} & \textbf{0.34}& \textbf{0.78} & \textbf{-13.70} & \textbf{-26.48}   \\
    \midrule
    \midrule
    \multirow{3}{*}{\textbf{Extract}} & AUDIT  & 6.67 & 0.07 & 1.97 & \textbf{0.45} &0.60 & -54.53 & -56.17 \\
    & M$^2$UGen  & 5.74 & 0.08 & 1.91 & 0.25 & 0.52 & -42.84 & -44.49 \\
    \cmidrule{2-9}
    & \textbf{Instruct-MusicGen}  & \textbf{4.96} & \textbf{0.11} & \textbf{1.36} & 0.40 & \textbf{0.78} & \textbf{-21.39} & \textbf{-23.03}  \\    
    \bottomrule
    \bottomrule
    \end{tabular}
    \caption{Comparison of text-based music editing models on the MoisesDB dataset.}
    \label{tab:moisesDB}
\end{sidewaystable}

I find that all models exhibit negative SI-SDR and SI-SDRi scores, which is a common occurrence when evaluating generative models on a signal level. These metrics are typically designed for source separation tasks and are not entirely fair to generative models, as they penalise even minor discrepancies between the generated and original signals. Generative models, like Instruct-MusicGen, often focus on producing perceptually plausible audio rather than perfectly matching the original signal at a technical level, causing the inherent gap between the model and the tasks.\footnote{Demos are available at: \url{https://bit.ly/instruct-musicgen}.}

\subsection{Subjective Experiments}

\subsubsection{Experimental Setup}

I conducted a subjective listening test to evaluate the model's performance. This test involved disseminating an online survey within the Music Information Retrieval (MIR) community and the broader research network, which resulted in the collection of 30 complete responses. The gender distribution of the participants was 23 males (76.7\%) and 7 females (23.3\%). Regarding the experience of musical training and experience, 4 participants (13. 3\%) had less than 1 year of experience, 13 (43.3\%) had between 1 and 5 years, and 13 participants (43.3\%) had more than 5 years of experience. 

For the data preparation, I randomly selected a subset of data points from the objective test dataset. Specifically, 6 audio samples were chosen consisting of 2 audio samples for each subtask (add, remove, extract). Each data point included results from the baseline models, the proposed models, and the ground truth from the dataset. The experiment has been approved by the ethics committee at Sony AI.

\subsubsection{Metrics}

\begin{enumerate}
    \item \textbf{Instruction Adherence (IA)} assesses how accurately the generated music follows the given editing instruction. In this experiment, participants rate the generated music on a scale from 1 to 5, where 1 indicates that the instruction was not followed at all, and 5 indicates that the instruction was followed perfectly. For example, if the instruction is ``Remove Drums," a rating of 1 would mean that the drums were not removed at all, while a rating of 5 would mean that the drums were completely removed.

\item \textbf{Audio Quality (AQ)} evaluates the overall audio quality of the generated music in comparison to the original music. Participants rate the audio quality on a scale from 1 to 5, where 1 represents very poor quality with significant degradation compared to the original music, and 5 represents excellent quality, as good as or better than the original music. This metric helps in understanding how the editing process affects the overall sound quality of the music.
\end{enumerate}

\subsubsection{Results}

\begin{table}[ht]
\small
    \centering
    \begin{tabular}{l|cc}
    \toprule
        \textbf{Model} & \textbf{Instruction Adherence}$\uparrow$ & \textbf{Audio Quality}$\uparrow$ \\
        \midrule
        AUDIT & 1.54 &	2.56 \\
        M$^2$UGen & 1.70 &	1.92 \\
        \midrule
        \textbf{Instruct-MusicGen} & \textbf{3.85} & \textbf{3.55}\\
        \midrule
        Ground truth & 4.36 & 4.21\\
        \bottomrule
    \end{tabular}
    \caption{The subjective experiment results. The model shows significant improvements on both the Instruction Adherence (IA) score and the Audio Quality (AQ) score over the baselines models.}
    \label{tab:subjective}
\end{table}

The results of the subjective experiments are summarised in Table \ref{tab:subjective}. I conducted two paired t-tests with Bonferroni correction, setting the significance level at $\alpha=0.05$. The results show that the proposed model demonstrates a significant improvement in both Instruction Adherence (IA) and Audio Quality (AQ) compared to the baseline models, AUDIT and M$^2$UGen. 

Specifically, the model achieves an IA score of 3.85 and an AQ score of 3.55, significantly outperforming AUDIT (IA: 1.54, AQ: 2.56) and M$^2$UGen (IA: 1.70, AQ: 1.92). These results suggest that the model not only better adheres to the given instructions but also maintains higher audio quality in the generated music.

\subsection{Ablations}

\subsubsection{Hyperparameter Selection}

The total trainable parameters include the input embedding layer, linear layers for joint embedding learning, and gate scalars, and the linear layers make up over 80\% of the total parameters. In this ablation, I want to investigate whether the parameters can be reduced by revising the architecture of the linear layers. 

The original linear layers have the shape of $(48, 2048, 2048)$, where 48 is the number of layers of the of MusicGen model, and $(2048, 2048)$ is a linear transformation from embedding to layer-wise representation. I replace a single layer with an auto-encoder architecture, consisting of two linear layers: $(2048, M)$ and $(M, 2048)$, where $M \ll 2048$. 

\begin{figure}[htbp]
    \centering
\includegraphics[width=\linewidth]{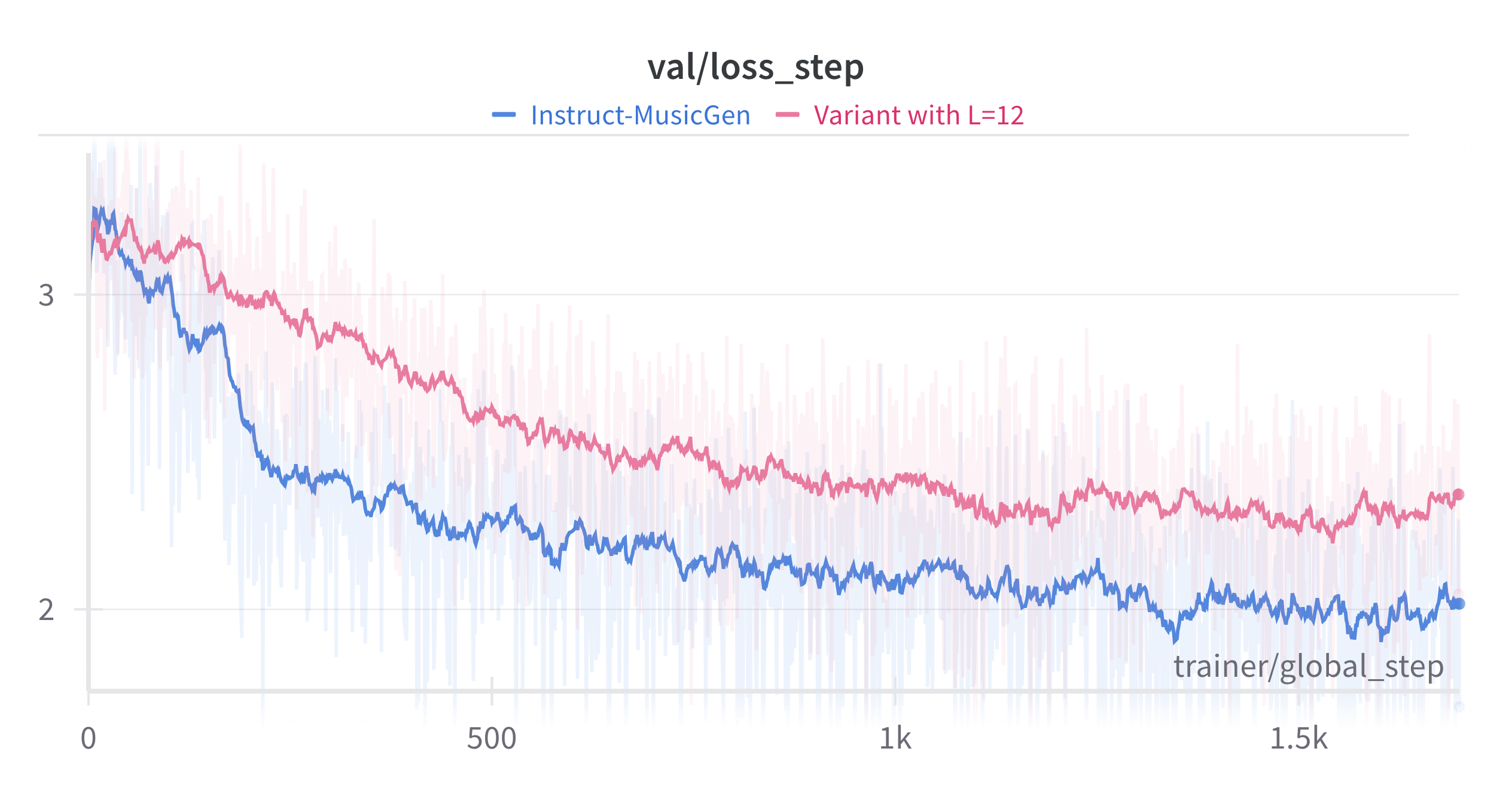}
    \caption{The comparison of L2 loss curve between the original Instruct-MusicGen and its variant with less parameters.}
    \label{fig:ablation_L}
\end{figure}

I record and compare the loss on the validation set between the variant and the original Instruct-MusicGen model, shown in Figure~\ref{fig:ablation_L}. The variation has $M=12$, which reduces the parameter size from 201M to 23M. From the experimental results, I find that although the paramater size is reduced, the instruction tuning still works for MusicGen, but the small parameter size limits the model's representation ability, making the model perform worse than the original version.

\subsubsection{Comparison with Original Text Encoder}

I use an audio fusion module to enable MusicGen to accept additional audio input, allowing the model to process instruction text input without adding a new adapter. By finetuning the cross-attention module between the text encoder and the MusicGen transformer, I can fuse instructional knowledge into text information. In this ablation study, I examine the necessity of finetuning the text encoder. Since the text encoder is already capable of accepting text input, can it directly understand the semantic meaning of editing instructions?

To investigate this, I revised the architecture of Instruct-MusicGen's text fusion module, configuring the music decoder to only accept cross-attention information from the original text encoder.

From Table~\ref{tab:ablation3}, I observe that removing the text fusion module and relying on the original text encoder to interpret the instructions significantly decreases the model's performance. Qualitative results from the predicted samples indicate that the original text encoder, which is designed for text description, struggles with the format of instructional commands. It fails to fully understand the semantics of keywords such as ``extract," ``remove," ``no," and ``only," which are not typically used to describe music but rather to indicate editing operations. Therefore, the presence of the text fusion module is crucial for MusicGen to correctly interpret the semantic meaning of instructions.

\begin{sidewaystable}[htbp]
    \centering
    \begin{tabular}{l|l|ccccccc}
    \toprule
    \toprule
    \textbf{Task} & \textbf{Models} & \textbf{FAD}$\downarrow$ & \textbf{CLAP}$\uparrow$ & \textbf{KL}$\downarrow$ & \textbf{SSIM}$\uparrow$ & \textbf{SI-SDR}$\uparrow$ & \textbf{SI-SDRi}$\uparrow$ \\
    \midrule
    \midrule
    \multirow{2}{*}{\textbf{Add}} & w/o. text fusion & 5.26  & 0.15 & 0.70 & 0.25 & - & - \\
    & \textbf{w. text fusion} & \textbf{3.75} & \textbf{0.23}  & \textbf{0.67}   & \textbf{0.26}   & - & -  \\
    \midrule
    \multirow{2}{*}{\textbf{Remove}} & w/o. text fusion & 5.58  &  0.09 & 1.23  &  0.31  & -10.32  & -12.25 \\
     & \textbf{w. text fusion} & \textbf{3.35} & \textbf{0.12}  & \textbf{0.66}  & \textbf{0.45}  & \textbf{-2.09}  & \textbf{-3.77}  \\
    \midrule
    \multirow{2}{*}{\textbf{Extract}} & w/o. text fusion & 7.58  &  0.04  & 1.87  &  0.42  & -23.06  & -20.21  \\
    & \textbf{w. text fusion} & \textbf{3.24} & \textbf{0.12}  & \textbf{0.54}  & \textbf{0.52}  & \textbf{-9.00}  & \textbf{-6.15}  \\
    \bottomrule
    \bottomrule
    \end{tabular}
    \caption{Performance comparison between models with and without text fusion module across different tasks. }
    \label{tab:ablation3}
\end{sidewaystable}

\subsubsection{Experiments Slakh Dataset (all stems)}

The results for the Slakh dataset with all stems are shown in Table~\ref{tab:slakh_full}, where the results from InstructME are reported. From the results, InstructME performs better than AUDIT and M$^2$UGen, and being superior to the proposed model in the ``remove" task. However, InstructME still \highlight{has some worse scores} in the ``add" and ``extract" task. \highlight{Additionally, in previous experiments, Instruct-MusicGen consistently achieved the highest or tied scores on the CLAP metric, but in this experiment, it performs slightly worse than M$^2$UGen.}

\begin{sidewaystable}[htbp]
    \centering
    \begin{tabular}{c|l|cccccc}
    \toprule
    \toprule
    \textbf{Task} & \textbf{Models} & \textbf{FAD}$\downarrow$ & \textbf{CLAP}$\uparrow$ & \textbf{KL}$\downarrow$ & \textbf{SSIM}$\uparrow$ & \textbf{SI-SDR}$\uparrow$ & \textbf{SI-SDRi}$\uparrow$ \\
    \midrule
    \midrule
    \multirow{4}{*}{\textbf{Add}} & AUDIT  & 6.30 & 0.08 & 1.27 & 0.14 & - & - \\
    & InstructME*  & 3.15 & - & - & - & - & - \\
    & M$^2$UGen  & 5.80 & \textbf{0.21} & 0.88 & 0.17 & - & - \\
    \cmidrule{2-8}
    & \textbf{Instruct-MusicGen}  & \textbf{3.07} & {0.20} &  \textbf{0.53}  & \textbf{0.21} & - & - \\
    \midrule
    \midrule
    \multirow{4}{*}{\textbf{Remove}} & AUDIT  & 12.19 & 0.06 & 2.58 & 0.27 & -45.40 & -49.32 \\
    & InstructME*  & \textbf{1.87} & - & - & - & - & - \\
    & M$^2$UGen  & 6.42 & 0.10 & 1.42  & 0.22 & -42.72 & -46.02 \\
    \cmidrule{2-8}
    & \textbf{Instruct-MusicGen} & 2.95 & \textbf{0.12} & \textbf{0.64} & \textbf{0.39} &\textbf{-3.01} & \textbf{-6.31}   \\
    \midrule
    \midrule
    \multirow{4}{*}{\textbf{Extract}} & AUDIT & 13.78 & 0.04 & 2.86 & 0.43 & -54.95 & -49.10 \\
    & InstructME* & 5.04 & - & - & - & - & - \\
    & M$^2$UGen  & 6.29 & \textbf{0.12} & 1.64 & 0.30 & -45.65 &  -42.53 \\
    \cmidrule{2-8}
    & \textbf{Instruct-MusicGen} & \textbf{3.25} & 0.08 & \textbf{0.85} & \textbf{0.45} & \textbf{-9.89} & \textbf{-6.77}  \\    
    \bottomrule
    \bottomrule
    \end{tabular}
    \caption{Comparison of Text-Based Music Editing Models on the Slakh dataset (full stems).}
    \label{tab:slakh_full}
\end{sidewaystable}

\subsection{Audio Samples}

\begin{figure}[htbp]
    \centering
    \subfigure[Input music.]{
        \includegraphics[width=0.3\textwidth]{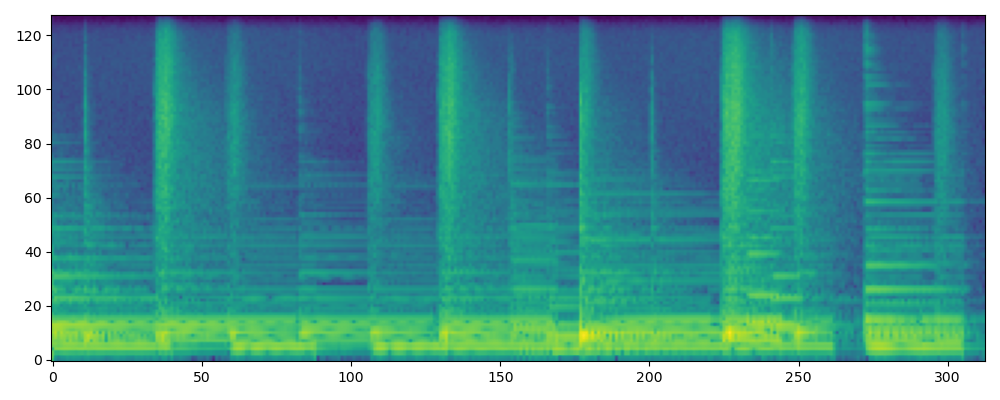}
    }
    \hfill
    \subfigure[Edited music output.]{
        \includegraphics[width=0.3\textwidth]{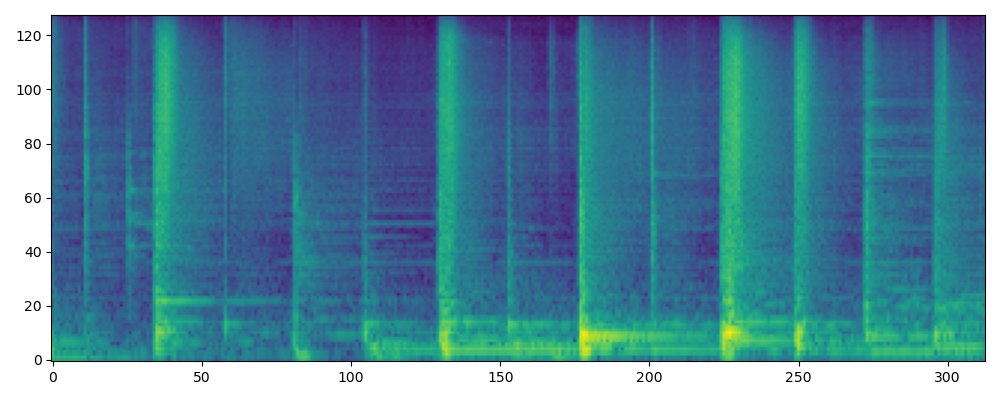}
    }
    \hfill
    \subfigure[Ground truth.]{
        \includegraphics[width=0.3\textwidth]{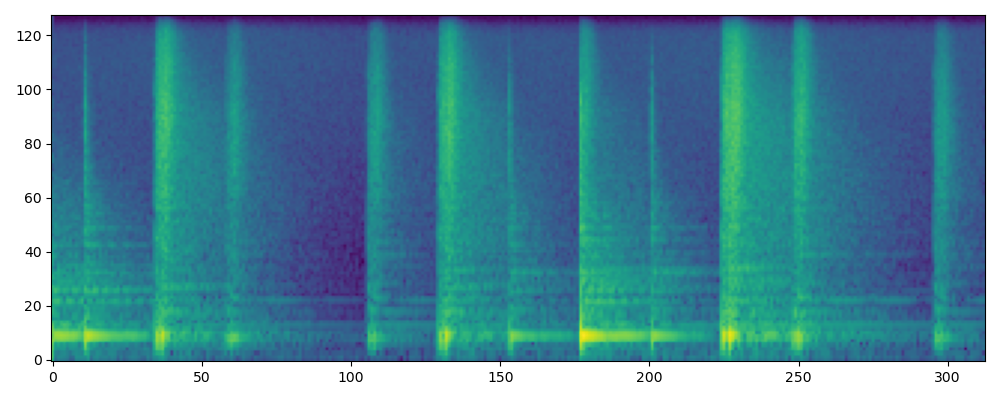}
    }
    
    \subfigure[Input music.]{
        \includegraphics[width=0.3\textwidth]{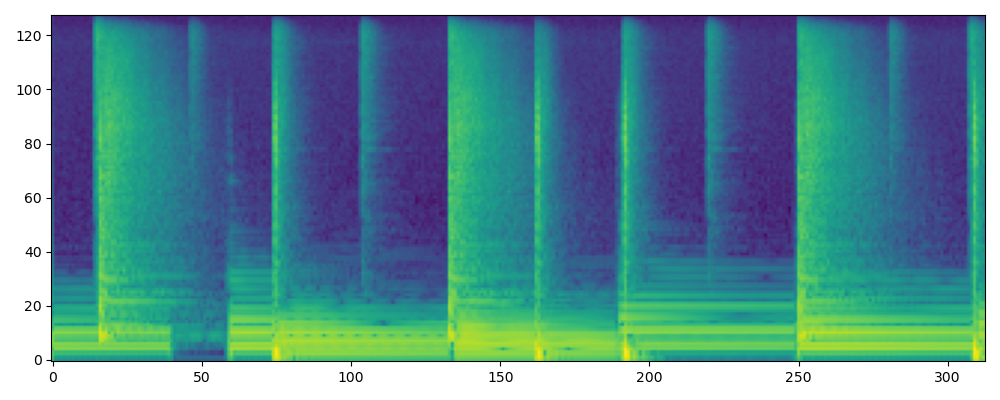}
    }
    \hfill
    \subfigure[Edited music output.]{
        \includegraphics[width=0.3\textwidth]{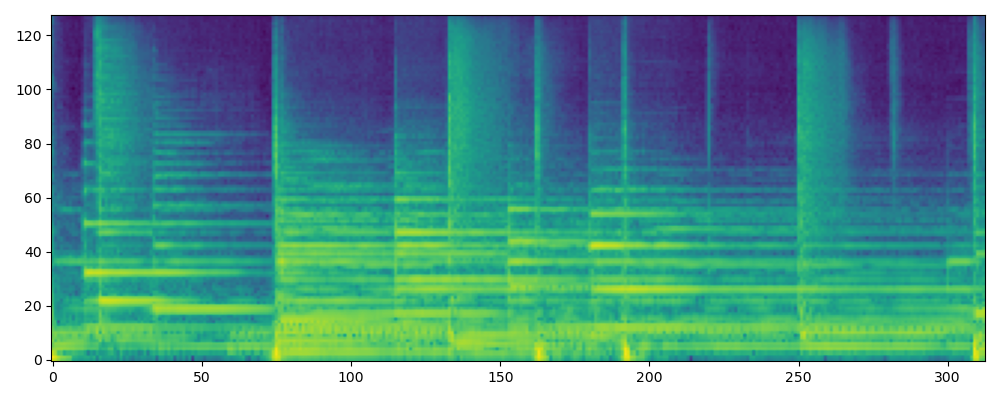}
    }
    \hfill
    \subfigure[Ground truth.]{
        \includegraphics[width=0.3\textwidth]{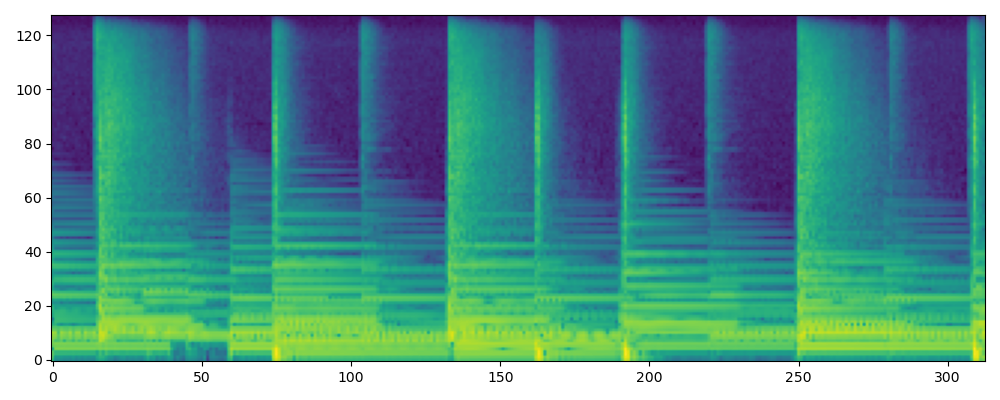}
    }
    
    \subfigure[Input music.]{
        \includegraphics[width=0.3\textwidth]{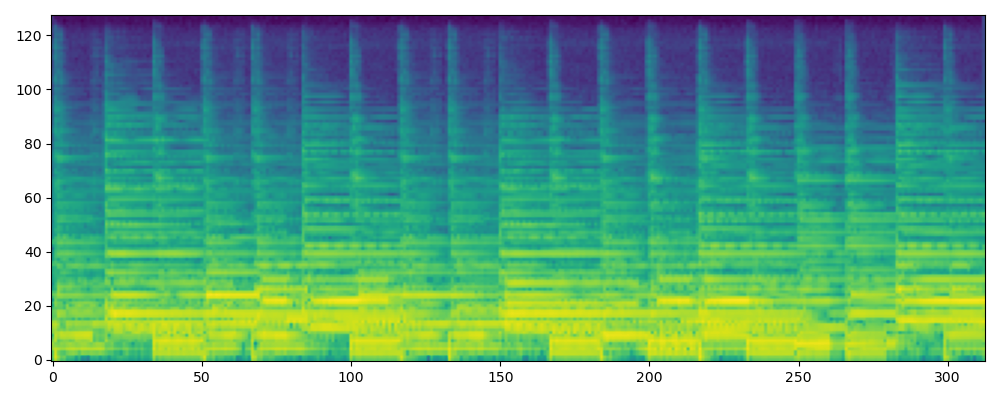}
    }
    \hfill
    \subfigure[Edited music output.]{
        \includegraphics[width=0.3\textwidth]{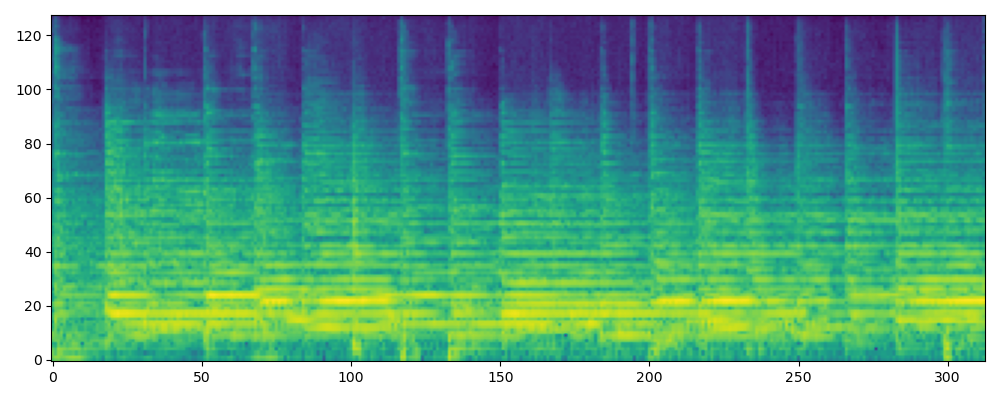}
    }
    \hfill
    \subfigure[Ground truth.]{
        \includegraphics[width=0.3\textwidth]{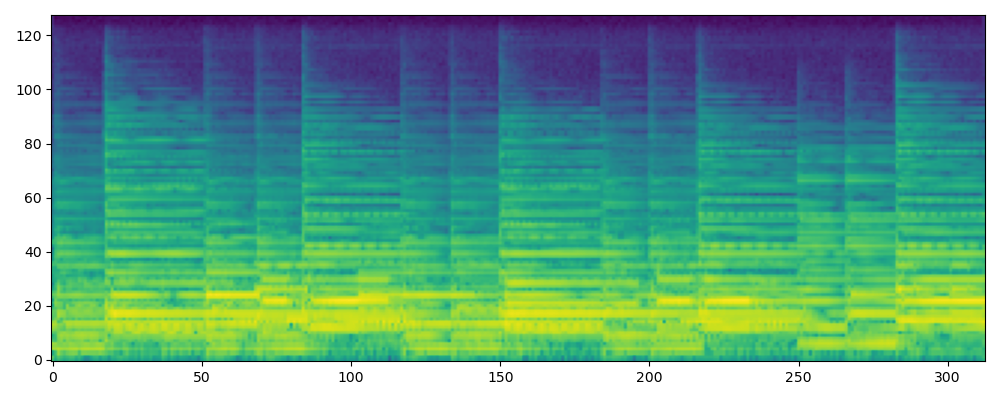}
    }
    
    \caption{Audio samples. \textbf{Fig (a-c):} Extracting a stem. Instruction: ``Extract drum."; \textbf{Fig (d-f):} Adding a stem. Instruction: ``Add piano."; \textbf{Fig (g-i):} Removing a stem. Instruction: ``Remove bass";}
    \label{fig:subfigures}
\end{figure}

I show some audio samples in Figure~\ref{fig:subfigures}. The generation results from Instruct-MusicGen show that the model effectively performs music editing tasks while largely preserving the original audio input. The spectrograms indicate that the model successfully adds, removes, or extracts instruments with minimal alteration to the rest of the audio, though some minor changes suggest potential detail loss, possibly due to codec compression. 

The P-Demucs score further reflects the model's capability, with a success rate of around 80\% in correctly adding the specified instrument. In cases where the model did not fully succeed, it still maintained the overall structure of the audio, indicating reliable performance with some room for improvement in refining the added elements.

\section{Limitation and discussion}

Instruct-MusicGen still has several limitations. Firstly, the generative editing tasks do not guarantee signal-level precision, as the process of predicting EnCodec tokens inherently involves some inaccuracy. Second, the finetuning process still relies on paired data, limiting the model's applicability to more complex editing tasks where such data is not readily available. These constraints highlight the need for further research to improve signal accuracy and expand the model's capabilities to handle a broader range of editing scenarios without heavily relying on paired datasets. Nonetheless, the ability to edit music with fewer computational resources opens up new avenues for creative expression and automation in music composition. Instruct-MusicGen represents a significant contribution to the field of text-to-music research, offering a robust and efficient tool for both musicians and researchers.

\section{Conclusion}\label{sec:conclusion}


In this chapter, I have examined the development and contributions of Instruct-MusicGen, a model designed to enable the editability for a pretrained text-to-music model. The foundation for Instruct-MusicGen was laid by Coco-mulla, a framework that introduced content-based controls through joint symbolic and acoustic embeddings. While Coco-mulla effectively addressed certain limitations inherent in text-only control models, Instruct-MusicGen has further advanced these capabilities by integrating sophisticated instruction-following mechanisms and dual-modality fusion processes.

The architectural enhancements in Instruct-MusicGen, including the introduction of text and audio fusion modules, allow for more precise and flexible music editing. This may enable a wider range of applications in AI-assisted music creation, particularly in scenarios requiring detailed manipulation of musical elements based on user instructions. However, the flexibility of the editing operations remains limited to a set of predefined tasks, such as adding, removing, or extracting specific musical stems. The evolution from Coco-mulla to Instruct-MusicGen reflects the ongoing efforts to refine and expand the capabilities of music language models, paving the way for future research in this domain.

In the context of this thesis, Instruct-MusicGen represents the culmination of a series of advancements aimed at improving the controllability and editability of AI systems in music production. The progression from Loop Copilot, which focused on iterative refinement and orchestration through a conversational interface, to MusicMagus, which introduced zero-shot text-to-music editing capabilities, has laid a solid foundation for the innovations presented in Instruct-MusicGen. Each system addressed specific challenges in AI-assisted music creation, progressively enhancing the precision and flexibility of music editing.

Instruct-MusicGen builds upon these prior contributions by integrating instruction tuning into the MusicGen model, enabling a higher degree of precision in editing tasks such as adding, removing, or modifying specific musical stems. This system not only achieves greater accuracy in edits but also broadens the applicability of music language models to more complex and dynamic production environments, offering a scalable and efficient solution. The incorporation of dual-modality fusion processes allows Instruct-MusicGen to effectively process both textual instructions and audio inputs, making it a robust tool for creative expression and automation in music composition.

However, Instruct-MusicGen still encounters limitations, particularly in signal-level precision and reliance on paired data during finetuning. Additionally, the experiments are confined to rigidly defined scenarios, limiting the full utilisation of the model's text processing abilities. The vocabulary and language of editing instructions are relatively simple, constrained by the dataset and template-based construction of the instruction editing dataset. These constraints highlight the need for further research to improve signal accuracy and expand the flexibility of editing operations. Future work should aim to reduce the dependence on paired datasets and explore more fundamental approaches to enhance the diversity of editing instructions, thereby fully leveraging the model's text processing potential.

\chapter{Conclusions and Future Work}
\label{ch:conclusions}

In this thesis, I have explored the landscape of AI-assisted music creation, focusing on enhancing the controllability and editability of text-to-music generation models. Throughout this journey, I introduced three systems—Loop Copilot, MusicMagus, and Instruct-MusicGen—\highlight{each addressing the limitations of its predecessors.} This chapter will provide a detailed review of the contributions made, discuss the limitations encountered, and propose future work to further advance this field.

\section{Summary of Contributions}
The overarching aim of this thesis was to enable iterative and dynamic control over music generation, addressing gaps in precision and flexibility. To achieve this, three key systems were developed:

\subsection{Loop Copilot: Conducting AI Ensembles for Music Generation and Iterative Editing}

Loop Copilot - built upon its preliminary work, COSMIC - is the first system developed in this thesis, aimed at addressing the need for iterative refinement during the music creation process. By leveraging a large language model (LLM), Loop Copilot was designed to orchestrate a set of specialised AI music models, enabling the user to guide the music generation through a conversational interface. This conversational framework allows users to iteratively refine their music in multiple rounds of dialogue, which reflects the natural workflow during collaborations among human composers and music producers.

A significant contribution of Loop Copilot is the Global Attribute Table (GAT), which records and maintains key musical attributes such as tempo, key, and mood. GAT ensures that iterative edits do not break the musical consistency, allowing for a coherent development of the music. The system architecture involved an LLM that interpreted user instructions, followed by task allocation to appropriate music models, each handling different aspects of the composition. This multi-model coordination is what makes Loop Copilot distinct from single-model music systems.

Loop Copilot was tested through an in-depth user study involving experienced and novice musicians. The study employed semi-structured interviews, usability testing, and a questionnaire focusing on metrics like user satisfaction, system coherence, and efficiency of the music creation process. One key observation was that users appreciated the interactive dialogue and the ease of modifying different attributes without technical intervention. However, the system struggled when users sought highly specific or unconventional edits, reflecting the LLM's limitations in interpreting more technical music jargon.

The key limitation of Loop Copilot lies in the precision of edits. While the LLM can effectively communicate broad instructions to specialised models, there was a notable gap in understanding detailed instructions such as fine adjustments to harmonic progressions or specific instrumentation. Moreover, the reliance on fixed models limited the flexibility when more complex edits were needed. This gap drove the development of more sophisticated solutions in the subsequent systems.

\subsection{MusicMagus: Zero-Shot Text-to-Music Editing via Diffusion Models}
MusicMagus was introduced as a response to the limitations in the precision of edits observed in Loop Copilot. Where Loop Copilot relied on orchestrating multiple models, MusicMagus shifted focus to zero-shot text-to-music editing via pretrained diffusion models. This system explored the concept of latent space manipulation to allow the modification of specific musical attributes, such as instrumentation, genre, or mood, without the need for re-training the underlying models. MusicMagus offers significant flexibility, allowing the system to operate on pre-existing music clips by transforming specific aspects while preserving the remaining aspects of the music.

The use of diffusion models enabled MusicMagus to implement intra-stem editing, where users could alter characteristics within individual musical stems, such as changing the instrumentation from a piano to a guitar. The latent space exploration method allowed for these edits to be made in a zero-shot manner, where new data is not required for training, thus enhancing the scalability of the system for different music styles and contexts.

\highlight{The evaluation of MusicMagus involved both subjective user testing and objective comparisons with baseline models, and it was found that MusicMagus provided a high level of stylistic coherence when modifying musical attributes.} Users especially valued its ability to maintain non-targeted elements, ensuring that the overall feel of the music was preserved even after edits. Additionally, quantitative evaluations showed that MusicMagus outperformed existing models in tasks such as genre and timbre transfer, specifically by maintaining the integrity of the original musical content.

However, limitations emerged in the form of handling more complex, multi-layered edits. For example, while MusicMagus works well on modifying single attributes (like switching an instrument or altering mood), it struggles with inter-stem editing tasks, such as adding a completely new instrument or modifying multiple aspects of a piece simultaneously. Another challenge was the diffusion model’s constraint on the length of the generated audio, which is typically shorter compared to music language models. This means that for long-form compositions, MusicMagus is less applicable, limiting its utility in real-world music production environments.

\subsection{Instruct-MusicGen: Unlocking Text-to-Music Editing for Music Language Models via Instruction Tuning}

Building on the limitations identified in MusicMagus, Instruct-MusicGen was developed to provide a more flexible and precise framework for text-to-music editing. This system introduces instruction tuning to the MusicGen model, enabling it to handle both audio inputs and text-based instructions simultaneously. By incorporating a text fusion module and an audio fusion module, Instruct-MusicGen enhances the precision of music edits, allowing for more complex tasks such as adding, removing, or modifying musical stems.

Unlike the zero-shot approach of MusicMagus, Instruct-MusicGen is fine-tuned using a triplet dataset for editing. This pairing allows for a high level of accuracy in following instructions, a key improvement over earlier models. Moreover, the system’s ability to process audio inputs directly enables it to function effectively in real-world production environments, where users often need to modify pre-existing tracks rather than generate new music from scratch.

Instruct-MusicGen introduced multi-modal control, where users could input a specific musical stem (e.g., vocals or drums) and provide textual instructions to modify it (e.g., ``add reverb to the vocals" or ``remove the drums after 20 seconds"). The dual-modality fusion allows for a seamless integration of text-based and audio-based instructions, thus providing precise control over the modifications. A notable feature is its multi-task editing capability, enabling users to combine multiple edits (such as adding an instrument and adjusting the tempo) within a single interaction.

The performance of Instruct-MusicGen was evaluated using both the Slakh2100 synthetic dataset and the MoisesDB out-of-domain dataset. In comparative evaluations, Instruct-MusicGen demonstrated superior performance over baseline models in terms of both instruction adherence and audio quality. 

However, limitations still exist. One challenge is the signal-level precision during certain edits. For example, when users attempt to modify a specific instrument or section of the music, the model may sometimes over-generalise, affecting nearby sections that were not meant to be altered. Additionally, the reliance on paired data for fine-tuning remains a bottleneck for scalability. Creating large, diverse datasets that cover all possible musical instructions is resource-intensive, limiting the system's applicability to broader, more diverse musical genres.

\section{Limitations}

While the systems developed in this thesis represent significant progress in AI-assisted music creation, several limitations emerged that highlight areas for future improvement. These limitations, observed across Loop Copilot, MusicMagus, and Instruct-MusicGen, point to technical, data-related, and conceptual challenges that must be addressed to push the field further.

\subsection{Precision in Music Editing}

Despite advances in controllability and flexibility, achieving fine-grained precision in music editing remains one of the most critical challenges for these systems. Each system progressively improved on the previous one, yet none fully addressed the need for micro-level editing accuracy.

Loop Copilot demonstrated that using a large language model (LLM) to coordinate multiple music models is effective for high-level, iterative edits. However, the system struggled to interpret complex or highly technical musical instructions, especially when users needed detailed modifications within specific sections of the music. For example, instructions such as ``slightly increase the volume of the high-hats in the second chorus" or ``make the strings sound more legato in the bridge" often resulted in oversimplified edits, where the system could not accurately parse and translate these fine distinctions into actionable changes.

MusicMagus, with its zero-shot text-to-music editing via diffusion models, offered an improvement in the precision of intra-stem edits, allowing users to modify specific musical attributes like instrumentation or mood. However, it struggled with multi-stem edits—tasks that involve adding, removing, or modifying multiple layers of a composition in tandem. The underlying diffusion process is not well-suited for highly localised edits in the time-axis, such as tweaking the timbre of just a few notes within a complex arrangement without affecting the surrounding elements. Moreover, MusicMagus exhibited some instability in maintaining musical coherence during fine-tuned edits, particularly when users attempted to make nuanced changes that affected harmonic or rhythmic structures.

Even Instruct-MusicGen, which introduces a multi-modal fusion approach to process both audio and text, showed limitations in signal-level precision. While it outperformed previous systems in handling more complex instructions, it still faced difficulties when users required granular, localised edits. The system struggled with tasks that demanded high fidelity, such as precisely modifying the dynamics or texture of an instrument within a busy mix, or making subtle adjustments to the timing and articulation of individual notes. These issues are particularly relevant in professional music production, where small changes can significantly impact the final output. The limitations in precision often resulted in edits that were either too broad or inadvertently affected neighboring elements of the composition, reducing the system's overall effectiveness in intricate workflows.

These precision challenges highlight the gap between broad-stroke control over generated music and the detailed, fine-tuned manipulations that human composers and producers typically require. Addressing this gap is critical for future advances in AI-assisted music tools, particularly in sound design, mixing, and mastering workflows, where detailed control is essential.

\subsection{Dependence on Paired Data}

One of the most significant limitations of Instruct-MusicGen and, to some extent, other systems is the reliance on paired datasets for fine-tuning. The development of high-quality paired data—where musical samples are precisely matched with corresponding text instructions—poses a significant challenge, both in terms of the cost and time required to curate such datasets. This limitation hampers the system's ability to scale and generalise across more diverse musical styles, genres, and editing tasks.

Paired data for music editing is relatively scarce compared to other AI domains like image or text generation. The specialised nature of music—where complex instructions must be linked to corresponding changes in musical attributes—means that even large datasets may not cover the full breadth of musical editing needs. For example, instructions like ``make the piano sound more melancholic" or ``add more swing to the drums" require subjective interpretations that are difficult to standardise in datasets.

The cost of creating these datasets further exacerbates the problem. Each instruction must be meticulously crafted and paired with an accurate edit in the corresponding audio sample. Given the variability in how different musicians might interpret similar instructions, the datasets must cover a wide range of possible musical edits. This results in a time-consuming and labor-intensive process that does not scale well.

Additionally, genre diversity is another concern. The systems in this thesis primarily focused on certain types of music (e.g., pop, classical, or jazz), but expanding to more niche or diverse genres (e.g., experimental, ambient, or world music) would require substantially more paired data to account for the unique attributes of each style. For instance, genres like jazz or electronic music might require more nuanced handling of improvisation, polyrhythms, or synthesised textures, which current datasets may not adequately represent.

The heavy reliance on paired data, especially for Instruct-MusicGen, thus limits the system's flexibility and scalability, restricting its utility to a subset of genres and tasks that have been explicitly trained. To make these systems more broadly applicable, new approaches—such as unsupervised learning, self-supervised learning, or few-shot learning—are needed to reduce this dependence, allowing the models to generalise better across diverse musical contexts.

\subsection{Handling Long-Form Music}

Another limitation lies in the systems’ capacity to handle long-form compositions, a critical need in real-world music production. The systems developed in this thesis were primarily designed for generating and editing relatively short music clips (typically within 30 seconds), which limits their applicability in professional environments where full-length tracks or extended compositions are common.

Loop Copilot and Instruct-MusicGen demonstrated the ability to make iterative edits to short loops, but their effectiveness diminishes when applied to longer pieces. This is particularly problematic when editing dynamic transitions between different sections of a song, such as the verse, chorus, and bridge, or when making structural changes to large compositions. Long-form music typically requires the system to handle temporal dependencies over extended periods, ensuring that edits made to one section maintain the musical coherence across the entire piece.

Temporal coherence is a major challenge. Long-form compositions often involve complex relationships between different musical sections, and making an edit to one part of the track (e.g., the introduction) might require corresponding adjustments in later sections (e.g., the outro). The current models are not equipped to manage these dependencies effectively, leading to disjointed edits where certain elements might become disconnected from the overall structure.

Another limitation in handling long-form music is related to memory constraints and the ability to retain contextual information over extended time spans. As these models process audio in shorter segments, they struggle to maintain an overarching understanding of the piece. This results in an inability to preserve the narrative arc or thematic development of a song, which is often key to musical storytelling.

\section{Future Work}

The advances made in Loop Copilot, MusicMagus, and Instruct-MusicGen mark significant progress in the field of AI-assisted music creation. However, several open questions remain that future research must address. Below, I outline four key areas for improvement, drawing from the challenges and insights encountered during this thesis.

\subsubsection{Improving Signal-Level Precision}
One of the persistent limitations across the systems presented in this thesis, particularly Instruct-MusicGen, is the difficulty in achieving signal-level precision. As demonstrated in the discussion, while Instruct-MusicGen can perform edits at the macro level, such as adding or removing entire musical stems, more refined control over micro-level attributes (such as detailed changes in timbre, dynamics, or specific sections of a track) remains a challenge. This is due in part to the limitations in how the system interprets text instructions and translates them into modifications at a granular audio level.

Future work could focus on incorporating techniques from neural audio coding and high-resolution audio models to enhance the system’s ability to make finer, more precise edits. For example, utilising neural audio synthesis for MusicMagus and Instruct-MusicGen could allow the system to manipulate individual audio features (e.g., the vibrato of a specific instrument) without affecting the rest of the composition. This would provide users with much greater control over their creative process, especially in professional contexts where even small changes can dramatically impact the final product.

Another avenue to explore is audio-based diffusion models that work at a higher resolution, allowing for more intricate adjustments. Current models, while effective for broad edits, often simplify or generalise finer details. High-resolution models could enable edits at the level of harmonic overtones, reverb tails, or subtle rhythmic variations, allowing for a new level of precision in music editing with generative models.

This approach could also benefit tasks like instrument isolation or track separation, where signal-level precision is critical for maintaining the quality of the original music while allowing detailed modifications. By improving signal accuracy, these models could be adapted for complex workflows in sound design, mastering, and post-production where every detail matters.

\subsubsection{Leveraging Unsupervised Learning}

A major bottleneck in the scalability of Instruct-MusicGen is the reliance on paired datasets for fine-tuning. Creating datasets where each music sample is paired with a detailed, natural language instruction is labour-intensive and costly, limiting the system's performance. As discussed earlier, this problem could be addressed by exploring unsupervised or weakly supervised learning techniques.

Unsupervised learning methods, such as contrastive learning, could allow the system to learn the relationships between textual instructions and musical changes without needing exact pairs. For instance, by training on larger, uncurated music editing datasets of music and associated metadata (genre, artist, or mood), models could learn how different attributes are represented in the latent space and apply this knowledge for text-guided editing tasks.

Incorporating few-shot learning could further enhance the model's flexibility, enabling it to learn new editing tasks with minimal labelled examples. For example, if the system encounters a new genre or style, it could adapt based on just a handful of examples rather than requiring extensive re-training. This would be particularly beneficial for tasks like style transfer, where there may be little or no existing data for specific combinations of genres or attributes. By leveraging the wealth of unlabelled music data available online, the system could reduce its dependence on specialised paired datasets, significantly broadening its applicability.

Moreover, self-supervised techniques could be employed to improve the model’s ability to understand more abstract or creative instructions, such as ``make this track more aggressive" or ``add a cinematic feel." These types of instructions are difficult to capture in a traditional paired dataset but could be learned through unsupervised methods that focus on capturing broader musical transformations.

\subsubsection{Scaling to Long-Form Compositions}
One of the key limitations noted across all the systems, particularly MusicMagus, is the handling of long-form compositions. While these models are capable of generating and editing short music clips (usually around 10-20 seconds), they struggle with larger, more complex musical structures that require edits over a longer timescale. This limitation is especially critical in real-world music production, where full-length tracks and compositions are the norm.

To address this, future research should explore hierarchical models that operate at multiple levels of abstraction. Such models would allow for edits at both the micro-level (e.g., tweaking a single note or instrument) and the macro-level (e.g., changing the structure of an entire piece or modifying transitions between sections). By structuring the model to handle both short-term dependencies (e.g., melodic changes) and long-term dependencies (e.g., large-scale form), it could manage longer, more complex compositions without losing coherence.

Additionally, techniques like memory-augmented networks~\citep{rag-audioldm, rag-music} could be introduced to ensure that the system maintains a sense of musical continuity over time. These networks could store and recall information about earlier sections of the piece, enabling edits that are contextually consistent even in longer compositions. This would be invaluable for film scoring, album production, or any creative work that involves intricate, evolving musical themes.

Another promising approach is to integrate models that specialise in temporal structures of music, such as models that capture rhythmic patterns, harmonic progressions, and dynamic shifts over time. This would enable not only better handling of long-form compositions but also more sophisticated edits that align with the larger musical narrative.

\subsubsection{Bridging the Interpretation Gap Between LLMs and Music Models}
The discussion highlighted the interpretation gap between LLMs and music models, which poses a challenge for systems like Loop Copilot and Instruct-MusicGen. While LLMs are highly effective at processing natural language instructions, they often struggle to accurately interpret and translate complex musical requests into actionable edits. For example, instructions like ``soften the transition between these two sections" or ``add a more jazzy feel to the bassline" may be ambiguous to the LLM and lead to incorrect or overly generalised edits~\citep{interpretation-gap}.

To address this, future research could focus on developing joint embeddings for text and music, which would allow the LLM and the music model to share a common understanding of musical instructions. By embedding both the textual instructions and the musical content into a shared latent space, the system could more accurately match the user’s intent with the corresponding musical transformation.

Another avenue for research is the use of multi-modal training techniques, where the LLM and music models are trained together to understand the nuances of music-related language. This could involve training on large datasets where musical edits are paired with corresponding natural language descriptions, allowing the system to learn the context-specific meaning of terms like ``groove," ``texture," or ``swing." Bridging this gap would lead to more intuitive and accurate interactions, making the system more accessible for musicians and producers who may not have technical expertise but want to control their creative output in detailed ways.

Additionally, feedback mechanisms could be integrated into the system, allowing users to refine their instructions iteratively. For instance, if the initial edit is not satisfactory, the user could provide feedback (e.g., ``make the guitar even softer") and the system would adjust accordingly. This type of iterative refinement process would mimic a human creative workflow, improving the alignment between the user’s intent and the system’s output.

\section{Broader implications}

The systems developed in this thesis have the potential to reshape the landscape of AI-assisted music creation. By combining large language models (LLMs) with specialised music generation models, this work opens up new possibilities for human-AI collaboration in creative domains, offering tools that allow users to interact with music in dynamic, iterative, and flexible ways.

\subsection{Expanding Creative Possibilities in Music}

\highlight{The advances made in text-guided music editing through systems like Loop Copilot and Instruct-MusicGen demonstrate that AI can go beyond simply generating music—it can act as a collaborative partner in the creative process. This shift from generation to collaboration has profound implications for the music production industry, enabling artists and producers to engage with music in new and innovative ways.}

\highlight{For example, music producers could explore using these systems to refine tracks in real time, experimenting with different arrangements or instrumentation without requiring deep technical expertise in music theory or production software. Similarly, songwriters could experiment with these tools to quickly prototype ideas, testing different lyrical or melodic concepts through simple text instructions. However, as these systems are still in their early stages, it is crucial to recognise that scaling and engineering them for real-world use remains a significant challenge. These tools, along with similar future developments, may eventually lower the barrier to entry for music editing, but much work is needed to realise their full potential for professional-level music production.}

\subsection{Implications for Other Creative Fields}
The methods developed here for text-to-music editing could inspire similar innovations in other creative fields. For instance, in film scoring, a director or composer could interactively shape the score by providing textual feedback, adjusting the emotional tone or timing of musical cues to match the narrative. Similarly, in game audio design, dynamic music systems could be fine-tuned through simple text commands to adapt the soundtrack in real time based on player actions or in-game events.

Live performance enhancement is another area with strong potential. AI-driven music models could be used to dynamically alter or generate music during live performances, allowing musicians to improvise together with the system. By interacting with the model through natural language or other inputs (e.g., gestures), performers could seamlessly adapt their performance to the audience or the mood of the event.




\subsection{Ethical Considerations and AI in Creative Fields}

The deployment of AI in music and other creative fields raises three major concerns: data copyright issues, questions about authorship, and the potential disruption of creative industries. 

First, data copyright issues are important to the ethical concerns surrounding AI in creative fields. Many AI systems rely on vast datasets, often containing copyrighted works, for training purposes. The use of these works without proper compensation or recognition raises significant concerns. In the music industry, for instance, many artists oppose the inclusion of their work in these datasets, as it infringes on their intellectual property rights and typically occurs without any form of reward~\footnote{AI chief quits over 'exploitative' copyright row. \url{https://www.bbc.co.uk/news/technology-67446000}}. Although some companies have started paying for data to mitigate legal risks, this practice does not fully address the broader ethical concerns. A comprehensive framework is needed to ensure that the rights of original creators are respected in the context of AI training.

Second, the issue of authorship becomes more complicated when AI systems contribute to the creation or editing of works. If an AI system plays a role in generating a musical piece, how should the contributions of both the human creator and the AI be recognised? The increasing autonomy of AI systems challenges traditional notions of creativity and ownership. Clear guidelines must be established to delineate human and AI contributions and to protect the rights of original creators, ensuring that both are fairly acknowledged in collaborative works involving AI.

Finally, the potential disruption of creative industries by AI must also be carefully managed. While AI systems can democratise access to professional-level tools, they also pose risks by automating tasks traditionally performed by human professionals. This disruption is evident in the music industry, where AI systems are already being used to generate music, raising concerns about job displacement. Moreover, the reliance on copyrighted material in training models has already led to legal challenges, such as lawsuits from major record labels~\footnote{Sony, Universal, Warner sue over AI music copyright violations. \url{https://www.bbc.co.uk/news/articles/ckrrr8yelzvo}}. The creative sector must find ways to integrate AI that complement rather than replace human creativity, ensuring that the contributions of human creators remain valued and protected.

\subsection{The Future of Human-AI Collaboration}

Looking ahead, the advances made in text-guided music editing set the stage for more seamless human-AI collaboration in creative processes. As models become more capable of understanding and executing complex instructions, their role in the creative process will evolve from simple assistants to co-creators. In this future, AI systems could serve as dynamic, adaptive partners, working alongside human artists to explore new creative directions, experiment with unconventional techniques, or streamline the iterative refinement process.

In the context of music, this could mean that composers and producers no longer need to rely solely on traditional digital audio workstations (DAWs) for editing tasks but can engage in an interactive dialogue with AI systems to achieve their artistic vision. Similarly, across other domains, such as fashion, architecture, or cinema, AI systems could provide new ways of exploring creative possibilities by interpreting high-level goals and iteratively refining designs based on human feedback.

Ultimately, systems such as 
 the ones developed in this thesis represent a step forward in the ongoing development of AI-assisted creativity. By enhancing the controllability and editability of music generation models, these systems offer a glimpse into the future of creative workflows, where human and AI creativity come together to unlock new era of artistic expression. As AI continues to evolve, it may play an increasingly integral role in the creative industries, shaping the way we think about and engage with art, music, and design in the years to come.

\appendix

\chapter{Appendix A: Loop Copilot}
\label{ch:appendix}

\section{SUS questionnaire}

\begin{enumerate}
    \item I think that I would like to use this system frequently.
    \item I found the system unnecessarily complex.
    \item I thought the system was easy to use.
    \item I think that I would need the support of a technical person to be able to use this system.
    \item I found the various functions in this system were well integrated.
    \item I thought there was too much inconsistency in this system.
    \item I would imagine that most people would learn to use this system very quickly.
    \item I found the system very cumbersome to use.
    \item I felt very confident using the system.
    \item I needed to learn a lot of things before I could get going with this system.
\end{enumerate}

\section{TAM questionnaire}

\begin{enumerate}
    \item I find Loop Copilot useful in music creation.
    \item Using Loop Copilot improves my experience in music creation.
    \item Loop Copilot enables me to accomplish tasks more quickly.
    \item I find that Loop Copilot increases my productivity in music creation.
    \item I find Loop Copilot easy to use.
    \item Learning to operate Loop Copilot is easy for me.
    \item I find it easy to get Loop Copilot to do what I want it to do.
    \item I find the interface of Loop Copilot to be clear and understandable.
    \item Given the chance, I intend to use Loop Copilot.
    \item I predict that I would use Loop Copilot in the future.
    \item I plan to use Loop Copilot frequently.
\end{enumerate}

\section{ChatGPT Prompts}

\begin{center}
    \centering
    \begin{longtable}{p{0.24\linewidth}|p{0.75\linewidth}}
    \toprule
    Tool & Prompt \\ 
    \midrule\endfirsthead
    \toprule
    Tool & Prompt \\ 
    \midrule\endhead
      System prefix & Loop Copilot is designed to be able to assist with a wide range of text and music related tasks, from answering simple questions to providing in-depth explanations and discussions on a wide range of topics. Loop Copilot is able to generate human-like text based on the input it receives, allowing it to engage in natural-sounding conversations and provide responses that are coherent and relevant to the topic at hand.

    Loop Copilot is able to process and understand large amounts of text and music. As a language model, Loop Copilot can not directly read music, but it has a list of tools to finish different music tasks. Each music will have a file name formed as ``music/xxx.wav", and Loop Copilot can invoke different tools to indirectly understand music. When talking about music, Loop Copilot is very strict to the file name and will never fabricate nonexistent files. 
    
    Loop Copilot is able to use tools in a sequence, and is loyal to the tool observation outputs rather than faking the music content and music file name. It will remember to provide the file name from the last tool observation, if a new music is generated.
    
    Human may provide new music to Loop Copilot with a description. The description helps Loop Copilot to understand this music, but Loop Copilot should use tools to finish following tasks, rather than directly imagine from the description.
    
    Overall, Loop Copilot is a powerful music dialogue assistant tool that can help with a wide range of tasks and provide valuable insights and information on a wide range of topics.

    TOOLS:
    ------
    
    Loop Copilot has access to the following tools: \\
\midrule
    System format& To use a tool, you MUST use the following format:

\texttt{\textbf{Thought}: Do I need to use a tool? Yes}

\texttt{\textbf{Action}: the action to take, should be one of [\{tool\_names\}]}

\texttt{\textbf{Action Input}: the input to the action}

\texttt{\textbf{Observation}: the result of the action}

When you have a response to say to the Human, or if you do not need to use a tool, you MUST use the format:

\texttt{\textbf{Thought}: Do I need to use a tool? No}

\texttt{\{ai\_prefix\}\: [your response here]}\\
\midrule
System suffix & You are very strict to the filename correctness and will never fake a file name if it does not exist.
You will remember to provide the music file name loyally if it is provided in the last tool observation.

Begin!

Previous conversation history:
\{chat\_history\}

Since Loop Copilot is a text language model, Loop Copilot must use tools to observe music rather than imagination. The thoughts and observations are only visible for Loop Copilot.

\texttt{\textbf{New input}: \{input\}}

\texttt{\textbf{Thought}: Do I need to use a tool? \{agent\_scratchpad\}}

You MUST strictly follow the format. \\
\midrule
\midrule
    Text to music & 
    \textbf{Name}: Generate music from user input text.
    
    \textbf{Description}: useful if you want to generate music from a user input text and save it to a file. like: generate music of love pop song, or generate music with piano and violin.
    
    The input to this tool should be a string, representing the text used to generate music.\\
    \midrule
    Drum pattern to music & 
    \textbf{Name}: Generate music from user input text based on the drum audio file provided.

    \textbf{Description}: useful if you want to generate music from a user input text and a previous given drum audio file. like: generate a pop song based on the provided drum pattern above.
    
    The input to this tool should be a comma separated string of two, representing the music\_filename and the text description.
    \\
    \midrule
    Music Imitation & 
    
    \textbf{Name}: Generate music from user input when the input is a title of music.

    \textbf{Description}: useful if you want to generate music which is silimar  and save it to a file. like: generate music of love pop song, or generate music with piano and violin.

    The input to this tool should be a comma separated string of two, representing the text description and the title.
    \\
    
    \midrule
    Stylistic rearrangement &  
    \textbf{Name}: Generate a new music arrangement with text indicating new style and previous music.

    \textbf{Description}: useful if you want to style transfer or rearrange music with a user input text describing the target style and the previous music. 
    
    Please use Text2MusicWithDrum instead if the condition is a single drum track. You shall not use it when no previous music file in the history. like: remix the given melody with text description, or doing style transfer as text described from previous music.
    
    The input to this tool should be a comma separated string of two, representing the music\_filename and the text description.\\
    \midrule
    Music variation generation & 
    \textbf{Name}: Generate a variation of given music.

    \textbf{Description}: useful if you want to generate a variation of music, or re-generate the entire music track. like: re-generate this music, or, generate a variant.
    
    The input to this tool should be a single string, representing the music\_filename.\\
    \midrule
    Add a track &  
    \textbf{Name}: Add a new track to the given music loop.

    \textbf{Description}: useful if you want to add a new track (usually add a new instrument) to the given music. like: add a saxophone to the given music, or add piano arrangement to the given music.
    
    The input to this tool should be a comma separated string of two, representing the music\_filename and the text description.\\
    \midrule
    Remove a track &  
    \textbf{Name}: Separate one track from a music file to extract (return the single track) or remove (return the mixture of the rest tracks) it.

    \textbf{Description}: useful if you want to separate a track (must be one of 'vocals', `drums', `bass', `guitar', `piano' or `other') from a music file. Like: separate vocals from a music file, or remove the drum track from a music file.
    
    The input to this tool should be a comma separated string of three params, representing the music\_filename, the specific track name, and the mode (must be `extract' or `remove').\\
    \midrule
    Re-generation/inpainting &  
    \textbf{Name}: Inpaint a specific time region of the given music.

    \textbf{Description}: useful if you want to inpaint or regenerate a specific region (must with explicit time start and ending) of music. like: re-generate the 3s-5s part of this music.
    
    The input to this tool should be a comma separated string of three, representing the music\_filename, the start time (in second), and the end time (in second).\\
    \midrule
    Add sound effects &  
    \textbf{Name}: Add a single sound effect to the given music. 

    \textbf{Description}: useful if you want to add a single sound effect, like reverb, high pass filter or chorus to the given music. like: add a reverb of recording studio to this music.
    
    The input to this tool should be a comma separated string of two, representing the music\_filename and the original user message.\\ 
    \midrule

    

    
    Pitch Shifting & 
    \textbf{Name}: Shift the pitch of the given music.

    \textbf{Description}: useful if you want to shift the pitch of a music. Like: shift the pitch of this music by 3 semitones.
    
    The input to this tool should be a comma separated string of two, representing the music\_filename and the pitch shift value.\\ 
    \midrule
    Speed Changing & 
    \textbf{Name}: Stretch the time of the given music.

    \textbf{Description}: useful if you want to stretch the time of a music. Like: stretch the time of this music by 1.5.
                    
    The input to this tool should be a comma separated string of two, representing the music\_filename and the time stretch value.\\ 
    \midrule
    Music captioning &  
    \textbf{Name}: Describe the current music.

    \textbf{Description}: useful if you want to describe a music. Like: describe the current music, or what is the current music sounds like.
    
    The input to this tool should be the music\_filename.\\ 

\bottomrule
\caption{List of system principles and task prompts. Each task features a unique name, description, and input parameter format for guiding the LLM.}
\label{tab:prompt}\\
    \end{longtable}

\end{center}

\bibliographystyle{plainnat}
\bibliography{refs}

\end{document}